\documentclass[a4paper,12pt]{article}
\usepackage{jheppub}

\usepackage[applemac]{inputenc}
\usepackage{amsmath}
\usepackage{slashed}
\usepackage{amscd,amstext,amsbsy,amsopn,amsthm,amsxtra,upref,amsfonts,amssymb,euscript,latexsym,graphicx,color}
\usepackage[dvipsnames]{xcolor}
\usepackage[all]{xy}
\usepackage[american]{babel}
\usepackage{multicol}
\usepackage{tikz}
\usepackage{tikz-3dplot}
\usepackage{mdframed}
\usepackage{todonotes}
\usetikzlibrary{patterns}
\usepackage{hyperref,pgfplots,mdframed,subcaption}
\usepackage{color}
\usepackage[normalem]{ulem} 

\usepackage{soul}

\usepackage{enumitem}

\usepackage{amsmath,array,cleveref,amsfonts,subcaption,graphicx,wrapfig,lscape,float,multirow,longtable,rotating,epstopdf}
\usepackage{bbm,youngtab,bbold}

\def\D{\Delta}
\def\G{\Gamma}

\tikzstyle{white}=[circle,draw,
inner sep=0pt,minimum size=2mm]
\tikzstyle{black}=[circle,fill=black,inner sep=0pt,minimum size=2mm]

\def\beqa{\begin{eqnarray}}
\def\eeqa{\end{eqnarray}}

\newcommand{\ov}[1]{\overline{#1}}
\newcommand\undermat[2]{%
  \makebox[0pt][l]{$\smash{\underbrace{\phantom{%
    \begin{matrix}#2\end{matrix}}}_{\text{\normalsize $#1$}}}$}#2}

\newcommand{\be}{\begin{equation}}
\newcommand{\ee}{\end{equation}}
\newcommand{\beq}{\begin{equation}}
\newcommand{\beql}[1]{\begin{equation}\label{#1}}
\newcommand{\eeq}{\end{equation}}
\newcommand{\ba}{\begin{array}}
\newcommand{\ea}{\end{array}}
\newcommand{\bea}{\begin{eqnarray}}
\newcommand{\beal}[1]{\begin{eqnarray}\label{#1}}
\newcommand{\eea}{\end{eqnarray}}
\newcommand{\ben}{\begin{enumerate}}
\newcommand{\een}{\end{enumerate}}
\newcommand{\bean}{\begin{eqnarray*}}
\newcommand{\eean}{\end{eqnarray*}}
\newcommand{\eref}[1]{(\ref{#1})}
\newcommand{\sref}[1]{\S\ref{#1}}

\newcommand{\fref}[1]{Figure \ref{#1}}
\newcommand{\btab}[1]{\begin{tabular}{#1}}
\newcommand{\etab}{\end{tabular}}

\newcommand{\comment}[1]{}

\def\a{\alpha}

\def\g{\gamma}
\def\G{\Gamma}
\def\D{\Delta}

\def\L{\Lambda}

\newcommand{\drawsquare}[2]{\hbox{%
\rule{#2pt}{#1pt}\hskip-#2pt
\rule{#1pt}{#2pt}\hskip-#1pt
\rule[#1pt]{#1pt}{#2pt}}\rule[#1pt]{#2pt}{#2pt}\hskip-#2pt
\rule{#2pt}{#1pt}}

\newcommand{\symm}{~\raisebox{-.5pt}{\drawsquare{6.5}{0.4}}\hskip-0.4pt%
        \raisebox{-.5pt}{\drawsquare{6.5}{0.4}}~}

\newcommand{\asymm}{~\raisebox{-3.5pt}{\drawsquare{6.5}{0.4}}\hskip-6.9pt%
        \raisebox{3pt}{\drawsquare{6.5}{0.4}}~}

\title{Dimers, Orientifolds and Anomalies}

\author[a]{Riccardo Argurio}
\author[b]{Matteo Bertolini}
\author[c,d,e]{Sebasti\'an Franco}
\author[a]{Eduardo Garc\'{\i}a-Valdecasas}
\author[b]{Shani Meynet}
\author[a]{Antoine Pasternak}
\author[f]{Valdo Tatitscheff}

\affiliation[a]{Physique Th\'eorique et Math\'ematique and International Solvay Institutes \\ Universit\'e Libre de Bruxelles; C.P. 231, 1050 Brussels, Belgium}

\affiliation[b]{SISSA and INFN, Via Bonomea 265; I 34136 Trieste, Italy}

\affiliation[c]{Physics Department, The City College of the CUNY \\ 160 Convent Avenue, New York, NY 10031, USA}

\affiliation[d]{Physics Program and $^e$Initiative for the Theoretical Sciences \\
The Graduate School and University Center, The City University of New York  \\
365 Fifth Avenue, New York NY 10016, USA}

\affiliation[f]{IRMA, UMR 7501, Universit\'e de Strasbourg et CNRS \\ 
7 rue Ren\'e Descartes 67000 Strasbourg, France}

\emailAdd{rargurio@ulb.ac.be, bertmat@sissa.it, sfranco@ccny.cuny.edu, eduardo.garcia.valdecasas@gmail.com, smeynet@sissa.it, antoine.pasternak@ulb.ac.be, valdotatitscheff@gmail.com}

\abstract{We study $4d$ $\mathcal{N}=1$ gauge theories engineered via D-branes at orientifolds of toric singularities, where gauge anomalies are cancelled without the introduction of non-compact flavor branes. Using dimer model techniques, we derive geometric criteria for establishing whether a given singularity can admit anomaly-free D-brane configurations purely based on its toric data and the type of orientifold projection. 
Our results therefore extend the dictionary between geometric properties of singularities and physical properties of the corresponding gauge theories.
}

\begin{document}

\maketitle


\newpage 

\section{Introduction}

D-branes at singularities provide an ideal setup for engineering interesting gauge theories in string theory. The multiple applications of such setups include: non-trivial generalizations of the original AdS/CFT correspondence \cite{Klebanov:1998hh,Klebanov:1999rd,Klebanov:2000nc,Klebanov:2000hb,Gauntlett:2004yd,Bertolini:2004xf,Benvenuti:2004dy,Cvetic:2005ft,Franco:2005sm,Butti:2005sw}, local string phenomenology \cite{Aldazabal:2000sa,Berenstein:2001nk,Verlinde:2005jr,Buican:2006sn,Malyshev:2007zz} and new perspectives, often geometric, on gauge theory dynamics and dualities \cite{Feng:2000mi,Feng:2001xr,Beasley:2001zp,Feng:2001bn,Feng:2002zw,Berenstein:2002fi}.

The correspondence between geometry and gauge theory is particularly well understood in the case of $4d$ $\mathcal{N}=1$ gauge theories on D3-branes probing toric Calabi-Yau (CY) 3-folds, for which the map is significantly streamlined by brane tilings (equivalently known as dimer models) \cite{Hanany:2005ve,Franco:2005rj,Franco:2005sm}.

Orientifolds \cite{Sagnotti:1987tw,Horava:1989vt,Dai:1989ua} of such singularities are extremely interesting for a variety of reasons. Among them, they expand the possible spectrum \cite{Gimon:1996rq,Douglas:1996sw,Franco:2007ii} (gauge groups and matter fields representations), break conformal invariance \cite{Argurio:2017upa}, play an important role in models with non-perturbative effects due to D-brane instantons \cite{Argurio:2007vqa,Bianchi:2007wy,Blumenhagen:2009qh} and are a key ingredient in certain models of phenomenological interest, including ones leading to metastable dynamical supersymmetry (SUSY) breaking \cite{Argurio:2007qk}.

Remarkably, when gauge theories are realized on D-branes at singularities, their dynamics and properties can often be connected to general properties of the underlying geometry. Salient examples include the links between: confinement and complex deformations \cite{Klebanov:2000hb}, runaway dynamical supersymmetry breaking and the obstruction of complex deformations \cite{Berenstein:2005xa,Franco:2005zu,Bertolini:2005di} and, more recently, newly found instabilities of otherwise SUSY breaking vacua and the presence of non-isolated singularities \cite{Buratti:2018onj,Argurio:2019eqb}. 

The ranks of the gauge groups on theories realized at D-branes at a singularity correspond to numbers of (wrapped) D-branes in the configuration. Roughly speaking, the cancellation of local anomalies in the gauge theories correspond to the cancellation of tadpoles in the string construction. When orientifolds are included, in some cases anomaly cancellation can only be achieved upon the addition of non-compact flavor D7-branes (see e.g.~\cite{Franco:2007ii,Bianchi:2013gka,Franco:2015kfa,Argurio:2017upa}), which give rise to (anti)fundamental matter, but there are instances where it is possible to cancel the anomalies even in the absence of extra flavors \cite{Franco:2007ii}.

This paper is devoted to the study of anomalies in gauge theories coming from D-branes at orientifolds of toric singularities, in the {\it absence} of flavor branes. We will introduce a new geometric algorithm for constructing anomaly-free theories and identify geometric criteria for the existence of such solutions. 
Remarkably, our results allow us to determine whether an orientifold singularity can admit anomaly-free D-brane gauge theories just by analyzing its geometric structure and avoiding any case-by-case analysis, which has been so far the only known approach for this class of theories. This geometric criterion is therefore a new addition to the list of connections between the geometry of singularities and general properties of the resulting gauge theories, some of which were mentioned above.

Our presentation is organized as follows. In \sref{Sec:DimerIntro} we review the dimer model description of gauge theories on D-branes at toric singularities and the role of zig-zag paths in solving anomaly cancellation conditions. In \sref{Sec:ACC}, we consider the more involved case of orientifolds, for which anomaly cancellation conditions generically correspond to non-homogeneous linear systems of equations due to the presence of tensor matter. In \sref{Sec:Algorithm}, we generalize the algorithm for solving anomaly cancellation conditions based on zig-zag paths to the case of orientifolds. This analysis will lead to the main result of the paper, which we present in \sref{Sec:Anomalies}: a purely geometric criterion for anomaly cancellation conditions in orientifold field theories just based on the toric data of the singularity. \sref{Sec:Conc} contains a summary of our results and an outlook.

\section{Dimers and Anomalies}\label{Sec:DimerIntro}

The $4d$ $\mathcal{N}=1$ gauge theories living on the worldvolume of D3-branes probing toric CY 3-folds are fully captured by bipartite graphs on $\mathbb{T}^2$ called {\it brane tilings} or {\it dimer models}, two terms we will use interchangeably.\footnote{More precisely, for every toric CY 3-fold there is at least one {\it toric phase}, namely a theory described by a dimer model. In general, Seiberg duality relates this theory to other phases, which are not describable by dimer models.} Brane tilings are indeed physical configurations of D5-branes suspended from an NS5-brane, which are connected to the D3-branes at the CY$_3$ singularity by T-duality. In this section we present a brief summary of some basic notions of D3-branes at toric singularities and their dimer model description. The versed reader may safely skip it. For thorough presentations, see e.g. \cite{Franco:2005rj,Franco:2005sm}. 

The map connecting brane tilings and the corresponding gauge theories is summarized in the dictionary presented in \Cref{Tab:Dict}.

\begin{table}
	\centering
	\begin{tabular}{|p{3.5cm}|p{9cm}|}
	\hline 
		\ \ \ {\bf Brane Tiling} &\ \ \ \ \ \ {\bf Quiver Gauge Theory} \\
		\hline \hline
		Face & $U(N_i)$ gauge factor \\ \hline
		Edge between faces $i$ and $j$ & Chiral superfield in the bifundamental representation of groups $i$ and $j$ (adjoint representation if $i = j$). The chirality, i.e.~orientation, of the bifundamental is such that it goes clockwise around black nodes and counter-clockwise around white nodes. \\ \hline
$k$-valent node & Superpotential term made of $k$ chiral superfields. Its sign is $+/-$ for a white/black node, respectively. \\ \hline
	\end{tabular}
\captionof{table}{Dictionary relating brane tilings to quiver gauge theories.\label{Tab:Dict}}
\end{table}	

\begin{figure}[h!]
	\centering
	\begin{subfigure}[t]{0.28\textwidth }
		\begin{center} 
			\includegraphics[width=\textwidth]{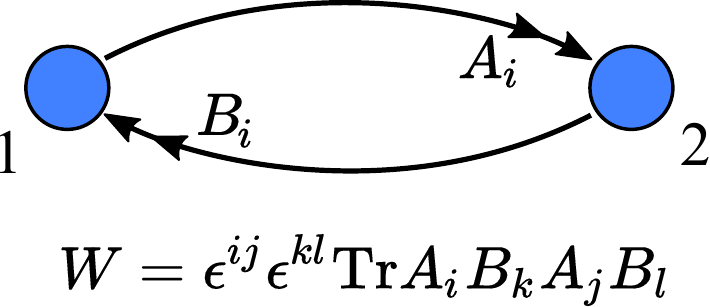}
			\caption{}
			\label{Fig:ConifoldQuiver}
		\end{center}
	\end{subfigure} \hspace{15mm}
	\begin{subfigure}[t]{0.30\textwidth } 
		\begin{center} 
			\includegraphics[width=\textwidth]{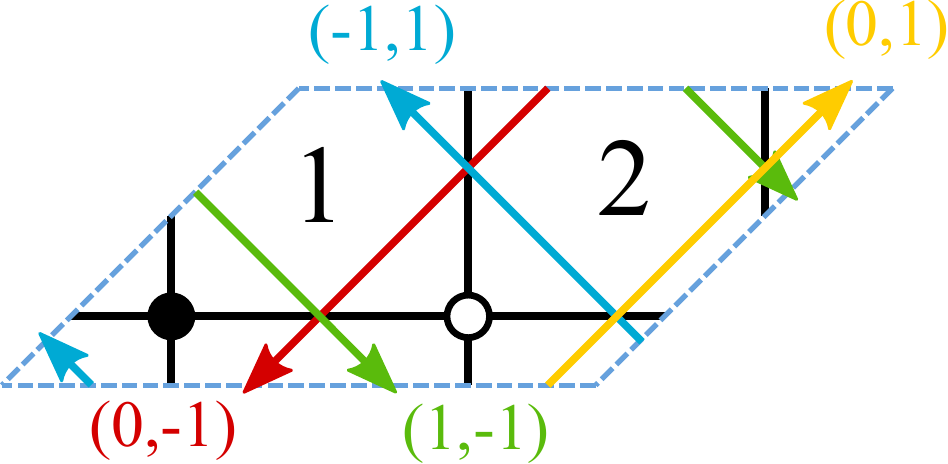}
			\caption{}
			\label{Fig:ConifoldZigZags}
		\end{center}
	\end{subfigure} \hspace{15mm}
	\begin{subfigure}[t]{0.18\textwidth } 
		\begin{center} 
			\includegraphics[width=\textwidth]{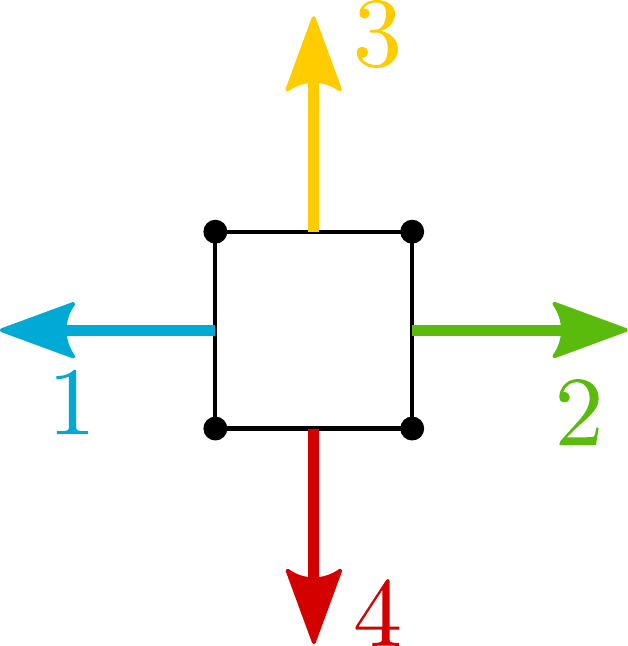}
			\caption{}
			\label{Fig:ToricDiagramConifold}
		\end{center}
	\end{subfigure}
	\caption{Conifold: (a) quiver diagram and superpotential, (b) dimer diagram with ZZPs and their windings, and (c) toric diagram with outward normals corresponding to ZZPs up to an $SL(2,\mathbb{Z})$ transformation. }
	\label{Fig:Conifold} 
\end{figure}

The gauge theory on D3-branes probing the conifold \cite{Klebanov:1998hh} is a prototypical example. This theory has gauge group $U(N_1) \times U(N_2)$, four bifundamental chiral fields $A_i = (\ov\square_1,\square_2)$ and $B_i = (\square_1, \ov\square_2) $, $i=1,2$, and superpotential $W = \epsilon^{ij}\epsilon^{kl} \text{Tr} A_i B_k A_j B_l$. The quiver, dimer and toric diagrams for the conifold are shown in \fref{Fig:Conifold}.

Brane tilings substantially streamline the connection between the probed toric CY 3-folds and the corresponding quiver theories thanks to powerful combinatorial tools, such as {\it perfect matchings} and {\it zig-zag paths} (ZZPs) \cite{Franco:2005rj,Hanany:2005ss,Feng:2005gw,Franco:2006gc}. We will be particularly interested in ZZPs, which are oriented paths on the dimer that go along edges and turn maximally right (resp. left) when they meet a white (resp. black) node. The ZZPs of the conifold dimer are shown in \Cref{Fig:ConifoldZigZags}. These paths close forming loops with no self-intersections that have non-zero homology around the $\mathbb{T}^2$ in which the diagram is embedded. The two holonomies are called mesonic charges, and are associated to two of the three $U(1)$ toric actions of the toric CY 3-fold, the remaining one being the $U(1)_R$ R-symmetry. Remarkably, ZZPs are in one-to-one correspondence with legs in the $(p,q)$ web diagram \cite{Hanany:2005ss,Feng:2005gw} (equivalently, the outward pointing vectors normal to the external edges of the toric diagram), where their $(p,q)$ labels are exactly the homology charges of the ZZPs. 

The ranks $N_i$ of the gauge groups associated to faces in the dimer reflect the configuration of branes at the singularity. These branes include both regular and fractional D3-branes. The latter correspond to higher dimensional branes wrapped on vanishing compact cycles \cite{Diaconescu:1997br}. D-branes source RR tadpoles that must be canceled, this being the geometric counterpart of anomaly cancellation  in the dual gauge theory. Tadpole cancellation amounts to cancelling the flux sourced by the branes in compact homology. The configuration $N_i = N$ corresponds to having $N$ regular D3-branes and no fractional branes. This configuration is always tadpole free since the CY is non-compact and the RR flux sourced by regular D3-branes can escape all the way to infinity. Configurations with unequal ranks are obtained adding fractional branes. The number of independent, anomaly-free fractional brane configurations is in one-to-one correspondence with the number of compact 2-cycles whose Hodge duals in the CY are non-compact 4-cycles. Indeed, D5 branes wrapped on these (and only these) 2-cycles are tadpole free since the RR flux can again escape to infinity. The number of non-compact 4-cycles, hence of consistent fractional branes, can be related to the number of ZZPs \cite{Butti:2006hc},
\begin{equation}
\#\text{ZZPs} - 3 = \# \text{Fractional Branes} \ . \label{Eq:NumberOfBranes}
\end{equation}
Note, in passing, that tadpole cancellation is equivalent to cancellation of non-abelian gauge anomalies even formally for gauge groups with zero rank \cite{Uranga:2000xp}. 

In this work we will use anomaly cancellation conditions 
 (ACC) and tadpole cancellation interchangeably. In fact, some $U(1)$ factors can be anomalous and become massive through a generalization of the Green-Schwarz mechanism. Even some of the non-anomalous $U(1)$'s can become massive, see \cite{Ibanez:1998qp}.
In the following, we will only consider the anomaly of the non-abelian part of the gauge groups. Henceforth we will often denote them as just $SU(N)$.

Cancellation of anomalies at a given node of the quiver corresponds to having the same number of incoming and outgoing arrows (weighted by the ranks of the nodes at their other endpoints). This is encoded in the (antisymmetric) matrix $A$ defined as $A = \tilde A - \tilde A^T$, where $\tilde A$ is the adjacency matrix of the quiver. The latter is a matrix whose elements $\tilde A_{ij}$ count the number of bifundamental chiral superfields $(\overline\square_i,\square_j)$ charged under the gauge group $i$ and the gauge group $j$. With an abuse of language in the following we will call $A$ the adjacency matrix, for simplicity. The matrix $A$ is only sensitive to the chiral content of the theories (e.g. it is zero for the conifold). Cancellation of anomalies amounts to solving the homogeneous system of equations defined by this matrix, that is, finding Ker$(A)$. Vectors in Ker$(A)$ are the ranks associated to regular and fractional branes. In this paper we will investigate how these conditions are modified once orientifold planes are introduced.

\subsection{Anomalies and Zig-Zag Paths}\label{construction}

Throughout this paper, we will find ZZPs to be particularly useful for our purposes. With this in mind, we now review a method for finding anomaly-free rank assignments of dimers based on ZZPs \cite{Butti:2006hc}.

\begin{figure}[h!]
	\centering
	\includegraphics[scale=0.3]{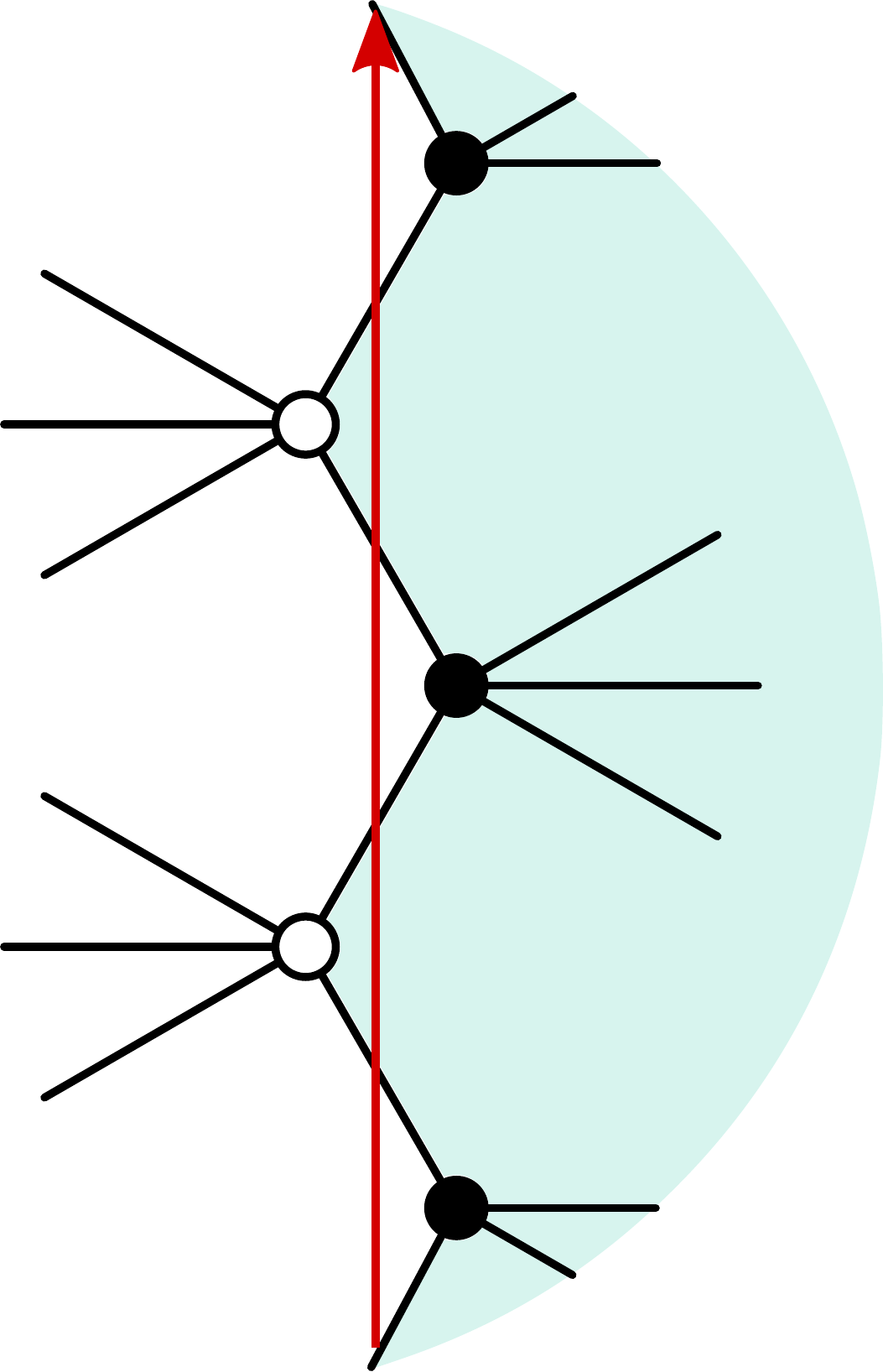}
	\caption{A ZZP as an anomaly wall.}
	\label{Fig:ZZPwall}
\end{figure}

We can regard every ZZP as defining an ``anomaly-free wall" on the dimer. This is because, due to its definition, every time a ZZP overlaps with a face in the dimer, it does so over exactly a pair of consecutive edges.\footnote{By overlapping with a face, we mean sharing an edge with it, not just touching it at a node.} These two consecutive edges correspond to an incoming and an outgoing arrow in the quiver for the gauge group associated to the face under consideration.\footnote{More generally, a ZZP might overlap with a given face more than once. Every overlap involves a pair of consecutive edges, so the previous discussion still applies. For an explicit example of this situation, see the yellow ZZP in the $dP_1$ dimer model shown in \fref{Fig:DimerdP1ZigZag2_0}.} 
This implies that  if we add some constant to the ranks of all the faces on one side of the ZZP, the ACC of the faces on the other side of the ZZP do not change, as illustrated in \Cref{Fig:ZZPwall}.

With this insight, one recovers the algorithm to construct anomaly-free rank assignments for dimer models outlined in \cite{Butti:2006hc}:
\begin{enumerate}
	\item The set of ZZPs is given by $\{(p_\G,q_\G) \vert \G=1,\dots, n \}$, where $p_\G$ and $q_\G$ are the winding numbers of the ZZP $\G$, with respect to a fixed unit cell. To every $(p_\G,q_\G)$ assign an integer coefficient $v_\G$.
	\item Choose one face and assign rank zero to it. 
	\item In going from face $a$ to an adjacent face $b$, the rank of the latter will be 
	\begin{equation}
	N_b=N_a+v_\G-v_\D \, , 
	\end{equation}
	where $v_\G$ is the coefficient of the ZZP moving to the left with respect to the path from $a$ to $b$, and $v_\D$ is the one in the opposite direction. 
	This operation is well defined since we are working on an oriented surface, which implies that we can consistently speak of ``right" and ``left" of a ZPP.

	\item Finally, one must impose two constraints which ensure that the rank assignment is single valued. Consider, for instance, moving along a loop along one of the two cycles of the fundamental cell. When returning to the initial face, the rank should be unchanged. This is granted by imposing
\beq
\Lambda=\sum_\G v_\G p_\G=0, \quad  M=\sum_\G v_\G q_\G=0 \, .
\label{Eq:TopConstraints}
\eeq
We will refer to these two conditions as the $\Lambda$ and $M$ {\it topological constraints}.
	
\end{enumerate}
Every choice of values for the $v_\G$'s consistent with the topological constraints \eref{Eq:TopConstraints} gives rise to an anomaly-free rank assignment. Moreover, notice that, by construction, every rank assignment is invariant under a global shift $v_\G \to v_\G+k$. One may use this freedom to fix one of the $v_\G$'s (equivalently, one of the ZZPs is not independent). 
There are thus two constraints and one redundancy to be fixed, reproducing the expected number of fractional branes in \eref{Eq:NumberOfBranes}.  In other words, this construction can account for the most general anomaly-free rank assignment, up to an overall shift of the ranks (i.e.~regular branes). Generically, this algorithm can produce negative ranks for some faces, which cannot be directly interpreted as ranks of gauge groups. Of course, one may always add regular branes until all ranks are positive.\footnote{It is worth noting that this procedure is closely related to the algorithm for constructing fractional brane rank assignments introduced in \cite{Benvenuti:2004wx}, in which the difference in the ranks between two nodes in the quiver is proportional to the baryonic $U(1)$ charge of the bifundamental field connecting them, with one independent vector for each baryonic $U(1)$. The relation between the two methods is through the correspondence between baryonic $U(1)$ symmetries and extremal perfect matchings \cite{Butti:2005ps} or, equivalently, ZZPs  (which correspond to differences between consecutive external perfect matchings). Our procedure is also equivalent to the one for labeling cluster variables associated to plabic (i.e. planar bicolored) graphs using ZZPs \cite{MR2205721}, and to the even more similar one in the context of cluster integrable systems \cite{goncharov2013dimers} that associates a divisor at infinity on the spectral curve, to each face of the bipartite fat graph under consideration \cite{fock2015inverse}.}

\subsection{Examples}

Let us illustrate the algorithm in \sref{construction} with two examples, to which we will return when discussing orientifolds.

\subsubsection{$dP_1$} \label{subsubsec:dP1NoOrienti}

\begin{figure}[h!]
	\centering
	\begin{subfigure}[t]{0.45\textwidth }
		\begin{center} 
			\includegraphics[width=\textwidth]{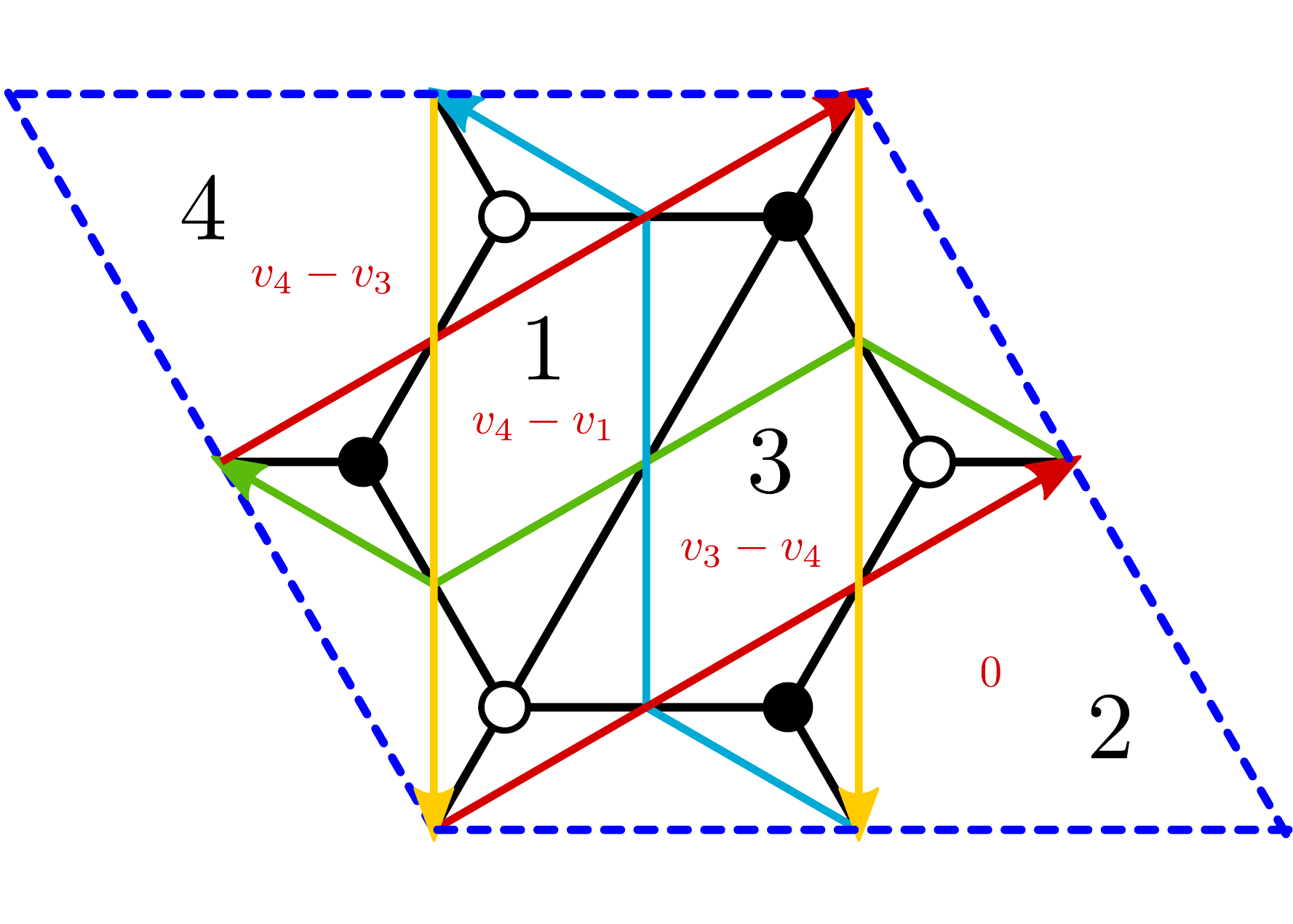}
			\caption{}
			\label{Fig:DimerdP1ZigZag2_0}
		\end{center}
	\end{subfigure} \hspace{15mm}
	\begin{subfigure}[t]{0.3\textwidth }
		\begin{center} 
			\includegraphics[width=\textwidth]{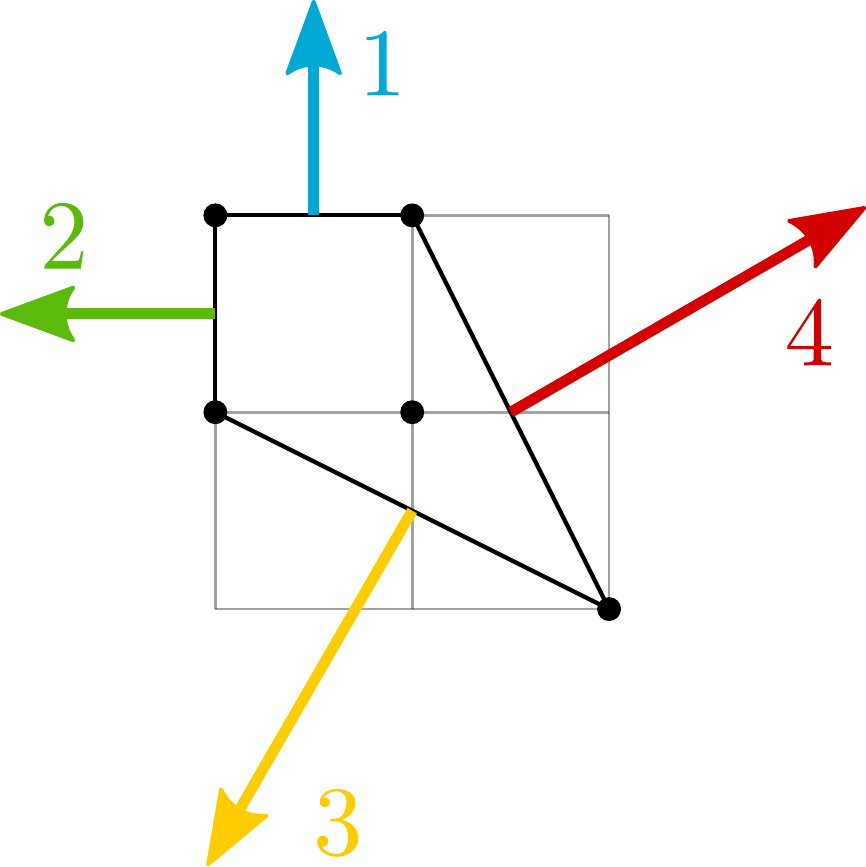}
			\caption{}
			\label{Fig:WebDiagramdP1_0}
		\end{center}
	\end{subfigure}
	\caption{(a) Dimer diagram for $dP_1$. We show the ZZPs and the rank assignments coming from them. (b) The toric/web diagram.}\label{Fig:dP1_0} 
\end{figure}

Consider the toric phase of the complex cone over $dP_1$, or $dP_1$ for short, which is shown in \Cref{Fig:dP1_0}. Let us apply the method described above for the computation of the fractional branes. 

Let us choose $N_2=0$. Applying the algorithm, the other faces are assigned the following ranks:
\begin{equation}
\left.\begin{array}{ccl}
N_1 & \leftrightarrow & v_4-v_1\\
N_2 & \leftrightarrow & 0\\
N_3 & \leftrightarrow & v_3-v_4\\
N_4 & \leftrightarrow & v_4-v_3\\
\end{array}\right.
\end{equation}

From the toric diagram we read the two topological constraints:
\begin{equation}
\begin{cases}
\begin{array}{ccccl}
\Lambda & = & - v_2 - v_3 + 2v_4 &=& 0\\
M & = & v_1 + v_4 - 2v_3 &=& 0
\end{array}
\end{cases} \Leftrightarrow \ \ \ 
\begin{cases}
\begin{array}{ccl}
v_1 & = & 2v_3-v_4 \\
v_2 & = & -v_3+2v_4
\end{array}
\end{cases}
\end{equation}
We can further use a global shift of the $v_i$ to set $v_4 =0$ and find the following rank assignment: 
\beq
\begin{array}{rcccl}
(v_1,v_2,v_3,v_4) & \equiv & \mathbf{v} & = & (2,-1,1,0)v_3 \, , \\
(N_1,N_2,N_3,N_4) & \equiv & \mathbf{N} & = & (-2,0,1,-1)v_3 \, .
\end{array}
\eeq
Which are the ranks corresponding to a well-known dynamical SUSY breaking fractional brane of $dP_1$ \cite{Berenstein:2005xa,Franco:2005zu,Bertolini:2005di}. We will return to $dP_1$ in 
\sref{subsubsec:dP1WithOrienti}.

\subsubsection{$PdP_4$} \label{subsubsec:PdP4NoOrientifold}

\begin{figure}[h!]
	\centering
	\begin{subfigure}[t]{0.30\textwidth }
		\begin{center} 
			\includegraphics[width=\textwidth]{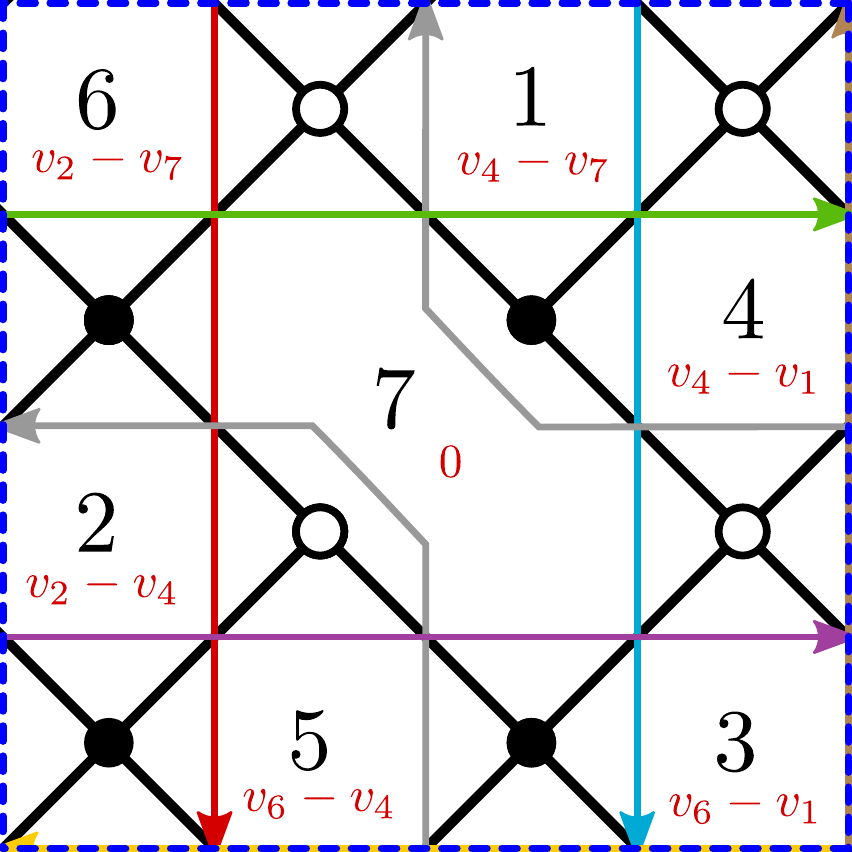}
			\caption{}
			\label{Fig:PdP4DimerZigZag_0}
		\end{center}
	\end{subfigure} \hspace{20mm}
	\begin{subfigure}[t]{0.30\textwidth } 
		\begin{center}
			\includegraphics[width=\textwidth]{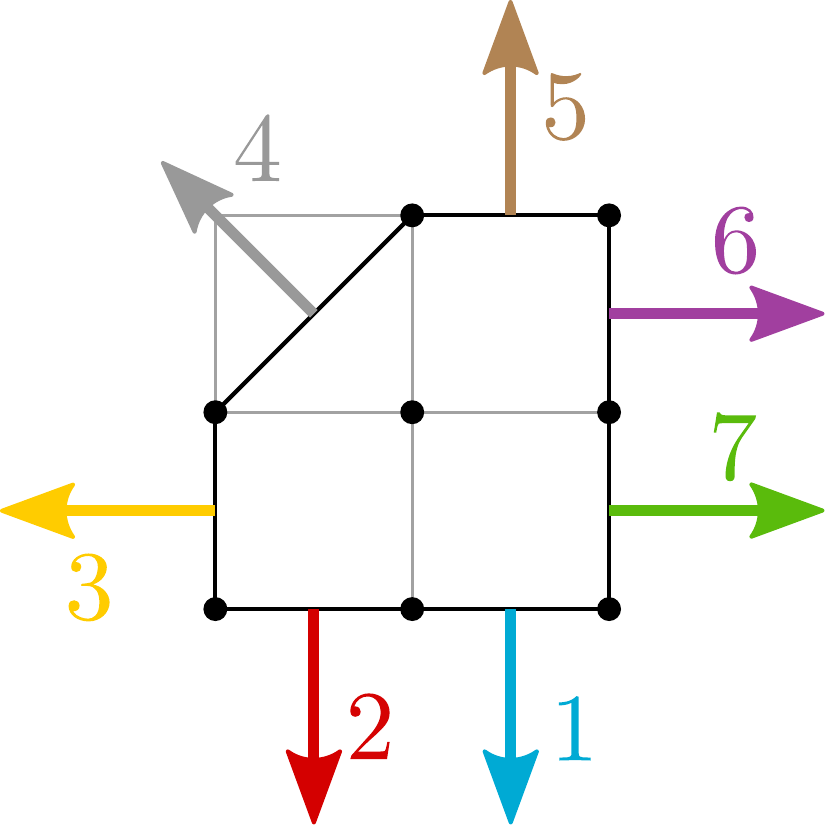}
			\caption{}
			\label{Fig:WebDiagramPdP4_0}
		\end{center}
	\end{subfigure}
	\caption{(a) Dimer diagram for $PdP_4$. We show the ZZPs and the rank assignments coming from them. (b) The toric/web diagram.}\label{Fig:pdp4_0} 
\end{figure}

As a slightly more complicated example, let us study the case of the $PdP_4$ singularity \cite{Feng:2002fv} in the toric phase considered in \cite{Argurio:2019eqb} and given in  \Cref{Fig:pdp4_0}.

From the toric diagram we read:
\begin{equation}
\begin{cases}
\begin{array}{ccl}
\Lambda & = &v_6+v_7-v_3-v_4  =0 \\
M & = & v_5+v_4-v_1-v_2 =0
\end{array}
\end{cases} \Leftrightarrow \ \ \ 
\begin{cases}
\begin{array}{ccl}
v_3 & = & v_6+v_7-v_4\\
v_5 & = & v_1+v_2-v_4
\end{array}
\end{cases}
\end{equation}
Since a global shift in the $v_i$ does not change the rank assignments, we can impose $v_4=0$. We then find the following rank assignment,
\begin{equation}
\mathbf{v} = (v_1,v_2,v_6+v_7,0,v_1+v_2,v_6,v_7) \ \  \to \ \ 
\mathbf{N} = (-v_7,v_2, v_6-v_1,-v_1,v_6,v_2-v_7,0) \, .
\end{equation}
We will return to this example in \sref{subsec:oline} upon orientifolding it.

\section{Anomaly Cancellation Conditions in Orientifolds}\label{Sec:ACC}

Determining whether an orientifolded theory admits anomaly-free solutions and, if so, finding them is a relatively straightforward task in a case by case basis. Indeed, writing down the set of anomaly equations for every gauge group and looking for solutions is not more complicated than for non-orientifolded models. In this section we systematize this calculation, introducing an algorithm for finding anomaly-free solutions in the presence of orientifolds. This, in turn, will allow us to relate the calculation to the one in the unorientifolded theory and, at a later stage, to extend the geometric determination of solutions in terms of zig-zag paths to orientifolds. 

In what follows, we will refer to the original, unorientifolded theory as the {\it mother theory}. Similarly, we will dub the orientifolded theory the {\it daughter theory}. As described in \sref{Sec:DimerIntro}  finding an anomaly-free rank assignment for the mother theory amounts to finding the kernel of its adjacency matrix (more precisely of the matrix $A= \tilde A - \tilde A^T$, with $\tilde A$ the proper adjacency matrix, see the discussion in \sref{Sec:DimerIntro}). Tensor matter in the daughter theory modifies the ACC, dovetailing the contribution of the O-planes to the RR-charges that must cancel in compact homology. In general, the anomaly/tadpole problem of orientifolded theories corresponds to a non-homogeneous linear system of the form:
\begin{equation}\label{Eq:NHLS}
\bar{A} \cdot N = f \, ,
\end{equation}
where $\bar{A}$ is the adjacency matrix of the daughter theory, and $f$ stands for the additional contribution of tensor matter. The difference between two solutions of the system \eref{Eq:NHLS} is a solution of the corresponding homogeneous one, i.e. it is in the kernel of $\bar{A}$. If one knows a particular solution $N_{\text{part}}$ of \eref{Eq:NHLS}, every solution $N$ can be expressed as: 
\beq
N = N_{\text{hom}} + N_{\text{part}} \, ,
\eeq
where $N_{\text{hom}}$ is a solution of the homogeneous system $\bar{A} \cdot N_{\text{ hom}}=0$. 

Remarkably, we will show that whether \eref{Eq:NHLS} has solutions or not can be directly determined from the toric diagram of the singularity under consideration. In other words, we will establish a geometric criterion for the satisfiability of the ACC in orientifolded theories.

\subsection{Dimers and Orientifolds}\label{section_dimers_and_orientifolds}

Before proceeding we need to recall a few basic features of orientifolds in dimer language. More details can be found in \cite{Franco:2007ii}. Related works include \cite{Imamura:2008fd,Garcia-Etxebarria:2015hua,Garcia-Etxebarria:2016bpb}.

\begin{figure}[h!]
	\centering
	\begin{subfigure}[t]{0.35\textwidth }
		\begin{center} 
			\includegraphics[width=\textwidth]{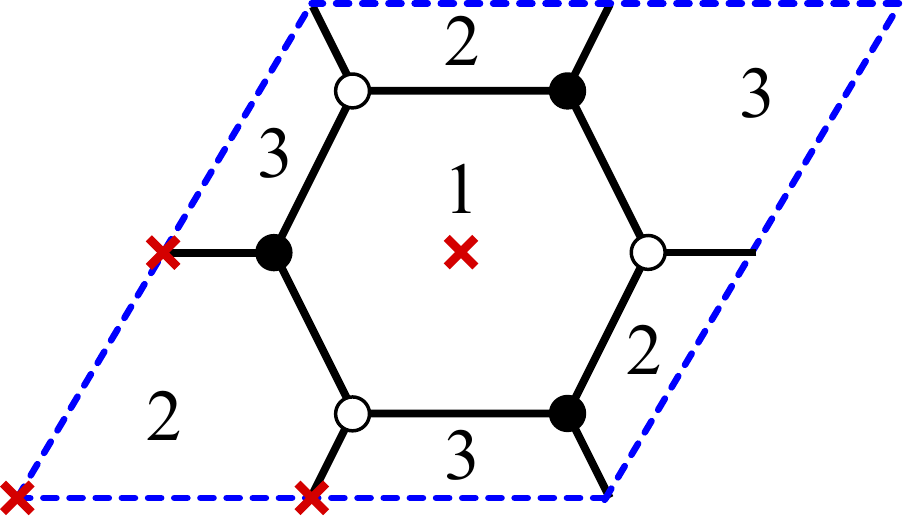}
			\caption{}
			\label{Fig:OrientifoldPoints}
		\end{center}
	\end{subfigure} \hspace{15mm}
	\begin{subfigure}[t]{0.20\textwidth } 
		\begin{center}
			\includegraphics[width=\textwidth]{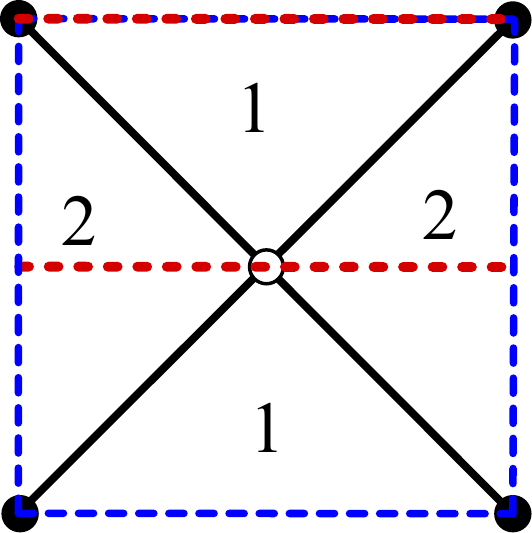}
			\caption{}
			\label{Fig:OrientifoldLine}
		\end{center}
	\end{subfigure} 
	\caption{(a) Orientifold of $\mathbb{C}^3/\mathbb{Z}_3$ with fixed points. (b) Orientifold of the conifold with fixed lines. Note that  we have changed the unit cell with respect to \Cref{Fig:ConifoldZigZags}, in order to get one with a $\mathbb{Z}_2$ symmetry.}\label{Fig:Orientifold} 
\end{figure}

The operation of orientifolding has been studied as a $\mathbb{Z}_2$ involution of the dimer diagram, leaving either fixed points or fixed line(s).\footnote{The possibility of $\mathbb{Z}_2$ involutions without fixed loci has not been explored in the literature.} Examples of both possibilities are shown in \fref{Fig:Orientifold}. The fixed points or fixed line(s) are assigned signs, which control the projection to the orientifolded theory. 

\paragraph{Fixed Points.} 
In an orientifold of this type, there are four fixed points in a unit cell (see \cite{Franco:2007ii}). In order to preserve SUSY, their four signs must satisfy the so-called {\it sign rule}: their product must be $(-1)^{n_W/2}$ where $n_W$ is the number of superpotential terms. Faces that sit on top of a fixed point are mapped to themselves and become $SO(N_i)$ or $USp(N_i)$ groups for $+$ and $-$ signs, respectively. Similarly, edges that sit on top of fixed points correspond to self-identified matter fields that are projected to symmetric $\symm$ (resp. antisymmetric $\asymm$) for $+$ (resp. $-$) signs. For faces and edges that are not self-identified, we keep any of the two mirror images in the projected theory. They correspond to $SU(N)$ gauge groups and bifundamental/adjoint matter. We refer the reader to \cite{Franco:2007ii} for details and subtleties regarding the orientifold projection.

For illustration, let us consider the fixed point orientifold of $\mathbb{C}^3/\mathbb{Z}_3$ shown in \Cref{Fig:OrientifoldPoints}. Using \Cref{Tab:Dict} we see that the superpotential has six terms, so the product of the signs for the four fixed points must be $-$. For concreteness, let us consider the signs $(+,-,-,-)$, where we have ordered them running counter-clockwise starting at the origin. The gauge symmetry and matter content of the orientifolded theory is
\begin{equation}
USp(N_1)_1 \times SU(N_2)_2 \quad \text{with} \quad 2 \asymm_2 + \symm_2 + 3 (\square_1, \ov\square_2) \ .
\end{equation}
The superpotential results from the projection of the original superpotential (see \cite{Franco:2007ii} for details and examples).

\paragraph{Fixed Lines.} 
Depending on the symmetry of the unit cell, there might be a single diagonal or two parallel fixed lines. We will often refer to the case of two parallel lines as horizontal/vertical lines due to their orientation. In either case, the signs of the fixed lines are unconstrained and can be chosen freely. The rule for projecting faces and fields is the same as with fixed points. 

As an example, let us consider the fixed line orientifold of the conifold shown in \Cref{Fig:OrientifoldLine}. Faces 1 are 2 are mapped into themselves, so both of them become either symplectic or orthogonal, depending on the signs of the fixed lines. In addition, the matter content consist of two bifundamentals $(\square_1, \square_2)$.\footnote{Since the representations of $SO(N)$ and $USp(N)$ are real, there is no distinction between fundamental and antifundamental representations.} As in the previous case, it is also straightforward to determine the projected superpotential.

\subsection{The Adjacency Matrix of Orientifolded Theories}\label{ACCmatricesorientifolded}

Consider a toric singularity and a corresponding dimer admitting a $\mathbb{Z}_2$ involution. We can divide the $n_g$ gauge groups of the mother theory into two sets: pairs of faces identified under the involution, and self-identified ones. Therefore, the  
adjacency matrix of the mother theory, $A_{IJ}$ with $I,J=1,\dots,n_g$, can be suitably rearranged as follows:
\beq
A = \left(\phantom{\begin{matrix} \\ B_{11} \\ \\ \hline \\ \\ B_{11} \\ \hline \\ \\ B_{11} \end{matrix}}
\right.\hspace{-1.5em}
\begin{array}{ccc|ccc|ccc}
& & & & & & & & \\
& B_{11} & & & B_{12} & & & B_{13} \\
& & & & & & & & \\
\hline
& & & & & & & & \\
& B_{21} & & & B_{22} & & & B_{23} \\
& & & & & & & & \\
\hline
& & & & & & & & \\
& B_{31} & & & B_{32} & & & B_{33} \\
\undermat{j}{& \text{\quad \quad} &} & \undermat{\large j+k}{& \text{\quad \quad} &} & \undermat{\large b}{& \text{\quad \quad} &} \\
\end{array}
\hspace{-1.5em}
\left.\phantom{\begin{matrix} \\ B_{11} \\ \\ \hline \\ \\ B_{11} \\ \hline \\ \\ B_{11} \end{matrix}}\right)\hspace{-1em}
\begin{tabular}{l}
$\left.\lefteqn{\phantom{\begin{matrix} \\ B_{11} \\ \hline \end{matrix}}}\right\}i$\\ \\
$\left.\lefteqn{\phantom{\begin{matrix} \\ B_{11} \\ \hline \end{matrix}}} \right\}i+k$\\ \\
$\left.\lefteqn{\phantom{\begin{matrix} \\ B_{11} \\ \hline \end{matrix}}} \right\}a$
\end{tabular} \, .
\eeq

\vspace{.85 cm}
Here faces $i,j=1,\dots,k$ are the surviving ones out of those in the pairs of faces that are mapped into each other (for every pair, we are free to keep any of the two faces). Faces $i+k,j+k$, with $i,j=1,\dots,k$, are their images. Finally, the remaining faces $a,b=1,\dots,n_g-2k$ are those that are self-identified. The $B$ matrices are the adjacency matrices between these different subsets. For example, $B_{13}$ is the adjacency matrix between surviving faces and self-identified faces, while $B_{23}$ is the adjacency matrix between the image faces and the self-identified ones. The matrix $A$ is by definition antisymmetric, which in terms of the submatrices $B$ implies that
\begin{align}
\begin{split}
B_{11} = - B_{11}^T \, , \quad B_{22} = - B_{22}^T \, , \quad B_{33} = - B_{33}^T \, , \\
B_{12} = - B_{21}^T \, , \quad B_{13} = - B_{31}^T \, , \quad B_{23} = - B_{32}^T \, . \label{Eq:Antisymmetry}
\end{split}
\end{align}

The $\mathbb{Z}_2$ symmetry of the phase under consideration endows it with further symmetry properties. The $\mathbb{Z}_2$ projection acts on the bifundamental fields as follows:
\begin{equation}\label{orie}
\begin{array}{ccc}
\text{Mother theory} & & \text{Daughter theory} \\[.1 cm]
(\overline\square_i,\square_j), \, (\overline\square_{j+k},\square_{i+k}) & \rightarrow & (\overline\square_i,\square_j) \\
(\overline\square_i,\square_{j+k}), \, (\overline\square_j,\square_{i+k}) & \rightarrow & (\overline\square_i,\overline \square_j) \\
(\overline\square_{i+k},\square_j), \, (\overline\square_{j+k},\square_i) & \rightarrow & (\square_i,\square_j) \\
(\overline\square_{a},\square_{i}), \, (\overline\square_{i+k},\square_{a}) & \rightarrow & (\square_a,\square_i) \\
(\overline\square_{i},\square_{a}), \, (\overline\square_{a},\square_{i+k}) & \rightarrow & (\square_a,\overline\square_i) \\
(\overline\square_{a},\square_{b}), \, (\overline\square_{b},\square_{a}) & \rightarrow & (\square_a,\square_b) 
\end{array} \, .
\end{equation}

These projections imply that:
\begin{equation}\label{Eq:Z2Symmetry}
\begin{split}
B_{11}=B_{22}^T \, ,  \quad  B_{12}=B_{12}^T \, ,   \quad   B_{21}=B_{21}^T\, ,\\
B_{31}=B_{23}^T \, ,  \quad   B_{13}=B_{32}^T ,  \,  \quad   B_{33} = B_{33}^T\, .
\end{split}
\end{equation}
We can apply \Cref{Eq:Antisymmetry,Eq:Z2Symmetry} together to find further relations between the $B$'s,
\begin{equation}
\begin{split}
B_{11} = - B_{22} \, , \quad B_{12} = - B_{21} \, , \quad \quad \quad \\
B_{13}= - B_{23}\, ,  \quad  B_{31}= - B_{32}\, ,  \quad  B_{33} = 0\, , \label{Eq:Z2relations}
\end{split}
\end{equation}
so that eventually the adjacency matrix is entirely determined by $B_{11}, B_{12}$ and $B_{13}$:
\begin{equation}
A = \left(
\begin{array}{ c    c   c}
B_{11} & B_{12} & B_{13}   \\  -B_{12} & -B_{11} & -B_{13}  \\  -B_{13}^T & B_{13}^T & 0 
\end{array}
\right)\ .
\end{equation}

\begin{figure}[h!]
	\centering
			\begin{center} 
			\includegraphics[width=0.32\textwidth]{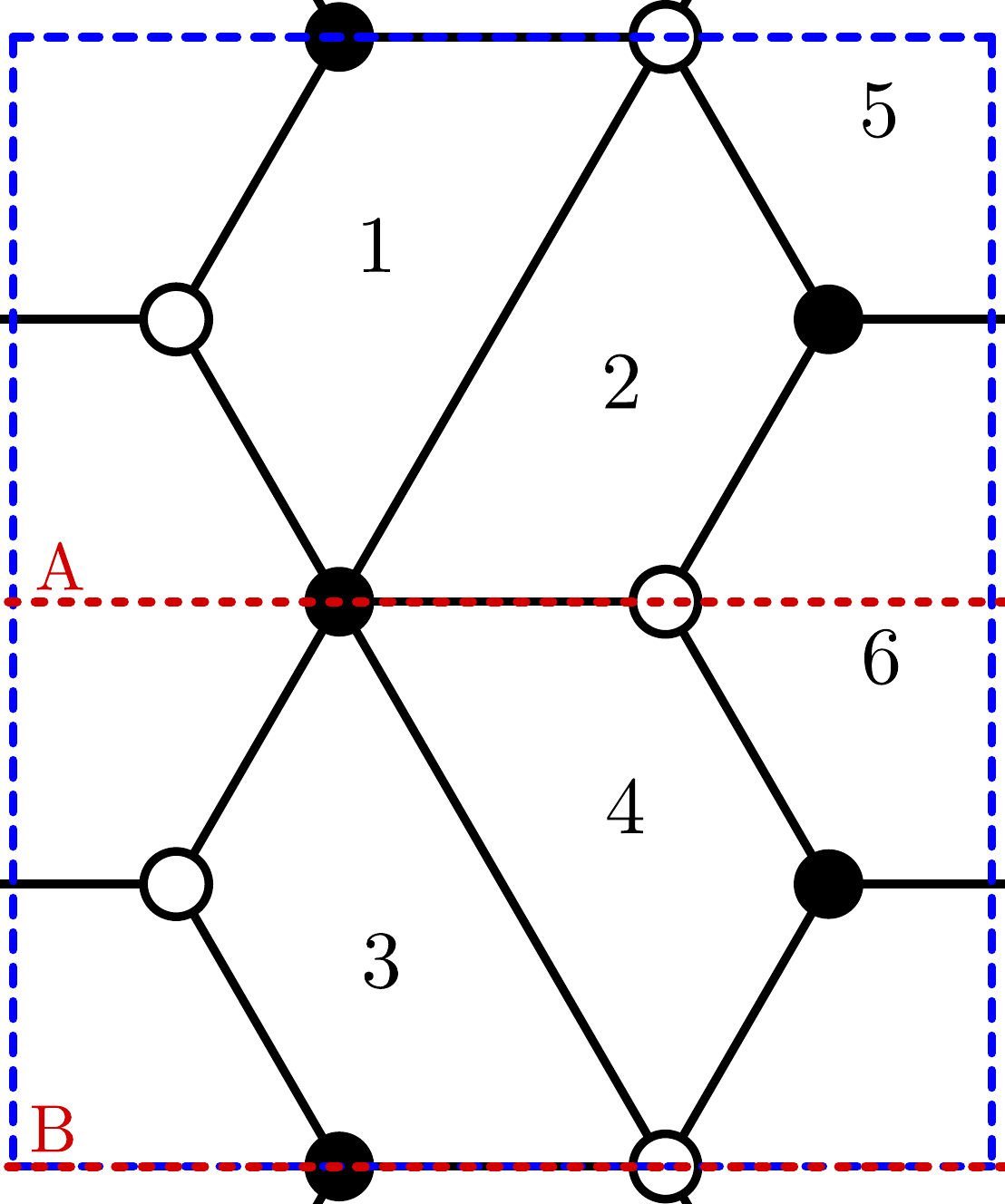}	
		\end{center}
	\caption{Dimer diagram for $PdP3_b$ with two horizontal fixed lines (dotted red).} 
\label{Fig:PdP3bDimer3}
\end{figure}

In order to illustrate these relations, let us consider the complex cone over $PdP_{3b}$, as studied in \cite{Argurio:2019eqb}. The dimer, which is shown in \fref{Fig:PdP3bDimer3}, admits a $\mathbb{Z}_2$ symmetry with two fixed lines. 
The numbering of the faces has already been chosen such that the adjacency matrix reads
\begin{equation}
A = \left(
\begin{array}{c c | c  c | c c}
0 & 1 & -1 & 0 & 1 & -1 \\ -1 & 0 & 0 & 1 & 1 & -1 \\ \hline 1 & 0 & 0 & -1 & -1 & 1 \\   0 & -1 & 1 & 0 & -1 & 1 \\ \hline -1 & -1 & 1 & 1 & 0 & 0 \\ 1 & 1 & -1 & -1 & 0 & 0
\end{array}
\right),
\end{equation}
which showcases the general structure in \Cref{Eq:Antisymmetry,Eq:Z2Symmetry,Eq:Z2relations}.

Let us now turn our attention to the daughter theory. To compute the ACC for the orientifolded theory we first note that $SO/USp$ groups are automatically anomaly-free and play no role. Further, for the ACC of the non-self-identified faces we have to take into account both the contributions from fields such as $(\overline\square_i,\square_j)$, that are counted by $B_{11}$, and fields such as $(\square_i,\square_j)$ and $(\overline\square_i,\overline\square_j)$ that are counted by $B_{12}$, see \eref{orie}. This leads to the homogeneous ACC for the projected theory given by
\begin{equation}
\ov{A} =\left(\begin{array}{ccc|ccc}
& & & & &  \\
& B_{11}+B_{12} & & & B_{13} &  \\
& & & & & 
\end{array}\right). \label{O-ACC}
\end{equation}
Applying this to the $PdP_{3b}$ example we get
\begin{equation}
\ov{A} = \left(
\begin{array}{c c | c c}
-1 & 1 & 1 & -1  \\ -1 &  1 & 1 & -1 \end{array}
\right)\, .
\end{equation}

\subsection{The Homogeneous Problem}\label{Sec:Homogeneous}

In the previous section, we have constructed the homogeneous part of the ACC for an orientifolded theory. We now show how solutions to the homogeneous problem, namely elements of $\mathrm{ker}(\ov{A})$, are obtained from symmetric rank assignments of the mother theory, which form a subspace of $\mathrm{ker}(A)$. This will allow us to extend the method explained in \sref{construction} to the homogeneous problem of orientifolded theories.

We say that a rank assignment of the mother theory $N^S_I$ is symmetric with respect to the $\mathbb{Z}_2$ involution if it satisfies 
\beq \label{Eq:SymmetricRank}
N^S_i = N^S_{i+k} \, , \quad N^S_a \,\,\, \text{free}\, .
\eeq
If this rank assignment is anomaly-free in the mother theory (i.e. if it is in the kernel of $A$), we have
\beq
A_{IJ} N^S_J = 0 \, , \label{Eq:InTheKernel}
\eeq
where here and henceforth, summation over repeated indices is understood.

Expanding this equation in terms of the $B$ matrices and exploiting the symmetry properties given in \eref{Eq:SymmetricRank}, it becomes
\begin{equation}\label{Eq:SymHomog}
\begin{split}
(B_{11} + B_{12})_{ij}N^S_j + (B_{13})_{ia} N^S_a = 0\, , \\
(B_{21} + B_{22})_{ij}N^S_j + (B_{23})_{ia} N^S_a = 0\, , \\
(B_{31} + B_{32})_{aj}N^S_j + (B_{33})_{ab} N^S_b  = 0\, .
\end{split}
\end{equation}
From \eref{Eq:Z2relations}, we conclude that the first two equations are actually one and the same, while the third equation is trivially satisfied for any symmetric rank assignment. From the first two equations we learn that any symmetric rank assignment $N^S_I$ in the mother theory which satisfies the ACC, defines a solution of the homogenous ACC system of the daughter theory given in \eref{Eq:NHLS}:
\begin{equation}
N_{\text{hom}} = ( N^S_i \vert N^S_a)\, .
\end{equation}
\Cref{Eq:SymHomog} indeed implies that such a vector satisfies:
\begin{equation}\label{Eq:Homog}
\ov{A} \cdot N_{\text{hom}} = 0 \, .
\end{equation}

Conversely, if one starts with a vector $( N^S_i \vert N^S_a)$ satisfying \eref{Eq:Homog}, the vector $( N^S_i \vert N^S_{i+k} \vert N^S_a)$ is a symmetric rank assignment of the mother theory. The definition of $\ov{A}$ in \eref{O-ACC} implies that the  equations in \eref{Eq:SymHomog} hold for $( N^S_i \vert N^S_{i+k} \vert N^S_a)$, i.e. that the latter satisfies the ACC of the mother theory. Hence, we have proved the following:

\begin{mdframed}[backgroundcolor=blue!20]
	Rank assignments in the daughter theory which satisfy the homogeneous ACC are in one-to-one correspondence with symmetric rank assignments in the mother theory which satisfy the ACC. 
\end{mdframed}

In the special case where tensors are absent from the daughter theory, the ACC are actually a homogeneous problem and the symmetric rank assignments in the mother theory provide directly the orientifold solutions. The regular brane is such a solution that always exists, and thus guarantees that an orientifold without tensors always admits a non-anomalous solution.

\subsection{The Non-Homogeneous Problem} \label{Sec:AntisymmetricRank}

Finding solutions to the ACC in orientifolded theories with tensors is not trivial because their very existence is not guaranteed, since the full system of ACC given in \eref{Eq:NHLS} has a non-homogeneous part coming from the tensor matter. The Rouch\'e-Capelli theorem gives us a criterion for its solvability: a non-homogeneous system,
\beq
\bar{A} \cdot N = f \, ,
\eeq
admits a solution if and only if
\beq
\mathrm{rank} (\bar{A}) =  \mathrm{rank} (\bar{A}\vert f ) \, ,
\eeq
where $(\bar{A}\vert f )$ is the matrix obtained appending the column $f$ to the matrix $\bar A$.
For us $f$ encodes the contribution to the ACC of the tensor matter, i.e.~of the self-identified chiral fields. 

In other words, every set of numbers $r_i$ such that
\begin{equation}\label{Eq:RowReduce}
r_i \bar{A}_{i\bar J} = 0 
\end{equation}
holds for all $\bar J=j,a$, must satisfy
\begin{equation}
r_i f_{i} = 0 \, 
\end{equation}
for the system to be solvable. In this section we show that the coefficients $r_i$ which satisfy \eref{Eq:RowReduce} correspond precisely to the antisymmetric rank assignments of the mother theory. 

Suppose that some coefficients $r_i$ satisfying \eref{Eq:RowReduce} exist. Using \eref{Eq:Z2relations} for $\bar J=j$, one can show that it implies 
\beq\label{Eq:RowRed1}
r_i (B_{11})_{ij} -r_i (B_{21})_{ij} = 0\, .
\eeq
Using \eref{Eq:Z2relations}, this is equivalent to
\begin{equation}\label{Eq:RowRed2}
r_i (B_{12})_{ij} -r_i (B_{22})_{ij} = 0\, .
\end{equation}
For $\bar J=a$, using \eref{Eq:Z2relations}, we find that
\begin{equation}\label{Eq:RowRed3}
r_i (B_{13})_{ia} -r_i (B_{23})_{ia} = 0\, .
\end{equation}
We write
\begin{equation}
N^A_I = (r_i \vert - r_i \vert 0) \, ,
\end{equation}
and equations \eref{Eq:RowRed1} to \eref{Eq:RowRed3} can be expressed as
\begin{equation}
 N^A_I A_{IJ}=0 = A_{JI}N^A_I \,  ,
\end{equation}
where the second equality merely uses the antisymmetry property of $A$. Hence, we have proved that any set of $r_i$ satisfying \eref{Eq:RowReduce} defines an antisymmetric rank assignment $N^A_I$ of the mother theory, which satisfies the mother ACC. 

Conversely, starting with an antisymmetric rank assignment $N^A_I$ in the mother theory
\begin{equation}
N^A_i = - N^A_{i+k}, \quad N^A_a = 0 \, ,
\end{equation}
which satisfies the ACC, one can use equations \eref{Eq:RowRed1} to \eref{Eq:RowRed3} backwards, and thus obtain a set of $r_i$ such that \eref{Eq:RowReduce} holds for all $\bar J=j,a$. 

Let us emphasize that while symmetric rank assignments in the mother theory are in one-to-one correspondence with solutions of the homogeneous system of ACC in the daughter theory (which by definition form the kernel of $\ov{A}$), the antisymmetric rank assignments in the mother theory correspond rather to the elements of the cokernel of $\ov{A}$, that we will see merely as technical tools. They are useful for determining whether a given daughter theory admits an anomaly-free rank assignment, since the elements in the cokernel of $\ov{A}$ encode the relations between the lines of $\ov{A}$, from which one can row-reduce $\ov{A}$.

\begin{mdframed}[backgroundcolor=blue!20]
	Coefficients of trivial linear combination of lines of $\ov{A}$ are in one-to-one correspondence with the anomaly-free antisymmetric rank assignments in the mother theory.
\end{mdframed}

To rephrase what we wrote at the beginning of the section, there are anomaly-free rank assignments in the daughter theory if and only if
\beq
N^A_i f_i = 0 
\eeq
for every antisymmetric solution $N^A_I$ of the mother theory, where $f$ is easily computed from the dimer and its orientifold. We call this the ``Rouch\'e-Capelli condition."

In general, note that any rank assignment $N_I$ can be split into a symmetric and an antisymmetric component,
\beq
\begin{split}
(N_i\vert N_{i+k} \vert  N_a) =  \frac{1}{2}(N_i+N_{i+k} \vert N_{i+k} +N_i \vert 2 N_a)   + \frac{1}{2}(N_i-N_{i+k} \vert N_{i+k} - N_i \vert 0) \, . \label{Eq:AntiAndSymmetric}
\end{split}
\eeq
Both parts are then half-integer valued. Multiplying such a possibly unphysical (in the case it is half integer-valued) rank vector by an even number yields a physical rank vector with the required (anti)symmetry. All of the reasoning of the last two subsections is pure linear algebra, and does not know about the need of integrality for rank assignments, which entirely comes from physics.

\section{A Zig-Zag Algorithm for Orientifolds} \label{Sec:Algorithm}

We will now generalize the procedure discussed in \sref{construction} to find (anti)symmetric rank assignments in orientifolded theories. The precise details of the algorithm depend on whether the $\mathbb{Z}_2$ involution leaves fixed lines or points. 
This difference comes from the fact that involutions with fixed lines map nodes to nodes of the same color, while involutions with fixed points map nodes to nodes with opposite color.

We illustrate this difference in \Cref{{Fig:ZigZagMap}}. There we can see that ZZPs around a node make a clockwise or counterclockwise  loop. If a node is mapped to a node of the same color it means that the orientation of the loop is preserved, while, in the opposite case, it is reversed.\footnote{We recall that under both involutions, a dimer is sent to a dimer with all ZZPs going in the opposite direction. The map between ZZPs is understood after additionally reversing the direction of every ZZP, as in \cite{Retolaza:2016alb}.} 
This observation will become crucial when we define (anti)symmetric rank assignment in both the case of fixed lines and points.
\begin{figure}[h!]
	\centering
	\begin{subfigure}[t]{0.30\textwidth }
		\begin{center} 
			\includegraphics[width=\textwidth]{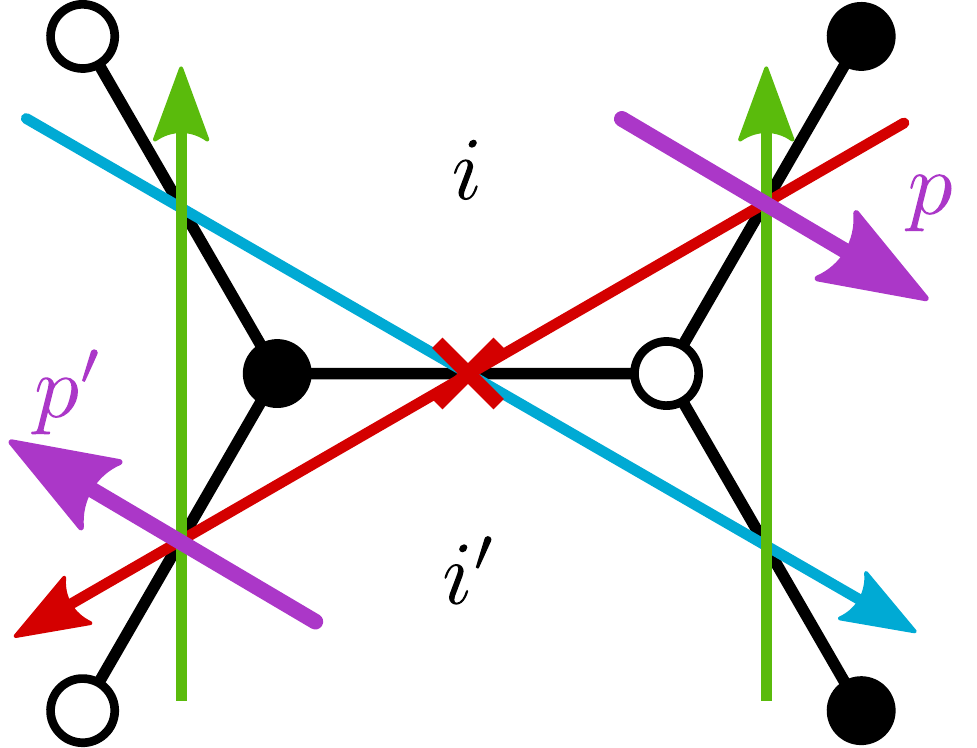}
			\caption{}
			\label{Fig:ZigZagMapPoint}
		\end{center}
	\end{subfigure} \hspace{15mm}
	\begin{subfigure}[t]{0.30\textwidth } 
		\begin{center} 
			\includegraphics[width=\textwidth]{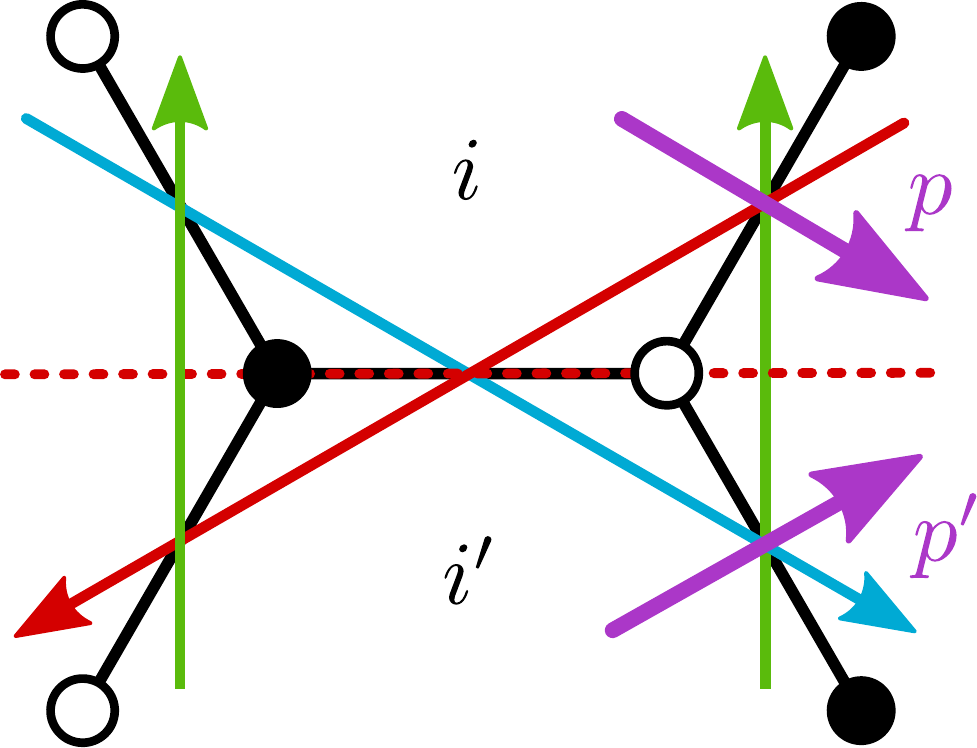}
			\caption{}
			\label{Fig:ZigZagMapLine}
		\end{center}
	\end{subfigure} 
	\caption{The orientifold actions with fixed points (a) and fixed lines (b). $p$ is a path from one face to an adjacent one, and $p'$ its image. In (a) the red and blue ZZPs are self-identified, while the green ones are mapped into each other. In (b), the red and blue ZZPs are mapped into each other, and the green ones are self-identified.}\label{Fig:ZigZagMap}
	\end{figure}

For the forthcoming analysis, we find it useful to introduce the notation $\{\Gamma\}=\{\a,\ov{\a},\g\}$ to describe the set of ZZPs: every pair ($\a$, $\ov{\a}$) corresponds to ZZPs mapped into each other under the orientifold projection, while $\g$ labels self-identified ZZPs.

\subsection{Fixed Line Orientifolds} \label{subsubsec:FixedLines}

\begin{figure}[h!]
	\centering
	\begin{subfigure}[t]{0.35\textwidth }
		\begin{center} 
			\includegraphics[width=\textwidth]{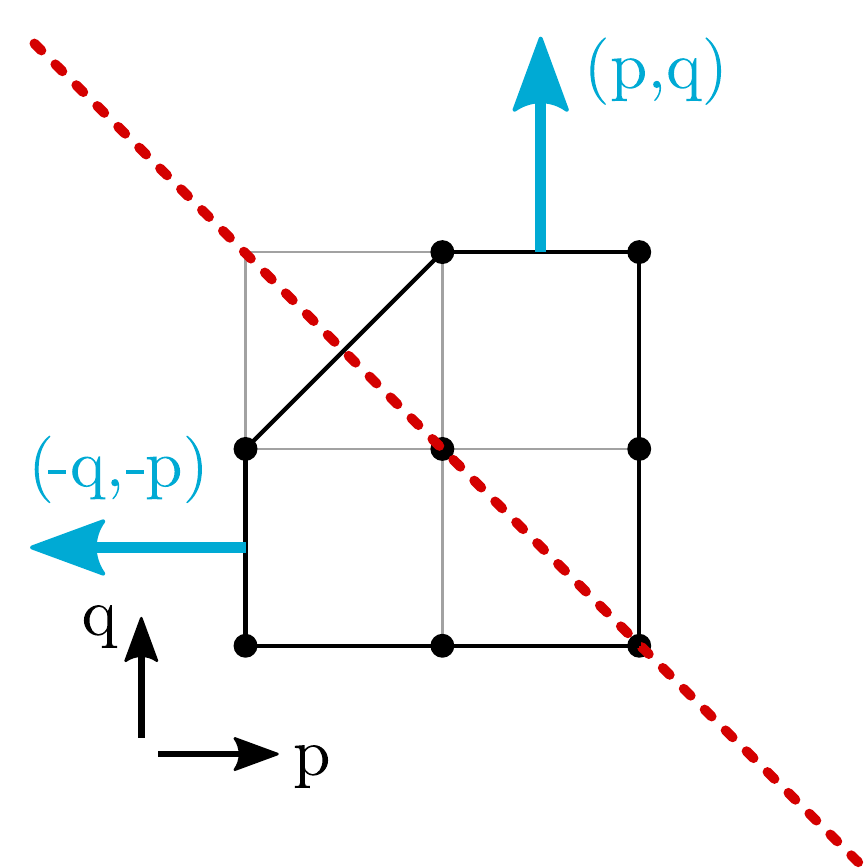}
			\caption{}
			\label{Fig:DiagonalOLine}
		\end{center}
	\end{subfigure} \hspace{15mm}
	\begin{subfigure}[t]{0.35\textwidth }
		\begin{center} 
			\includegraphics[width=\textwidth]{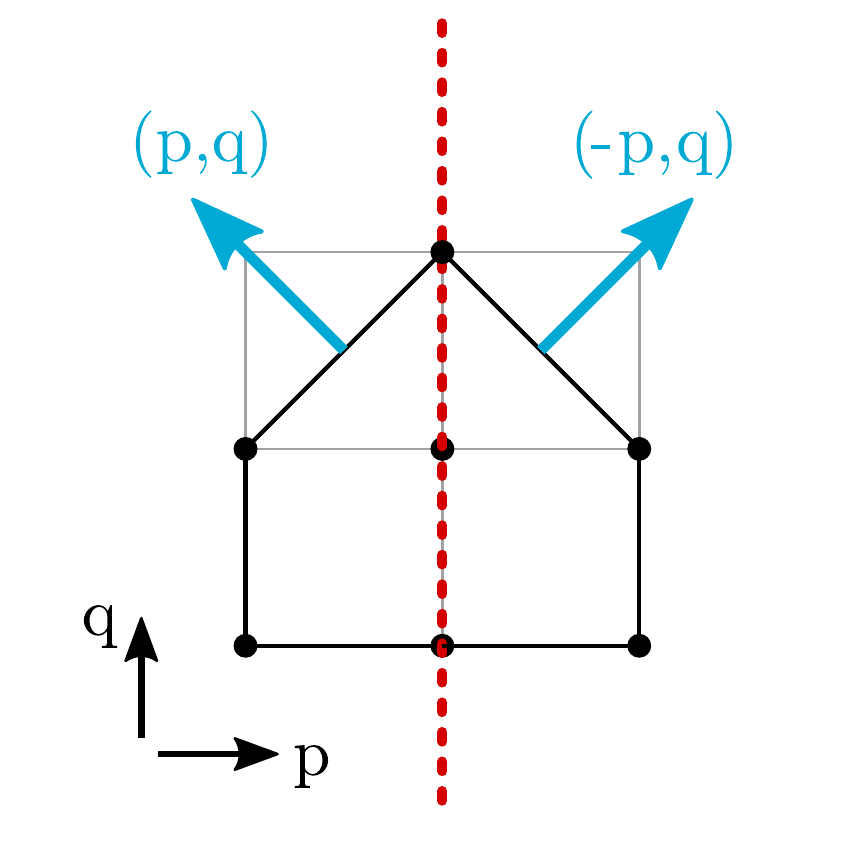}
			\caption{}
			\label{Fig:HorizontalOLine}
		\end{center}
	\end{subfigure}
	\caption{The toric diagrams for fixed line orientifolds have an axis of reflection symmetry. The corresponding axes for: (a) diagonal and (b) horizontal O-lines. In both cases we show in blue a generic ZZP and its image.}\label{Fig:DiagAndHorizontalLines} 
\end{figure}

Due to how they act on ZZPs, orientifolds with fixed lines in the dimer correspond to toric diagrams with axes of $\mathbb{Z}_2$ reflection symmetry.\footnote{We will refer to such lines of reflection symmetry in the toric diagram as {\it axes} in order to avoid confusion with the fixed lines in the dimer (which we also call O-lines).} \fref{Fig:DiagAndHorizontalLines} illustrates the general structure of such axes and the map between a ZZP and its image in the cases of orientifolds with  diagonal and horizontal O-lines (which is analogous to the case with vertical O-lines). Let us elaborate on this kind of figure. Naively, the orientation of the reflection axis in these toric diagrams can be modified by an $SL(2,\mathbb{Z})$ transformation, potentially eliminating the distinction between the diagonal and vertical/horizontal O-line cases. However, the toric diagram after such $SL(2,\mathbb{Z})$ transformation would no longer be symmetric with respect to the axis. Alternatively, we can think about the toric diagrams with reflection axes as coming from specific dimers with fixed lines. In this context, an $SL(2,\mathbb{Z})$ transformation translates into a change of the unit cell of the dimer. But the unit cell is fixed by the specific orientifold under consideration: not any $SL(2,\mathbb{Z})$ transformation is permitted once we have chosen an orientifold identification. In other words, the orientifold obstructs $SL(2,\mathbb{Z})$ transformations. 

\paragraph{Symmetric rank assignments.} 
For $\mathbb{Z}_2$ involutions with fixed lines, symmetric rank assignments correspond to symmetric ZZP value assignments:
\beq
v_{\a} = v_{\ov{\a}}, \quad v_{\g} \text{ free}
\eeq

First, recall from \sref{construction} that the difference between the ranks of any two faces in the dimer is equal to a sum (with signs) of the values of the ZZPs one crosses as one goes between the two faces. ACC at each face of the dimer ensure that the value of this sum is invariant under smooth (homological) deformations of the path one follows. Furthermore, the topological constraints guarantee that the value of the sum is independent of the homology class of the path on the torus.

Consider two faces $i$ and $j$ and a path $p$ connecting them, and $i'$, $j'$ and $p'$ their respective images under the $\mathbb{Z}_2$ symmetry. 
Every time $p$ crosses a ZZP $\alpha$, its image $p'$ crosses $\alpha'$, and these two crossings have the same sign, since the orientation is preserved. From this, it is clear that, if the ZZP value assignment is symmetric, the rank assignment generated by the method in \sref{construction} is also symmetric. 

\begin{mdframed}[backgroundcolor=blue!20]
	In the case of dimer models with involutions fixing lines, symmetric rank assignments correspond bijectively to symmetric ZZP value assignments (up to the global shift in the values, and such that the topological constraints are satisfied). 
\end{mdframed}

For symmetric value assignments, the topological constraints read:

\begin{itemize}
	\item \underline{Diagonal line} ($p_{\bar{\alpha}}=-q_\alpha, q_{\bar{\alpha}}=-p_\alpha$):
	\begin{align}
	0= \L = \sum_{\a} v_\a (p_\a - q_\a) +\frac12 \sum_{\gamma} v_\gamma (p_\gamma - q_\gamma) = -M=0  \, .
	\end{align} 
	
	\item \underline{Vertical lines} ($p_{\bar{\alpha}}=-p_\alpha, q_{\bar{\alpha}}=q_\alpha$):
	\begin{align}
	& M = 2\sum_{\a} v_\a  q_\a +\sum_{\gamma} v_\gamma q_\gamma = 0  \, , \nonumber \\
	& \Lambda = 0 \, ,
	\end{align} 
	The case of horizontal lines follows exchanging $p_\Gamma$ with $q_\Gamma$ and $\Lambda$ with $M$. 
	
\end{itemize}

We can now compute the total number of symmetric rank assignments. If the dimer under consideration has $n$ ZZPs, symmetric rank assignments correspond to a choice of $v_\Gamma$, such that $v_\alpha=v_{\ov \alpha}$, and such that topological constraints hold. We also have the freedom to shift the rank of all gauge groups, since regular branes respect the required symmetry. Putting all this together, the number of independent symmetric rank assignments modulo some (possibly half-integer) number of regular branes is 
\beq
\label{Eq:DimSymFL}
\text{dim}(\text{ker}(\ov{A})) -1 = \frac{1}{2}(n+n_{s}) - 2 \, ,
\eeq
where $n_s$ is the number of self-identified ZZPs.

\paragraph{Antisymmetric rank assignments.} 

Antisymmetric rank assignments are found in a similar fashion, by imposing the antisymmetry explicitly on the ZZP values, i.e. $v_{\G} = -v_{\G}$, or equivalently
\beq \label{antizzp}
v_\a = - v_{\ov{\a}}, \quad v_\g = 0 \, .
\eeq
This follows from the same reasoning as in the symmetric case: due to the geometric action of the symmetry, it is clear that antisymmetric ZZP value assignments lead to antisymmetric rank assignments in the dimer. Furthermore, if the ZZP value assignment is not antisymmetric \textit{up to a shift}, it is straightforward to see that the rank assignment cannot be antisymmetric either. 

In this case there is a subtlety that was not present in the symmetric case. First, the ZZP value method only knows about differences of ranks in the dimer. Equivalently, it only describes anomaly-free rank assignments up to some (half-integer) number of regular branes. The relevant point here is that regular branes are not antisymmetric. Hence, starting from an antisymmetric value assignment, it can well be that the rank assignment one constructs is not antisymmetric per se, but merely antisymmetric after having added some number of regular branes (we will see examples of this later). Then, in the method of \sref{construction}, a global shift of the ZZP values does not change the resulting rank assignment. The global shift does not preserve antisymmetry, so among the family of value assignments corresponding to a given rank assignment (modulo regular branes), there is a special representative which is an antisymmetric value assignment. 
Thus instead of focusing on the bijection between the set of antisymmetric rank assignments up to a (half-integer) number of regular branes, and the set of ZZP value assignments which satisfies the topological constraints, and which can be transformed into antisymmetric value assignments thanks to the global shift, one can consider the only representative of such a class of ZZP value assignments, which is antisymmetric. We have proven the following:

\begin{mdframed}[backgroundcolor=blue!20]
	In the case of dimer models with involutions fixing lines, antisymmetric rank assignments correspond bijectively to antisymmetric ZZP value assignments which satisfy the topological constraints.
\end{mdframed}

When combined with \eref{antizzp}, the topological constraints $\L = M = 0$ again merge into a single constraint, regardless of the type of fixed line orientifold. 
The surviving combination however depends on the nature of the fixed lines:
\begin{itemize}
	\item \underline{Diagonal line}:
	\begin{equation}
	\L = \sum_\a v_{\alpha} (p_\a+q_\a) = - M =0 \, .
	\end{equation} 
	\item \underline{Vertical lines}:
	\begin{align}
	&\L =2 \sum_\a v_{\alpha} p_\a = 0 \nonumber \, , \\
	&M=0 \, .
	\end{align} 
	For horizontal lines we merely need to exchange $p_\alpha$ with $q_\alpha$, and $\Lambda$ with $M$.
\end{itemize}

The number of antisymmetric rank assignments is easily computed to be
\begin{equation}\label{Eq:DimAsymFL}
\text{dim}(\text{coker}(\ov{A})) = \frac{1}{2}(n-n_s) - 1 
\end{equation}
Adding \eref{Eq:DimSymFL} and \eref{Eq:DimAsymFL}, we find that the total number of independent either symmetric or antisymmetric anomaly-free rank assignments is $n-3$, as should be the case since it is the number of anomaly-free rank assignments in the mother theory, up to (half) regular branes. 

Below we illustrate these ideas with a few explicit examples, containing both diagonal and vertical/horizontal fixed lines.

\subsubsection{No Anomaly-Free Solution: $dP_1$ with Diagonal Fixed Line}  
\label{subsubsec:dP1WithOrienti}

\begin{figure}[h!]
	\centering
	\begin{subfigure}[t]{0.45\textwidth }
		\begin{center} 
			\includegraphics[width=\textwidth]{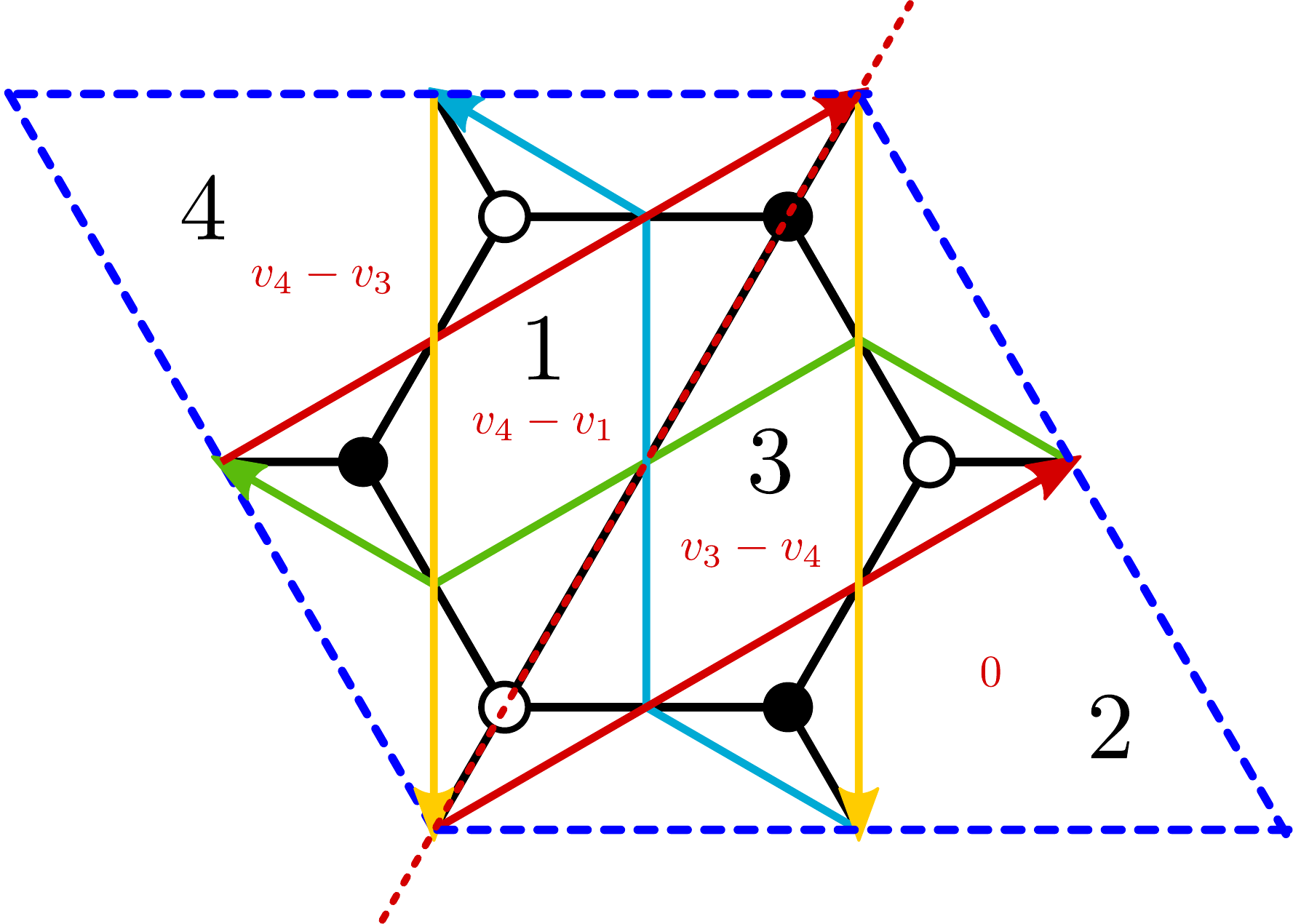}
			\caption{}
			\label{Fig:DimerdP1ZigZag2}
		\end{center}
	\end{subfigure} \hspace{15mm}
	\begin{subfigure}[t]{0.3\textwidth }
		\begin{center} 
			\includegraphics[width=\textwidth]{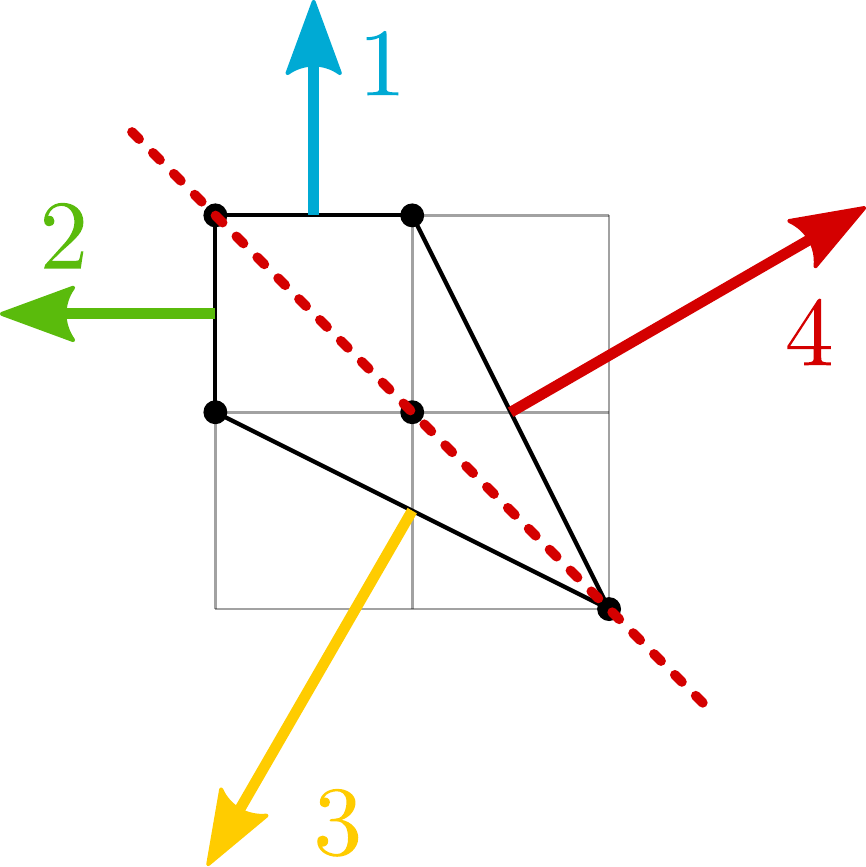}
			\caption{}
			\label{Fig:WebDiagramdP1}
		\end{center}
	\end{subfigure}
	\caption{(a) Dimer diagram for $dP_1$ with a diagonal fixed line (dotted red). We show the ZZPs and the rank assignments coming from them. (b) The toric/web diagram with the corresponding symmetry axis. }
\label{Fig:dP1} 
\end{figure}

Consider the complex cone over $dP_1$, which we discussed in \sref{subsubsec:dP1NoOrienti} as an example of the ZZP method for the mother theory. It admits a $\mathbb{Z}_2$ involution with a fixed line, as shown in \fref{Fig:dP1}. 

\paragraph{Adjacency matrices.} 
The adjacency matrix of the mother theory is easily read from the dimer, and it is given by
\begin{equation}
A = \left(
\begin{array}{c c | c c}
0 & 2 & -1 & -1 \\ -2 & 0 & -1 & 3 \\ \hline 1 & 1 & 0 & -2 \\ 1 & -3 & 2 & 0  
\end{array}
\right)
\end{equation}
For concreteness, let us consider the case of a positive O-line. The adjacency matrix for the orientifolded theory is found using \eref{O-ACC}. It is supplemented with the inhomogeneous part and becomes
\begin{equation}\label{matrixdP1}
(\bar{A} \vert f) = \left(
\begin{array}{c c | c}
-1 & \; \; 1 & +4 \\ -3 & \; \; 3 & -4 \end{array}
\right)\ .
\end{equation}
We will later discuss how to determine systematically the $f_i$s. Here it is sufficient to see that since the O-line has a $+$ sign, both tensors are symmetric. The sign of $f_i=\pm4$ has to be correlated with the sign of the diagonal elements of $\bar A$, so that in the ACC we eventually find $\pm(N_i+4)$ for symmetric tensors and $\pm(N_i-4)$ for antisymmetric ones (recall that $f$ is on the right hand side of the ACC equations \eref{Eq:NHLS}).

One may directly solve the simple system \eref{matrixdP1}, but we will rather use the algorithm we developed. In the dimer in \Cref{Fig:DimerdP1ZigZag2} we indicate the linear combination of ZZPs that corresponds to every face (we have chosen face $2$ to have rank $0$). In \sref{subsubsec:dP1NoOrienti} we studied the anomaly-free rank assignments in the mother theory and found a one parameter family (besides the regular brane):
\beq
\mathbf{N}= (-2,0,1,-1)v_3 \ ,  
\eeq
which can be decomposed into symmetric and anti-symmetric parts,
\beq
(-2,0,1,-1)v_3 = - \frac{1}{2}(1,1,1,1)v_3 + \frac{1}{2}(-3,1,3,-1)v_3
\eeq
We see that there is one antisymmetric rank vector $(-3,1,3,-1)$ and no symmetric one (except the regular brane). We now show how to find them directly from the ZZPs.

The antisymmetric rank vector is found by imposing $v_1 = -v_2$ and $v_3 = -v_4$. We cannot use the global shift, since it is not antisymmetric. The periodicity constraints are $\L = M = v_1 - 3v_3 = 0$. We thus find the antisymmetric rank assignment $\mathbf{N}=(-3,1,3,-1)v_3$. In the daughter theory this vector is $\mathbf{\ov{N}}=(-3,1)v_3$. However, it is not in $\text{ker}(\ov{A})$, but in the cokernel. We can use it to row reduce $\ov{A}$ and study whether the linear system $(\ov{A} | f)$ has solutions. Denote $\mathbf{f} = (+4,-4)^T$ the inhomogeneous part of $(\ov{A} | f)$. If $\mathbf{\ov{N}} \cdot \mathbf{f} \neq 0$, the theory is anomalous.  
This is indeed the case in this example, so we conclude that the daughter theory does not admit an anomaly-free solution.

\subsubsection{No Anomaly-Free Solution: $PdP_4$ with Diagonal Fixed Line} \label{subsec:oline}

\begin{figure}[h!]
	\centering
	\begin{subfigure}[t]{0.3\textwidth }
		\begin{center} 
			\includegraphics[width=\textwidth]{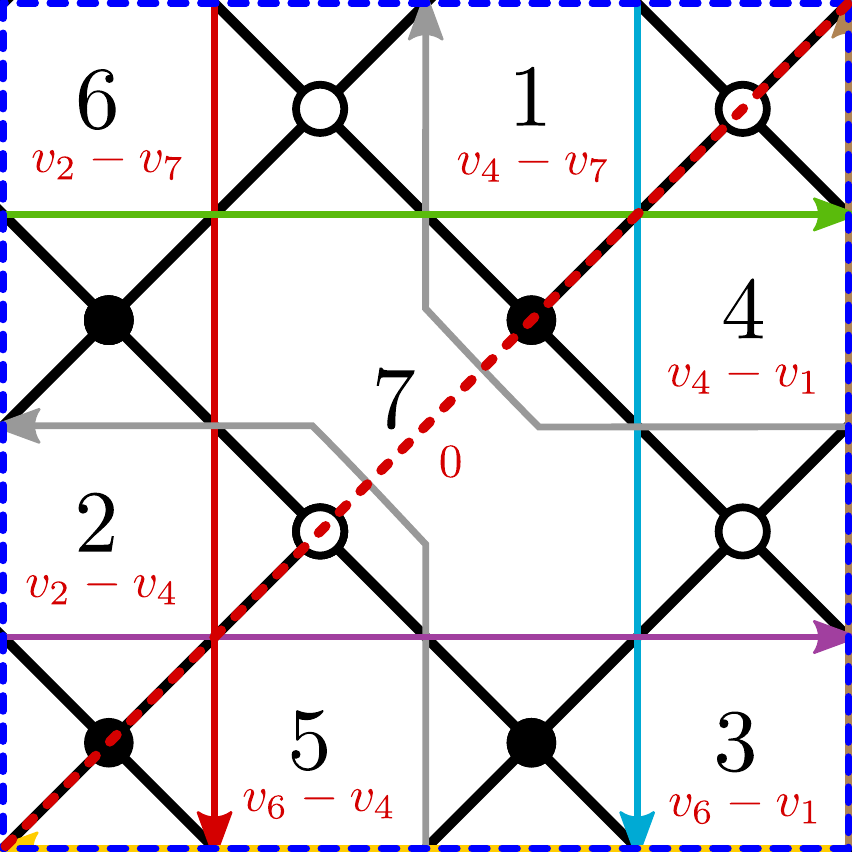}
			\caption{}
			\label{Fig:PdP4DimerZigZag2}
		\end{center}
	\end{subfigure} \hspace{20mm}
	\begin{subfigure}[t]{0.30\textwidth } 
		\begin{center} 
			\includegraphics[width=\textwidth]{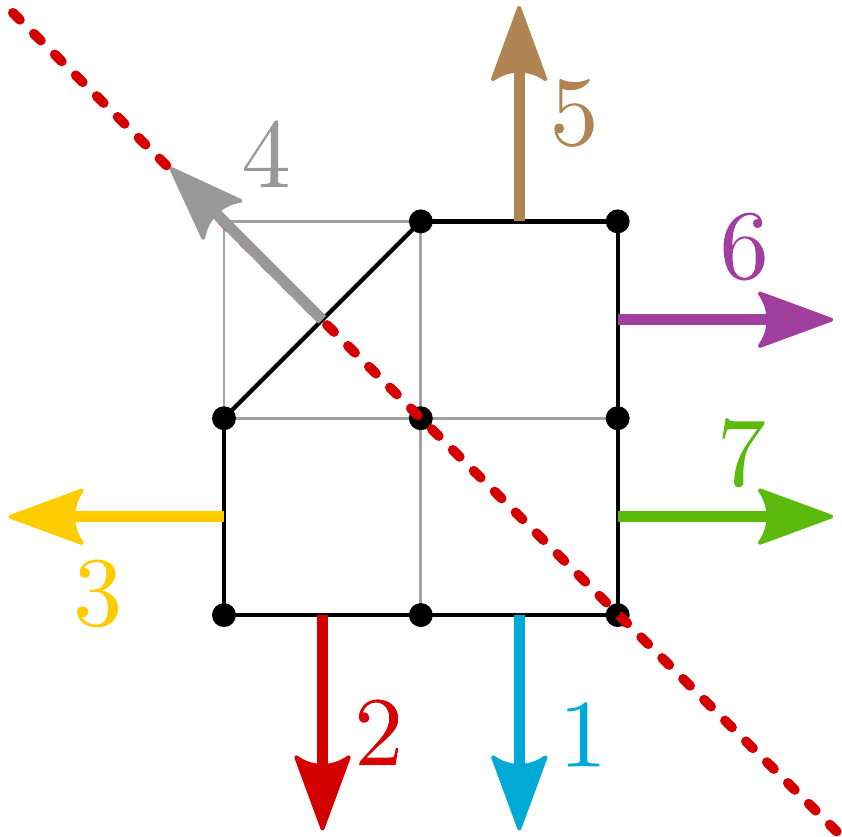}
			\caption{}
			\label{Fig:WebDiagramPdP4}
		\end{center}
	\end{subfigure}
	\caption{(a) Dimer diagram for $PdP_4$ with a diagonal fixed line (dotted red). We show the ZZPs and the rank assignments coming from them. (b) The toric/web diagram with the corresponding symmetry axis.  }\label{Fig:PdP4} 
\end{figure}

Consider now $PdP_4$, which we previously discussed in \sref{subsubsec:PdP4NoOrientifold}. \fref{Fig:PdP4} shows the dimer and toric diagram for the orientifold under consideration. In \sref{subsubsec:PdP4NoOrientifold} we saw that anomaly-free rank assignments of the mother theory are given by:
\begin{equation}
\mathbf{N} = (-v_7,v_2, v_6 - v_1, - v_1, v_6, v_2 - v_7, 0 ) \, .
\end{equation}
The topological constraints are:
\begin{align}
\L &: v_4 + v_3 = v_6 + v_7 \ , \\
M &: v_4 + v_5 = v_1 + v_2  \, . 
\end{align}

\paragraph{Adjacency matrices.}
The adjacency matrices of the mother and daughter theories are:
\begin{equation}
A = \left(
\begin{array}{c c c | c  c c | c }
0 & 0 & 1 & -1 & -1 & 0 & 1 \\
0 & 0 & 1 & -1 & -1 & 0 & 1  \\
-1 & -1 & 0 & 0 & 0 & 1 & 1 \\
\hline  
1 & 1 & 0 & 0 & 0 & -1 & -1 \\ 
1 & 1 & 0 & 0 & 0 & -1 & -1 \\ 
0 & 0 & -1 & 1 & 1 & 0 & -1 \\ 
\hline 
-1 & -1 & -1 & 1 & 1 & 1 & 0 
\end{array}
\right), \quad \quad (\bar{A} \vert f) = \left(
\begin{array}{c c c | c | c}
-1 & -1 & 1 & 1 & -4 \\ -1 & -1 & 1 & 1 & -4 \\ -1 & -1 & 1 & 1 & +4 \end{array}
\right) \ , 
\end{equation}
where, for concreteness, we have assumed that the sign of the orientifold line is negative.

\paragraph{Symmetric rank assignments.} 
Impose $v_3 = v_5, \, v_2 = v_6, \, v_1 = v_7$. The constraints $M=0, \, \L = 0$ combine into $v_4 = v_1 + v_2 -v_3$. We can use the global shift freedom to set $v_4=0$, which leads to $\mathbf{v^S} = (v_1,v_2,v_1+v_2,0,v_1+v_2,v_2,v_1)$. The resulting symmetric rank assignments in the mother and daughter theories are
\beq
\begin{array}{ccl}
\mathbf{N^S} & = & ( -v_1,v_2, v_2-v_1, -v_1,v_2,v_2-v_1,0) \\
\mathbf{\ov{N}^S} & = & ( -v_1,v_2, v_2-v_1,0) \ .
\end{array}
\eeq
Note that $\mathbf{\ov{N}^S}$ should be understood as the column vector whose first three elements refer to the faces 1--3 that have an image, while the last refers to the self-identified face 7. When considered as a row vector, one should drop the last element.

\paragraph{Antisymmetric rank assignments.} 
Impose $v_1 = -v_7, \, v_2 = -v_6, \, v_3 =- v_5 = 0, \, v_4=0$. We also need to impose the constraint $v_1+v_2 = -v_3$ with no global shift freedom. We then find a two-parameter family of antisymmetric assignments for the $v_\G$, $\mathbf{v^A} = (v_1, v_2, -v_1 - v_2, 0 , v_1+v_2, -v_2, -v_1)$. The corresponding antisymmetric rank assignment is 
\beq
\mathbf{N^A}  =  ( v_1,v_2, -v_1-v_2, -v_1,-v_2,v_1+v_2,0)\ .
\eeq
In the daughter theory, this rank assignment gives rise to the two row vectors
\beq
\mathbf{\ov{N}^A_1} = (1,0,-1)v_1, \quad \mathbf{\ov{N}^A_2} = (0,1,-1)v_2
\eeq
Let us denote by $\mathbf{f} = (-4,-4,4)^T$ the inhomogeneous part of $(\ov{A} | f)$. We find $\mathbf{\ov{N}^A_1} \cdot \mathbf{f}= -8$ and $\mathbf{\ov{N}^A_2}\cdot \mathbf{f} = -8$. We conclude that anomalies cannot be cancelled in this theory.

This example and the previous one consist of orientifolds with a diagonal fixed line. Both cases turned out to lead to theories in which anomalies cannot be cancelled. In \sref{Subsec:DiagonalLines} we will present a more detailed general analysis and discuss under which conditions such orientifolds can admit anomaly-free solutions. 

\subsubsection{An Anomaly-Free Example: $PdP_{3b}$ with Two Fixed Lines} \label{subsubsec:PdP3Oline}

\begin{figure}[h!]
	\centering
	\begin{subfigure}[t]{0.3\textwidth }
		\begin{center} 
			\includegraphics[width=\textwidth]{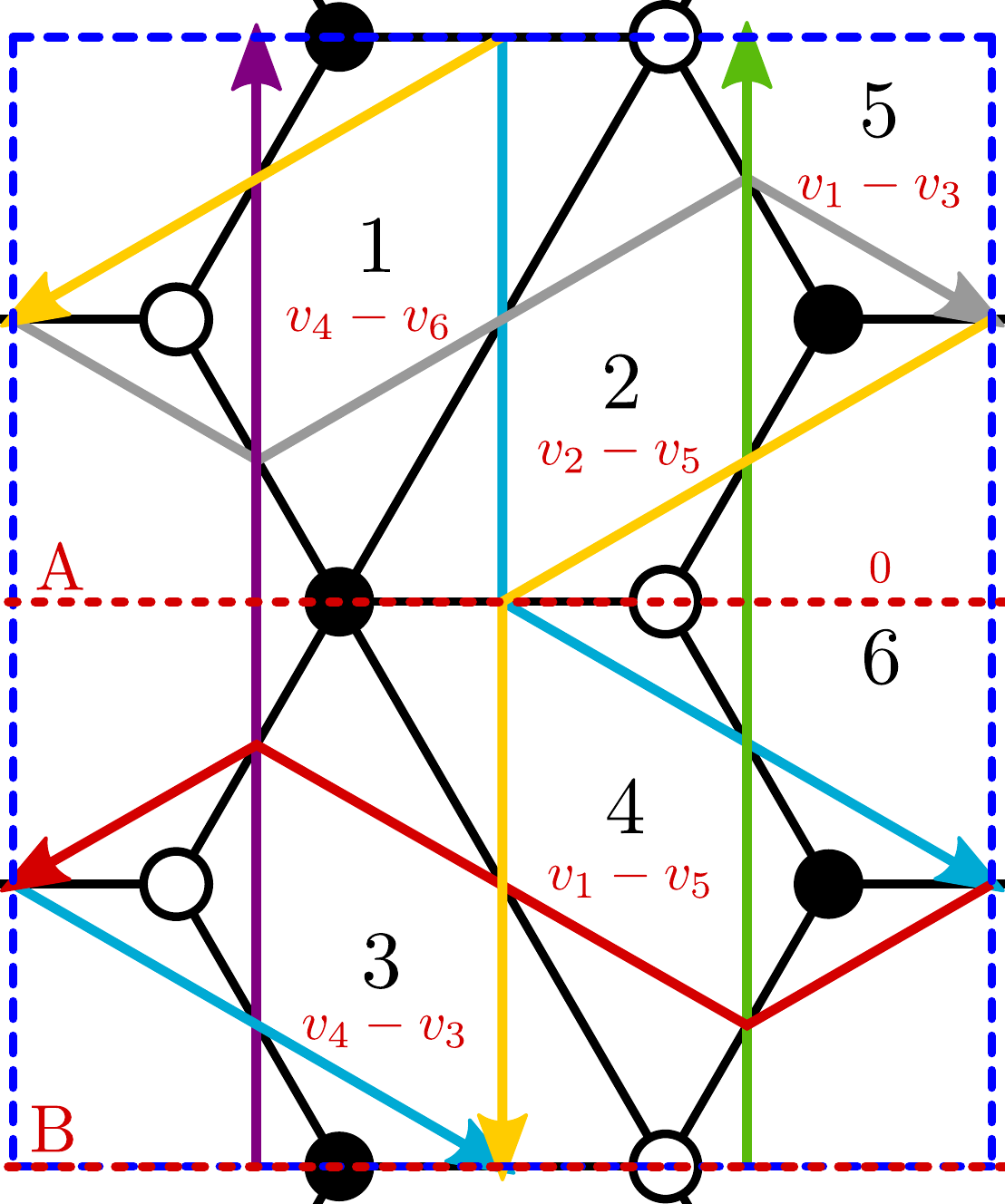}
			\caption{}
			\label{Fig:PdP3bDimer}
		\end{center}
	\end{subfigure} \hspace{15mm}
	\begin{subfigure}[t]{0.3\textwidth } 
		\begin{center}
			\includegraphics[width=\textwidth]{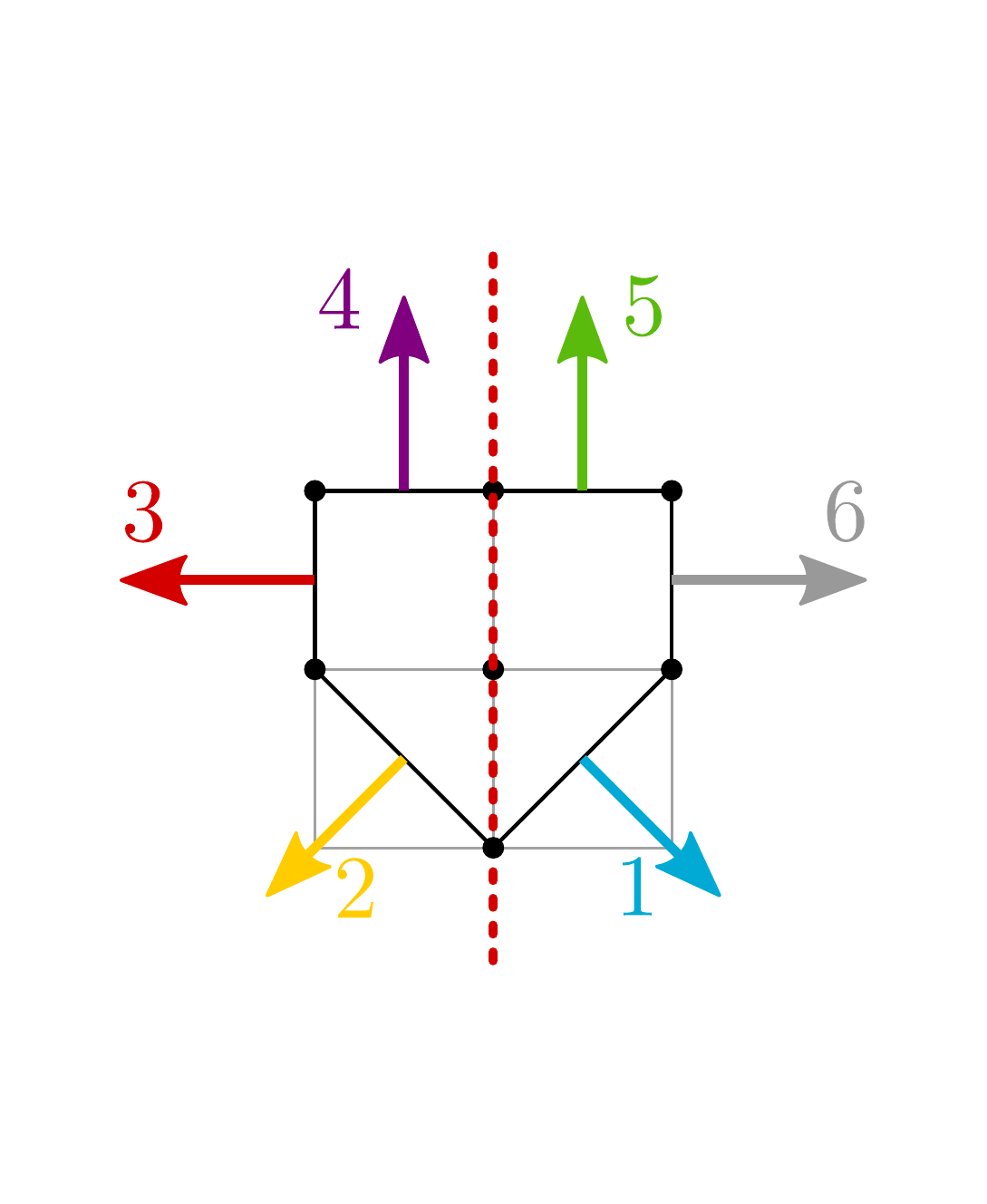}
			 \caption{}
			\label{Fig:PdP3bToricDiagram}
		\end{center}
	\end{subfigure}
	\caption{(a) Dimer diagram for $PdP_{3b}$ with two horizontal fixed lines (dotted red). We show the ZZPs and the rank assignments coming from them. (b) The toric/web diagram with the corresponding symmetry axis. }
\label{Fig:PdP3b} 
\end{figure}

\fref{Fig:PdP3b} shows the dimer and toric diagram for an orientifold of $PdP_{3b}$ with two fixed lines. This theory was studied in \cite{Argurio:2019eqb}, where it was shown that the daughter theory admits an anomaly-free rank assignment if the two O-lines have opposite signs. Note that the horizontal fixed lines in the dimer correspond to a vertical axis of symmetry in the toric diagram. 

\paragraph{Adjacency matrices.} 

The adjacency matrices of the mother and daughter theories are:
\begin{equation}\label{Eq:adjPdP3b}
A = \left(
\begin{array}{c c | c  c | c c}
0 & -1 & 1 & 0 & -1 & 1 \\ 1 & 0 & 0 & -1 & -1 & 1 \\ \hline -1 & 0 & 0 & 1 & 1 & -1 \\   0 & 1 & -1 & 0 & 1 & -1 \\ \hline 1 & 1 & -1 & -1 & 0 & 0 \\ -1 & -1 & 1 & 1 & 0 & 0
\end{array}
\right), \quad \quad (\bar{A} \vert f) = \left(
\begin{array}{c c | c c | c}
1 & -1 & -1 & 1 & -4 \cdot \text{sign}(B) \\ 1 &  -1 & -1 & 1 & +4 \cdot \text{sign}(A) \end{array}
\right) \ ,
\end{equation}
where sign$(A)$, sign$(B)$ are the signs of the two O-lines. Let us now turn to the study of symmetric and antisymmetric rank assignments.

\paragraph{Symmetric rank assignments.} 
Let us impose $v_2 = v_1, \, v_6 = v_3$. The constraint $\L=0$ is trivially satisfied, while $M = 0$ becomes (keeping $v_1$ and $v_3$):
\beq
2v_1 - v_4 = v_5 \, .
\eeq
Setting $v_4=0$, we get
\beq
\mathbf{v^S}=(v_1,v_1,v_3,0,2v_1,v_3) \, ,
\eeq
giving in turn
\beq
\mathbf{N^S} = (-v_3,-v_1,-v_3,-v_1,v_1-v_3,0)\, .
\eeq
Projecting down this vector, we obtain the solutions to the homogeneous problem in the daughter theory.

\paragraph{Antisymmetric rank assignments.} 
We now impose $v_2 = -v_1, \, v_3 = -v_6, \, v_4 = v_5 = 0$. As expected, $M$ is trivially satisfied and one just needs to impose $\L=0$, which reads $v_3 = v_1$. Remember that the global shift has already been fixed. We then find a one-dimensional family of antisymmetric assignments for the $v_\G$:
\begin{equation}
\mathbf{v^A}=(v_1,-v_1,v_1,0,0,-v_1) \, .
\end{equation}
The corresponding antisymmetric rank assignment is $\mathbf{N^A} = (v_1,-v_1,-v_1,v_1,0,0)$. In the daughter theory this rank assignment gives $\mathbf{\ov{N}^A} = (1,-1)v_1$. One may now use it to row reduce $\ov{A}$. Denote by $\mathbf{f} = (-4 \cdot \text{sign}(B),+4 \cdot \text{sign}(A))^T$ the inhomogeneous part of $ (\ov{A} | f)$. We find  $\mathbf{\ov{N}^A} \cdot \mathbf{f} = -4 \cdot \text{sign}(B) - 4 \cdot \text{sign}(A)$.  If $\mathbf{\ov{N}^A} \cdot \mathbf{f} \neq 0$, the theory is anomalous, so we need $\text{sign}(A) = -\text{sign}(B)$, as anticipated.

\paragraph{Anomaly-free rank assignments.} 

As explained in the introduction of the current section, since we have a parametrization of the symmetric rank assignments, we merely need a single solution of the tadpole-cancellation system to write all of them. 

Looking at the adjacency matrix of the daughter theory in \eref{Eq:adjPdP3b} with $\text{sign}(A)=+$ and $\text{sign}(B)=-$, a straightforward solution to the rank assignment is $N_1=4$ and $N_2=N_5=N_6=0$ (in the daughter theory we keep faces $1,2,5$ and $6$). This gives the following three-parameter family of solutions to the ACC, where we {have added $N+v_1+v_3$ regular branes: 
\begin{equation}
\left\{
\begin{array}{ccl}
	N_1 & = & N+v_1+4 \\
	N_2 & = & N+v_3\\
	N_5 & = & N+2v_1 \\
	N_6 & = & N + v_1+v_3\ . \\
	\end{array}
\right. 
\end{equation}

\subsection{Fixed Point Orientifolds}\label{subsubsec:FixedPoints}

In orientifolds with fixed points, every ZZP is mapped to a ZZP with the same winding numbers \cite{Retolaza:2016alb}. The image of a ZZP can therefore be either itself or another ZZP, if more than one ZZP with the same winding numbers exist.

Contrarily to the cases with fixed lines, in fixed point orientifolds nodes in the dimer are mapped to nodes of the opposite color. In analogy with the case of line orientifolds, let us consider a path $p$ going from a face $i$ to a face $j$, and its image $p'$ going from the image of $i$ to the image of $j$. If $p$ crosses a ZZP $\a$, then $p'$ crosses its image $\a'$, but since the color of the nodes is inverted in the image, the signs of the crossings are opposite. This implies that a symmetric, respectively antisymmetric, rank assignment is associated to an antisymmetric, respectively symmetric, value assignment for the ZZP. We therefore have:

\begin{mdframed}[backgroundcolor=blue!20]
	In dimer models with fixed point involutions, symmetric rank assignments up to (half)-regular branes correspond bijectively to antisymmetric ZZP value assignments which satisfy the topological constraints. Similarly, antisymmetric rank assignments correspond bijectively to symmetric ZZP value assignments which satisfy the topological constraints and up to a global shift.
\end{mdframed}

We have seen that in the cases of fixed point orientifolds, symmetric rank assignments correspond to ZZP value assignments such that:
\begin{equation}
v_{\a} = - v_{\ov{\a}}, \quad v_{\g}=0 \, .
\end{equation}
One can easily verify that the topological constraints are always satisfied by this choice of $v_\Gamma$, hence the number of symmetric rank assignment is:
\begin{equation}
\text{dim}(\text{ker}(\ov{A})) = \frac{1}{2}(n-n_s)  \, .
\end{equation}

Antisymmetric rank assignments, conversely, correspond to:
\begin{equation}
v_{\a} =  v_{\ov{\a}}, \quad v_{\g}= \text{free} \, .
\end{equation}
In this case both topological constraints $\Lambda$ and $M$ are not trivial:
\begin{equation}
\begin{cases}
\displaystyle \L =  \sum_\a p_\a v_\a + \sum_{\ov{\a}} p_{\ov{\a}}v_{\ov{\a}} +\sum_{\gamma} p_{\g} v_{\gamma} =2\sum_\a p_\a v_\a + \sum_{\g} p_{\g} v_{\g}=0  \\
\displaystyle M =  \sum_\a q_\a v_\a + \sum_{\ov{\a}} q_{\ov{\a}} v_{\ov{\a}}+\sum_{\gamma} q_{\g} v_{\gamma}=2 \sum_\a q_\a v_\a +\sum_{\gamma} q_{\g} v_{\gamma}=0
\end{cases}
\end{equation} 

This leads to:
\begin{equation}
\text{dim}(\text{coker}(\ov{A})) = \frac{1}{2}(n+n_s) -3 \, . \label{Eq:DimantiSym}
\end{equation} 

Upon summing the contributions of symmetric and antisymmetric rank assignments, we retrieve the total number of fractional branes, $n-3$, modulo regular branes.

\subsubsection{An Example: $PdP_{3b}$}
Let us return to $PdP_{3b}$, already studied in \sref{subsubsec:PdP3Oline} but now with fixed points instead of lines. The dimer is shown in \Cref{Fig:PdP3bDimerPoint}. Note that we have changed the unit cell and face numbering with respect to \Cref{Fig:PdP3b} to make it consistent with fixed point reflections. 

\begin{figure}[h!]
	\centering
	\includegraphics[width=0.3\textwidth]{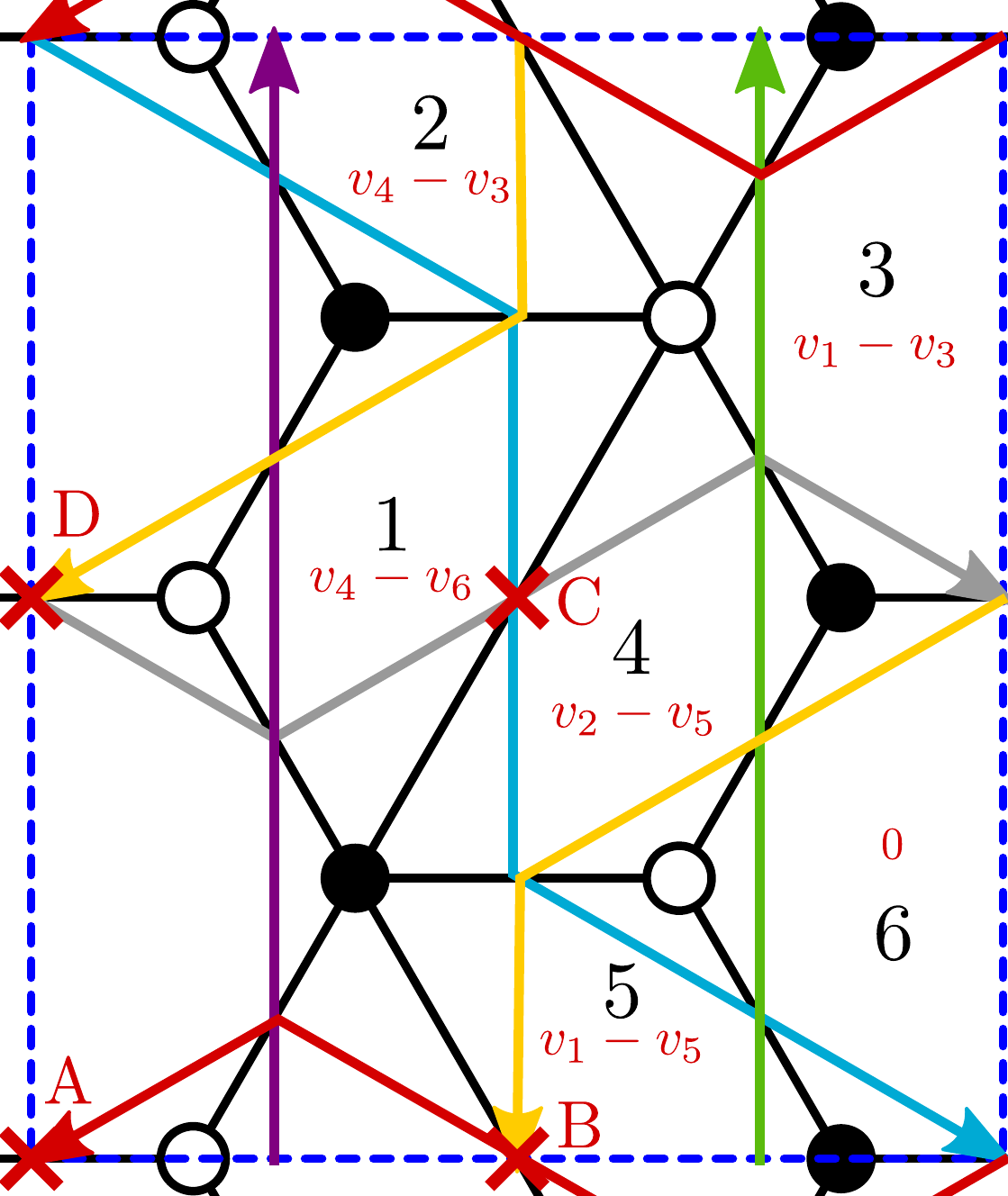}
	\caption{Dimer diagram for $PdP_{3b}$ with fixed points. We show the ZZPs and the rank assignments coming from them.}\label{Fig:PdP3bDimerPoint} 
\end{figure}

\paragraph{Adjacency matrices.} The adjacency matrices of the mother and daughter theories are: 
\begin{equation}\label{Eq:adjPdP3bPoint}
	A = \left(
	\begin{array}{c c  c | c  c c}
		0 & 1 & -1 & -1 & 0 & 1 \\ -1 & 0 & 1 & 0 & 1 & -1 \\  1 & -1 & 0 & 1 & -1 & 0 \\  \hline 1 & 0 & -1 & 0 & -1 & 1 \\  0 & -1 & 1 & 1 & 0 & -1 \\ -1 & 1 & 0 & -1 & 1 & 0
	\end{array}
	\right), \quad \quad (\bar{A} \vert f ) = \left(
	\begin{array}{c c c | c}
		-1 & 1 & 0 & +4 \cdot \text{sign}(C) \\ -1 & 1 & 0 & - 4 \cdot \text{sign}(B) \\ 2 & -2 & 0 & - 4 \cdot \text{sign}(A) + 4 \cdot \text{sign}(D) \end{array}
	\right) \ ,
\end{equation}
where sign$(A)$ to sign$(D)$ are the signs of the O-points. Note that the ZZPs $4$ and $5$ are interchanged by the projection, while all other ZZPs are mapped to themselves.  Let us now turn to the study of antisymmetric and symmetric rank assignments.

\paragraph{Symmetric rank assignments.} 
This time we start with antisymmetric ZZP assignments, since for point orientifolds they provide symmetric rank assignments. Let us impose $v_4 = -v_5$,  $v_1=v_2=v_3=v_6 = 0$. As already said, the topological constraints are both trivially satisfied. Note that there is no global shift to fix. We obtain a one-parameter family of $v_\G$ assignments:
\begin{equation}
	(0,0,0,1,-1,0)v_4 \, .
\end{equation}
The corresponding rank assignment is:
\begin{equation}
	\mathbf{N^S} = (1,1,0,1,1,0)v_4 \, ,
\end{equation}
which is symmetric, as expected. Projecting down this vector, one obtains the solutions to the homogeneous problem in the daughter theory.

\paragraph{Antisymmetric rank assignments.} 

We now turn to symmetric ZZP assignments, responsible for the antisymmetric rank assignments. We only need to impose $v_4 = v_5$. We further fix the global shift by choosing $v_4 = 0$. The topological constraints become: 
\begin{align}
	\begin{split}
		\L: v_3 = v_1-v_2+v_6 \ ,\\
		M: v_2 = -v_1 \ .
	\end{split} 
\end{align}
We find a two-dimensional family of symmetric assignments for the $v_\G$:
\begin{equation}
	(1,-1,2,0,0,0)v_1 + (0,0,1,0,0,1)v_6 \, .
\end{equation}
The corresponding antisymmetric rank assignments are:
\begin{equation}
	\mathbf{N}^A = (0,-2,-1,-1,1,0)v_1 + (-1,-1,-1,0,0,0)v_6 \ .
\end{equation}
Which, up to half regular branes is equal to:
\begin{equation}
	\mathbf{N}^A = (1,-3,-1,-1,3,1)\frac{v_1}{2} + (-1,-1,-1,1,1,1)\frac{v_6}{2} \, ,
\end{equation}
which is antisymmetric, as expected. Let us split it into two vectors and project them down to the daughter theory to obtain,
\begin{equation}
	\mathbf{\ov{N}}^A_1 = (1,-3,-1)\frac{v_1}{2}, \quad \mathbf{\ov{N}}^A_2 = (-1,-1,-1)\frac{v_6}{2}\ .
\end{equation}
Again, let us use these rank assignments to row reduce $\ov{A}$ by denoting $\mathbf{f} = (+4 \cdot \text{sign}(C),-4 \cdot \text{sign}(B),-4 \cdot \text{sign}(A) +4 \cdot \text{sign}(D))^T$. One finds that, for the theory to admit non-anomalous solutions, one must satisfy,
\begin{align}
	\begin{split}
		\mathbf{\ov{N}}^A_1 \cdot \mathbf{f} =& \frac{v_1}{2} \left(\text{sign}(C) + 3\text{sign}(B) +\text{sign}(A)-\text{sign}(D)\right)=0\ ,\\
		\mathbf{\ov{N}}^A_2 \cdot \mathbf{f} =& \frac{v_6}{2} \left(-\text{sign}(C) + \text{sign}(B) +\text{sign}(A)-\text{sign}(D)\right)=0\ .
	\end{split}\label{Eq:TadpoleFixedPoint}
\end{align}

\paragraph{Anomaly-free rank assignments.} 
The solution to \eref{Eq:TadpoleFixedPoint} depends on the sign choices for the four fixed points. Consider for example
\begin{align}
	\text{sign}(A) = \text{sign}(C) = +, \quad  \text{sign}(B) = \text{sign}(D)= - \, ,
\end{align}
which is consistent with the sign rule for fixed point orientifolds. In this case, we go back to \eref{Eq:adjPdP3bPoint} to find a two-parameter family of solutions:
\begin{equation}
	\left\{
	\begin{array}{ccl}
		N_1 & = & N+v_4\\
		N_2 & = & N+v_4+4 \\
		N_3 & = & N\ . \\
	\end{array}
	\right. 
\end{equation}

\section{General Criteria for Anomaly-Free Orientifolds}
\label{Sec:Anomalies}

In this section we present a general study of the solutions to the non-homogeneous system of ACC of the daughter theory. Remarkably, we can exploit the algorithm of the previous section to determine the existence of such solutions directly from toric data, regardless of the particular phase of the theory. This gives a purely geometric criterion determining whether an orientifolded theory may admit a toric phase with non-anomalous rank assignments.

\subsection{Diagonal Line Orientifolds}\label{Subsec:DiagonalLines}

Let us consider orientifolds with a diagonal fixed line. Without loss of generality, we assume that the fixed line has winding numbers $(1,1)$ in the fundamental cell of the dimer. The mapping of ZZPs in this kind of orientifolds has been studied in \cite{Retolaza:2016alb} and we presented a preliminary discussion in \sref{subsubsec:FixedLines}. The diagonal fixed line in the dimer translates into a reflection symmetry axis in the toric diagram with slope $-1$, as we already illustrated in \fref{Fig:DiagonalOLine}. This 90$^{\circ}$ rotation of the symmetry axis of the toric diagram with respect to the fixed line in the dimer was explained in \cite{Yamazaki_2008}.

Reflection with respect to the axis of the toric diagram maps a ZZP with winding $(p,q)$, to a ZZP with winding $(-q,-p)$. \fref{setdiagtadobs} shows an example of a generic toric diagram with a diagonal line orientifold.
\begin{itemize}
	\item Let $l$ be the number of pairs $\lbrace v_\alpha,v_{\ov \alpha}\rbrace$ ,with $\alpha=1,..,l$, of ZZPs mapped one to another, which are not parallel to the symmetry axis of the toric diagram.
	\item Let $l_\parallel$ be the number of self-identified ZZPs $\lbrace v_\gamma \rbrace$ for $\gamma=1,...,{l_\parallel}$, which are parallel to the symmetry axis of the toric diagram. 
\end{itemize} 

\begin{figure}[h!]
	\centering
	\includegraphics[scale=0.45]{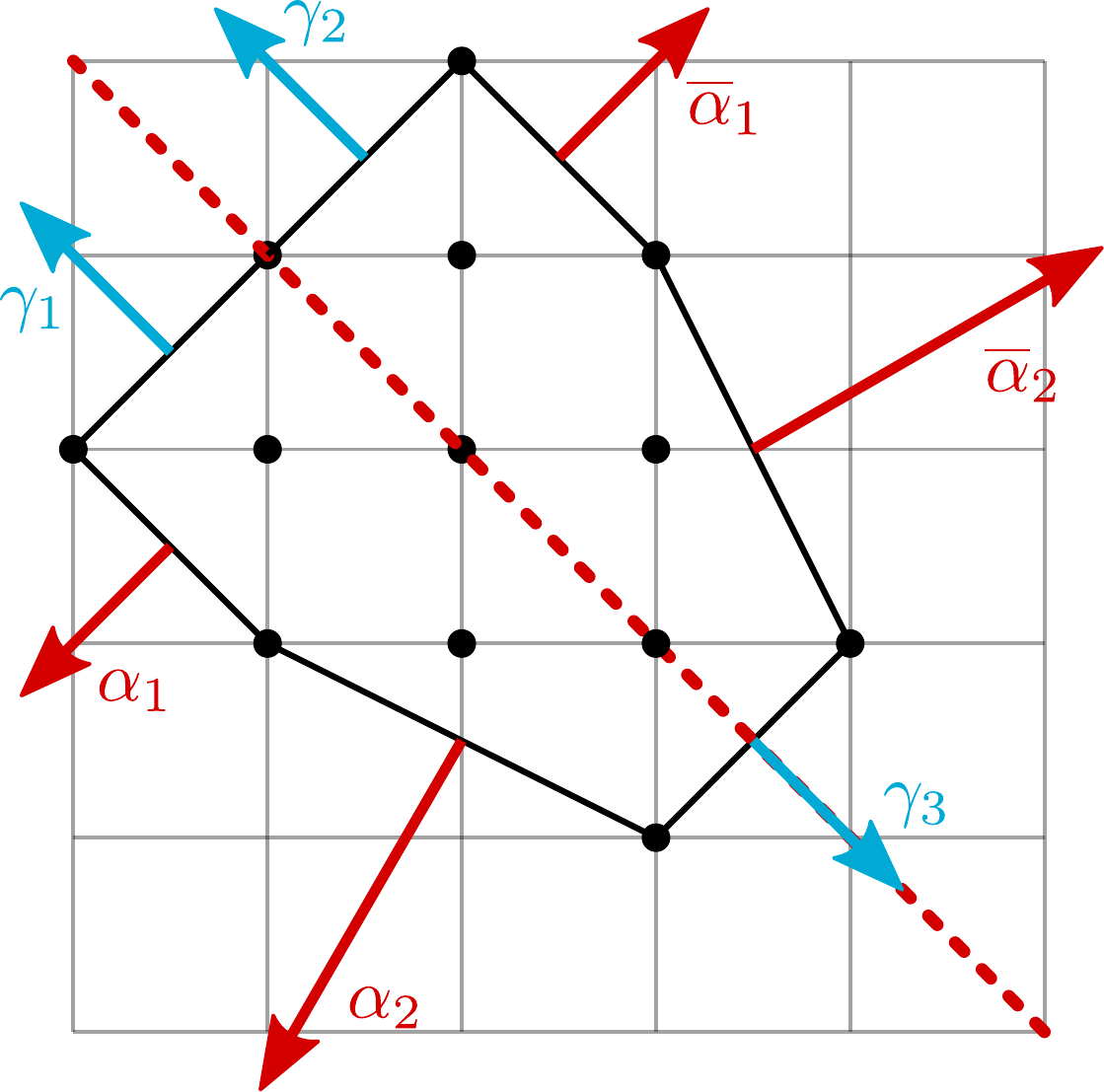}
	\caption{A generic toric diagram with a diagonal axis symmetry.}
	\label{setdiagtadobs}
\end{figure}

From the previous section, we know how to produce the coefficients of the trivial linear combinations of rows. They are the ranks of the projected $SU$ groups that result from imposing the following conditions on the $v_{\Gamma}$:
\begin{eqnarray}\label{vantisym}
v_{\alpha} &=& - v_{\overline{\alpha}} \, , \nonumber \\
v_{\gamma} &=& 0 
\end{eqnarray}
for all $\alpha$ and $\gamma$'s. The topological constraints $\Lambda$ and $M$ are given by:
\begin{align}\label{antiLambda}
\Lambda = \sum_{\alpha, \ov{\alpha}}(v_\alpha p_\alpha + v_{\ov \alpha} p_{\ov \alpha})
= \sum_{\alpha}v_\alpha (p_\alpha + q_{ \alpha})  =-M
\end{align}
where we used $p_{\ov \alpha}=-q_{\alpha}$.

We now recall the Rouch\'e-Capelli theorem: A non-homogeneous linear system has solution \textit{iff} the rank of the associated homogeneous matrix is equal to the rank of the matrix associated to the full system. A trivial linear combination of rows of the homogenous matrix is still trivial when considering the matrix associated to the full system. This can be stated as:
\begin{equation}\label{RC}
\sum_i N_i f_i=0 
\end{equation}
where $f_i$ is the non-homogeneous contribution to the ACC matrix of the orientifolded theory, coming from the tensor matter. 

We now  need to derive an expression for $N_i$ in terms of the $v_{\Gamma}$. The Rouch\'e-Capelli theorem tells us that the ACC system admits a solution \textit{iff} \eref{RC} holds for every value of $v_\Gamma$ consistent with the topological constraints.

\subsubsection{Faces with at Most One Tensor}

Let us first focus on the simpler case where every gauge group has at most one tensor field. This result will be easily extended later to cases with more tensors. Consider a face of the mother theory with an edge on top of a fixed line. The rank assignment providing the coefficients for row reduction is given by the condition $N_i=-N_{i+k}$, $v_\alpha=- v_{\ov \alpha}$, and the difference between the ranks of two adjacent faces is given by $N_i-N_{i+k}=v_{\alpha}-v_{\ov \alpha}$. Combining these two results, we obtain
\begin{eqnarray}
2N_i=N_i-N_{i+k} \, = \, v_\alpha - v_{\ov \alpha}=2 v_\alpha \, . \label{antisymmetricranks}
\end{eqnarray}

Let us now determine the $f_i$ from the toric data. The method we are going to discuss below can be regarded as a generalization to orientifolds of the algorithm for finding the (minimal) matter content of a quiver in terms of basic knowledge of the $(p,q)$ winding numbers of its ZZPs (equivalently of the external legs of the $(p,q)$ web dual to the toric diagram). The intersection number between a given ZZP and the fixed line is\begin{equation}\label{cross}
\text{det} \begin{pmatrix}
p & 1 \\
q & 1 
\end{pmatrix} 
=p-q \, .
\end{equation}
At every such crossing this ZZP, if not self-identified, will intersect its image on the line. The edge on which they cross will produce a tensor or conjugate tensor field, depending on the orientation of the crossing. This is depicted in \fref{ZZPs_diagonal_line}. 

\begin{figure}[h!]
	\centering
	\includegraphics[width=3cm]{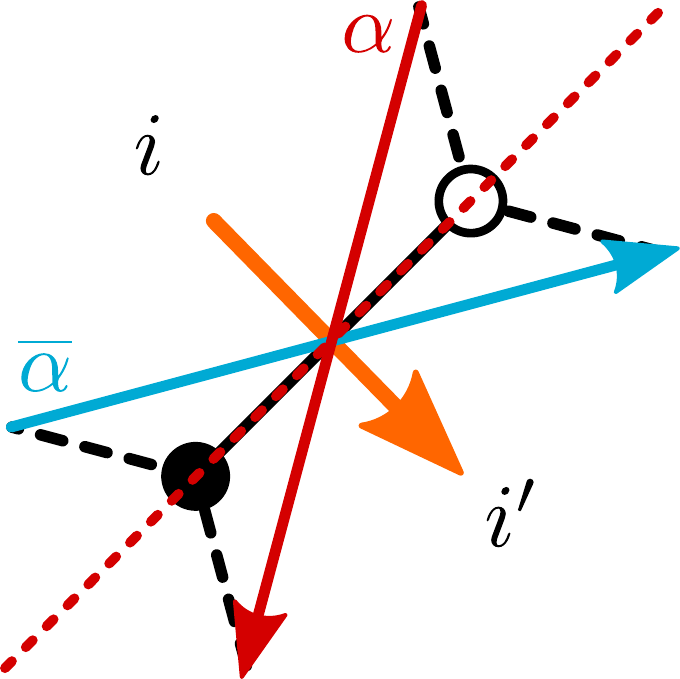}
	\caption{Crossing between a ZZP (and its image) over an edge on top of a diagonal fixed line. We show the corresponding bifundamental field in the mother theory.}
	\label{ZZPs_diagonal_line}
\end{figure}

From the discussion above, it is clear that the non-vanishing components of $f_i$ are exactly those corresponding to the faces with a tensor, for which we have just determined the rank in terms of the ZZP values. Taking into account that the same ZZP can be related to $p_\alpha-q_\alpha$ tensors,
this allows us to write \eref{RC} as
\begin{equation}\label{RCvi}
\sum_i N_i f_i=(\pm 4)\sum_{\alpha} v_\alpha (p_\alpha - q_\alpha) =0 \, ,
\end{equation}
where we have factorized the choice of sign for the diagonal O-line.

It is worth noting that the intersection with sign is a topological quantity that counts the minimal number of intersections of the ZZP with the fixed line in the dimer. This is, in fact, a homological invariant. In principle, more intersections are allowed, but they will come in pairs, one with positive and one with negative intersection, as shown in \fref{ZZPs_triple_intersection_diagonal_line}. When computing the total contribution they cancel, leaving us with \eref{RCvi}, which does not depend on the particular phase we are considering.
\begin{figure}[h!]
	\centering
	\includegraphics[width=5.5cm]{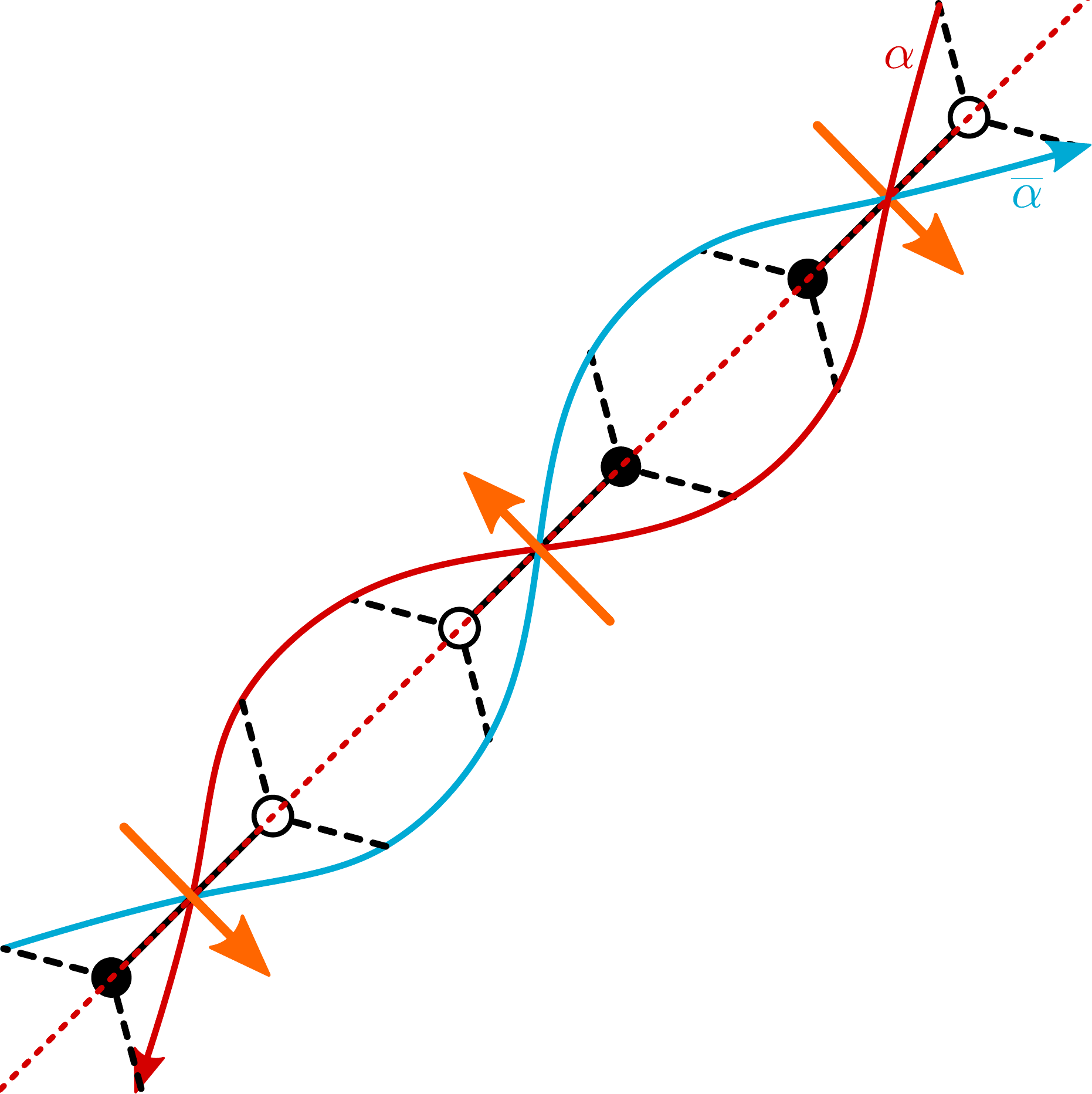}
	\caption{When ZPPs are deformed, additional intersections are added in pairs. We show the corresponding bifundamental fields in the mother theory.}
	\label{ZZPs_triple_intersection_diagonal_line}
\end{figure}

We use the topological constraint \eref{antiLambda} to express the value assigned to $v_1$, as
\begin{equation}
v_{1}=- \frac{1}{p_1 + q_1}\sum_{\alpha \neq 1}v_{\alpha}(p_{\alpha}+q_{\alpha}) \, .
\end{equation}
Plugging this expression into \eqref{RCvi} and rearranging the terms, we reach the following equality:
\begin{equation}\label{RCvib}
\sum_{\alpha \neq 1} v_{\alpha}\left(p_\alpha q_1-p_1q_\alpha\right) =0 \, .
\end{equation}
Then, the Rouch\'e-Capelli theorem can be satisfied for generic $v_\alpha$ \textit{iff}
\begin{equation}
p_\alpha q_1-p_1q_\alpha\equiv\text{det} \begin{pmatrix}
p_\alpha & p_1 \\
q_\alpha & q_1 
\end{pmatrix}=0\ ,
\end{equation}
which implies that $p_\alpha=p_1, q_\alpha=q_1$ for every $\alpha$. We dub the corresponding class of toric diagrams the \textit{trapezoids}. An example of such a trapezoid is shown in \fref{trapezoid}. 
Among trapezoids, we of course include also \textit{triangles}.

\begin{figure}[h!]
	\centering
	{\includegraphics[scale=0.45]{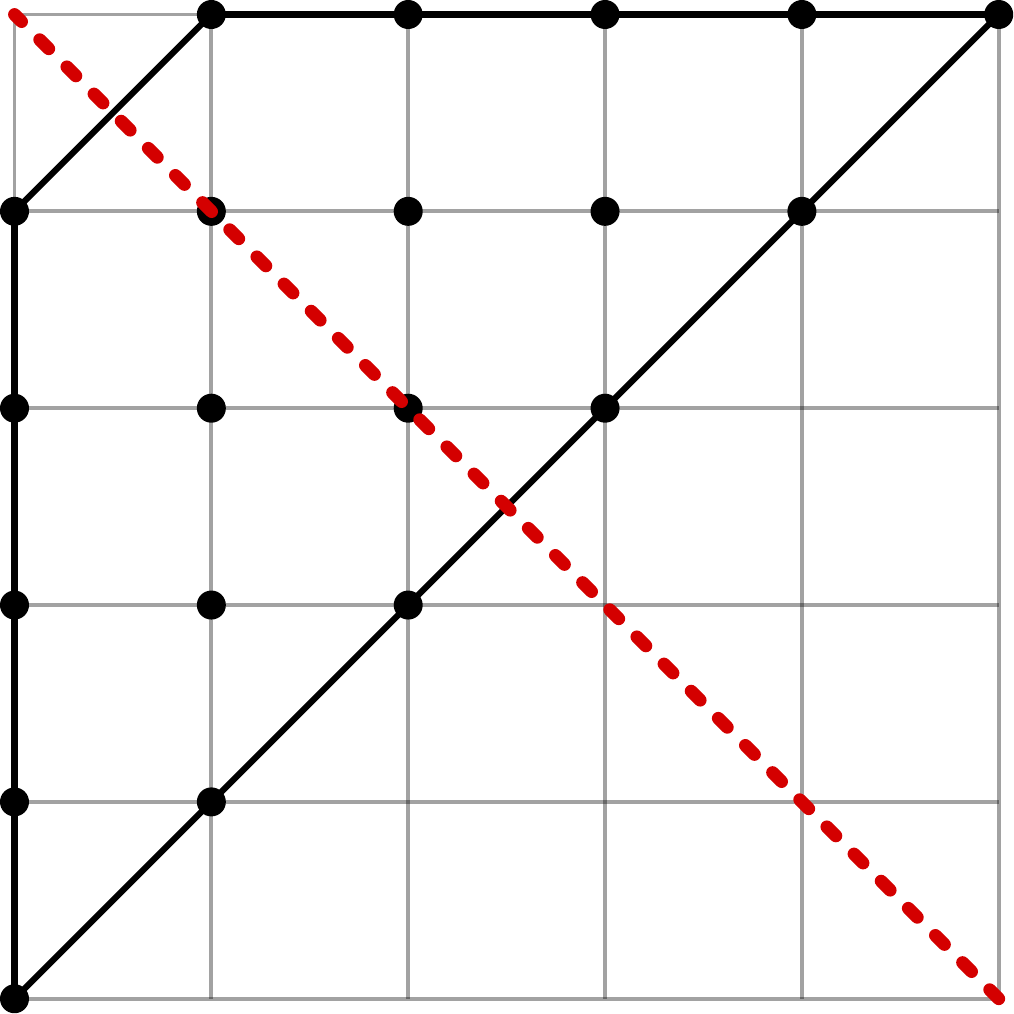}}
	\caption{An example of a trapezoid for which you can find a non-anomalous diagonal line orientifold.}
	\label{trapezoid}
\end{figure}

Note also that there is a subset of trapezoids for which \eqref{RCvi} is trivially satisfied. They have $p_\alpha = q_\alpha$ for every $\alpha$ so we refer to them as the \textit{rectangles}, and describe orbifolds of $F_0$. See \fref{rectangle} as an example. We remark that rectangles are the toric diagrams that give rise to line orientifolds without tensors in the spectrum. Thus, we recover the result that the latter always admit a non-anomalous solution.
\begin{figure}[h!]
	\centering
	\includegraphics[scale=0.47]{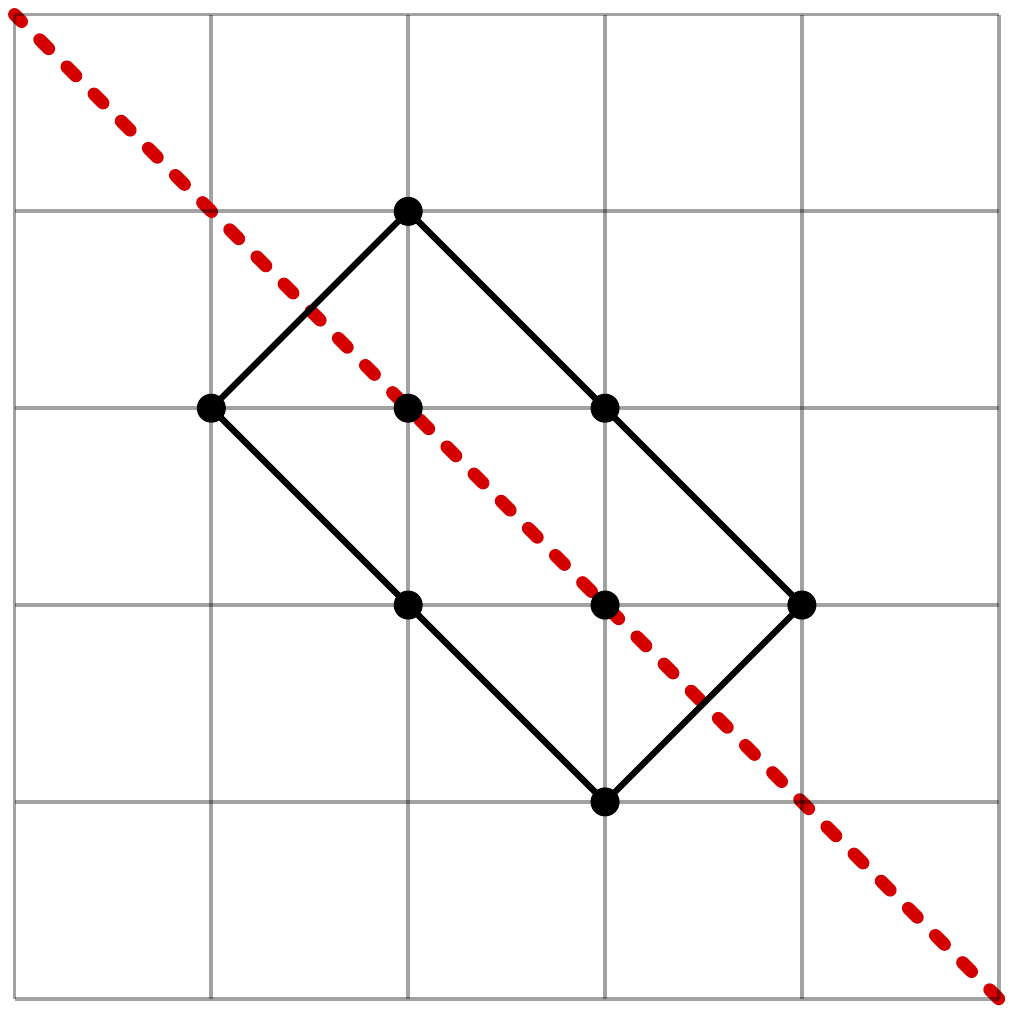}
	\caption{An example of a rectangle toric diagram with its  diagonal axis of symmetry. }\label{rectangle}
\end{figure}

\begin{mdframed}[backgroundcolor=blue!20]
	\textbf{Preliminary result for diagonal line orientifolds}: Unless the toric diagram of the singularity under consideration is a trapezoid, any orientifold theory obtained from a dimer symmetric with respect to its diagonal, \textit{and in which every face has at most one edge along this diagonal}, does not admit anomaly-free solutions.
\end{mdframed}

\subsubsection{Faces with Multiple Tensors}

Faces with multiple tensors arise in examples as simple as the conifold or $\mathbb{C}^2 / \mathbb{Z}_{2n+1}$ orbifolds, upon orientifolding with respect to a diagonal line. We now discuss how the previous discussion is extended to these cases. We start by considering how multi-tensor faces may be embedded in the dimer. We will see that there are restrictions on the number of tensors a face can have. Moreover, their existence is non-trivial and imposes constraints on the toric diagrams. The analysis of this case is slightly different from the one in the previous section but will lead to the same result.

Interestingly, it is possible to find an upper bound on the number of tensors a face in the dimer can have. \fref{massterm} shows a face with two self-identified edges on the same side of the O-line. 
If they were adjacent, they would be connected at a 2-valent node, which corresponds to a mass term and then they could be integrated out. Naively, we might imagine that this can be avoided by introducing additional structure between the two edges, which is represented as a blob in \fref{massterm}. But the ZZPs generating the edges on the line are the only ones that participate in the blob. In other words, the orange and purple ZZPs in \fref{massterm} must be identified with the blue and green ZZPs, with the precise identification depending on the number of intermediate edges. Therefore, the blob can only correspond to a sequence of edges connected by mass terms. After integrating them out, we are left with either zero or one tensor for an even or odd number of edges, respectively. This implies that a given face can only support more than one tensor in two cases: if they belong to different O-lines or if they belong to the same O-line but are coming from different copies of the unit cell as illustrated in \fref{manytensors}. In both cases, the previous analysis applies to each instance that the face touches a fixed line, so we conclude that the maximum possible number of tensors at a given face is two. The total number of tensors in the full theory is, however, unrestricted. 

\begin{figure}[h!]
	\centering
	\includegraphics[width=5.4cm]{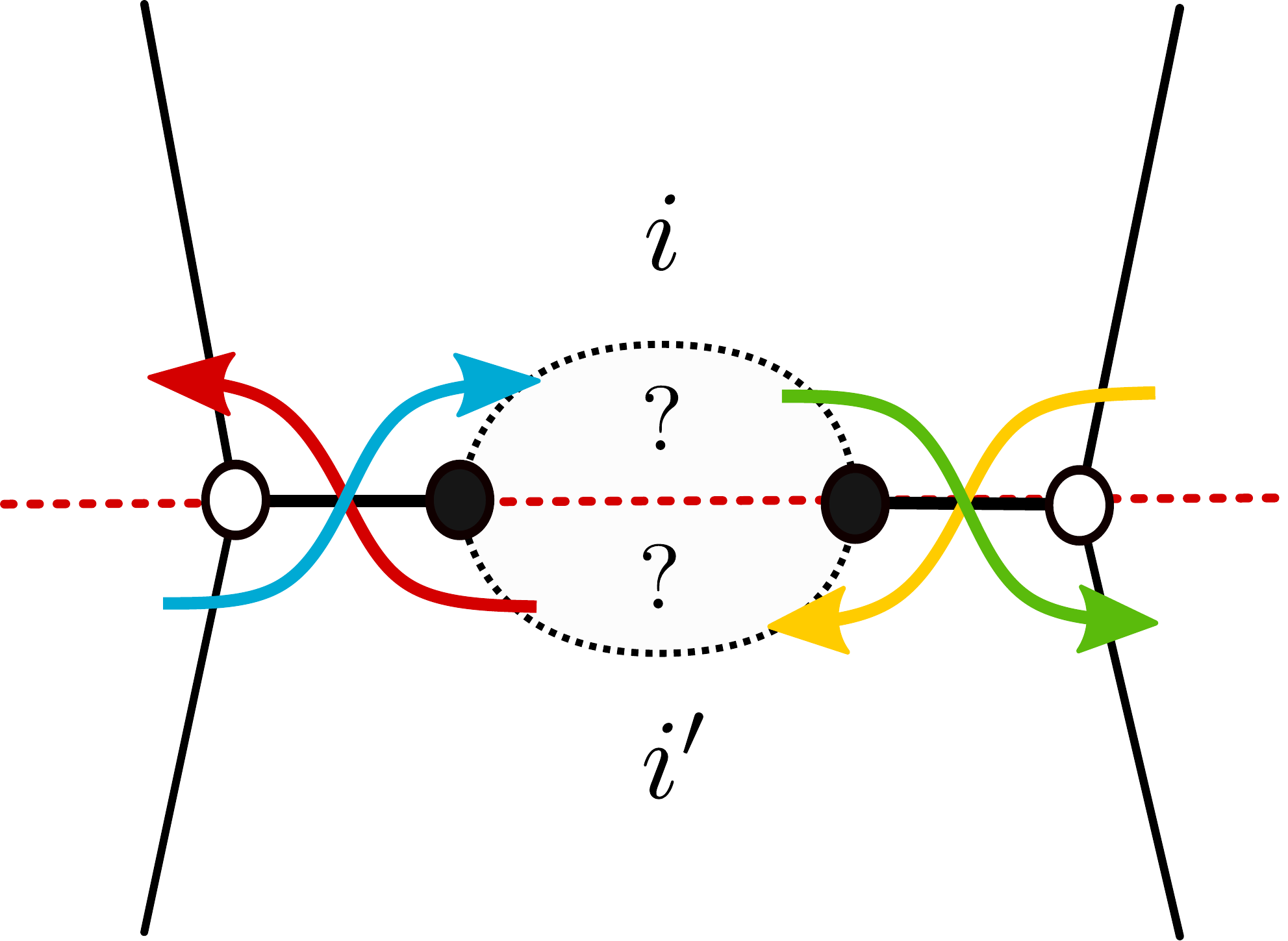}
	\caption{Two edges of a given face on a fixed line, separated by a general structure.}
	\label{massterm}
\end{figure}

\begin{figure}[h!]
	\centering
	\includegraphics[width=7.1cm]{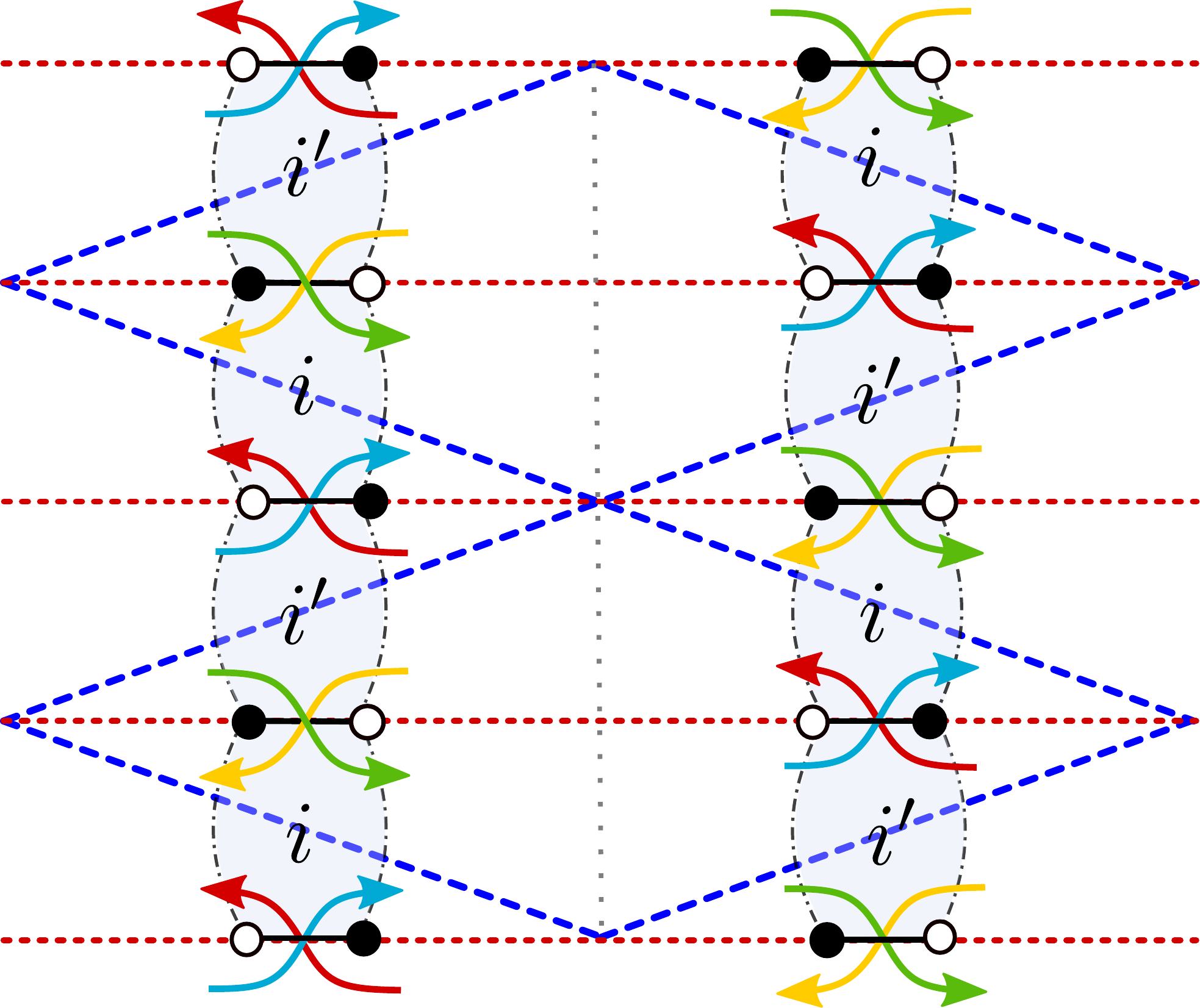}
	\caption{Faces with edges on top of the fixed line at different copies of the unit cell.}
	\label{manytensors}
\end{figure}


From \fref{manytensors}, we see that there can be three types of ZZPs: ZZPs parallel to the fixed line, which are forbidden since they would have to go through the face with two tensors, spoiling it; ZZPs orthogonal to the fixed line, i.e. self-identified ZZPs, which do not give rise to tensors; finally, ZZPs which intersect in pairs on self-identified edges giving rise to tensors. Thus, the singularity can only have self-identified ZZPs, those of the $\g$ kind, and at most two couples of ZZPs of the $\a$ kind. Moreover, the $(p,q)$ numbers of the latter are also subject to constraints. They cannot cross faces $i$ and $i^\prime$ otherwise than passing by the O-lines, so they can intersect the grey dotted axis in \fref{manytensors} at most twice if only one couple of ZZPs $\a$ is involved:
\begin{equation}
\vert p_\alpha +q_\alpha \vert \leq 2 \quad \text{for $\alpha =1$} \, ,
\end{equation}
and once in the case of two couples:
\begin{equation}
\vert p_\alpha +q_\alpha \vert  = 1 \quad \text{for $\alpha =1,2$}\, . \label{twotensorsconstraint}
\end{equation}
Those relations apply both for ZZPs $\alpha$ and $\bar{\alpha}$, for which the sums are respectively negative and positive.

If there is only one couple, the singularity corresponds to a trapezoid as the ones discussed in the previous section. Indeed, we have only one couple of ZZPs of the $\a$ kind and the topological constraint imposes $v_1=0$ for them, turning the RC condition into a trivial equation. 

For two couples, the topological constraints and \eqref{twotensorsconstraint} impose
\begin{equation}
v_1 = - v_2 \, .\label{antisymmetrictwotensors}
\end{equation}
This is the counterpart of the fact that faces $i$ and $i^\prime$ in \fref{manytensors} have to be of opposite ranks, following \eqref{antisymmetricranks}. Now, we can write the RC condition allowing faces to support one or two tensors in terms of $v_1$ only:
\begin{equation}\label{RCvi2}
\sum_i N_i f_i=(\pm 4) (v_1 (p_1 - q_1) - v_1 (p_2 - q_2)) =0 \, .
\end{equation}
Knowing \eqref{twotensorsconstraint}, the only solution is $(p_1,q_1)=(p_2,q_2)$ so that we recover trapezoids. Let us note that the last equation considered with \eqref{antisymmetrictwotensors} can be brought to the form of \eqref{RCvib} for two couples of ZZPs $\alpha$, hence it is not surprising that a subset of trapezoids appears again as solutions in this context. For instance, the conifold does not provide a non-anomalous diagonal line orientifold while $\mathbb{C}^2 / \mathbb{Z}_{2n+1}$ orbifolds do.

We conclude with a  general result for diagonal line orientifolds:
\begin{mdframed}[backgroundcolor=blue!20]
	\textbf{Diagonal line orientifolds}: Unless the toric diagram of the singularity is a \textit{trapezoid}, any orientifold theory obtained from a dimer with a diagonal O-line is anomalous.
\end{mdframed}

See \fref{3trapezoids} for more examples.

\begin{figure}[h!]
	\centering
	{\includegraphics[width = 1.6in]{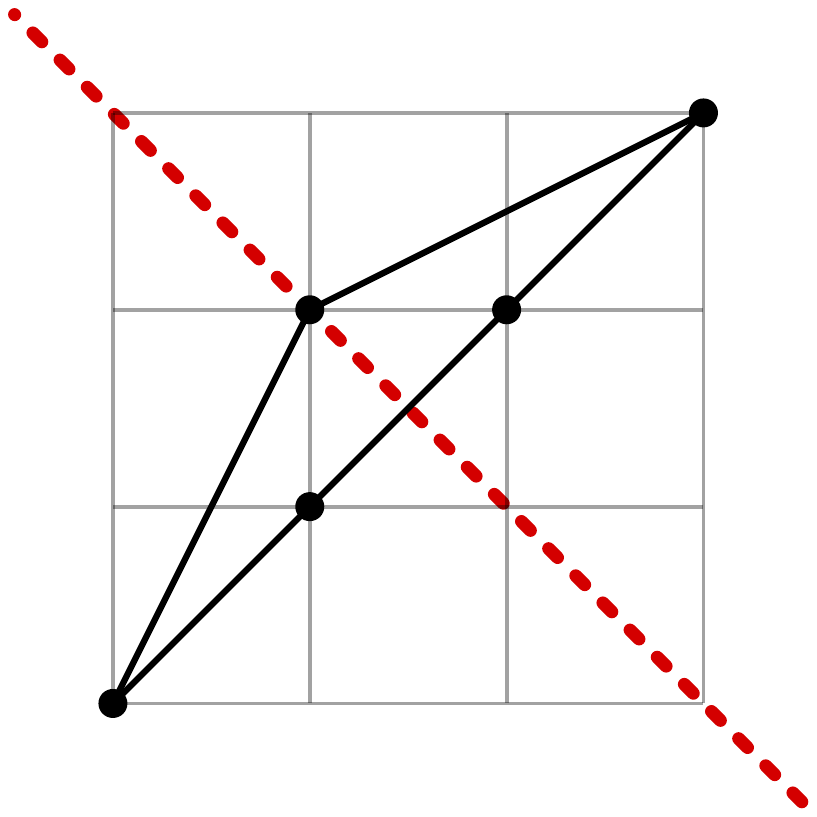}}\hskip 15pt 
	{\includegraphics[width = 1.6in]{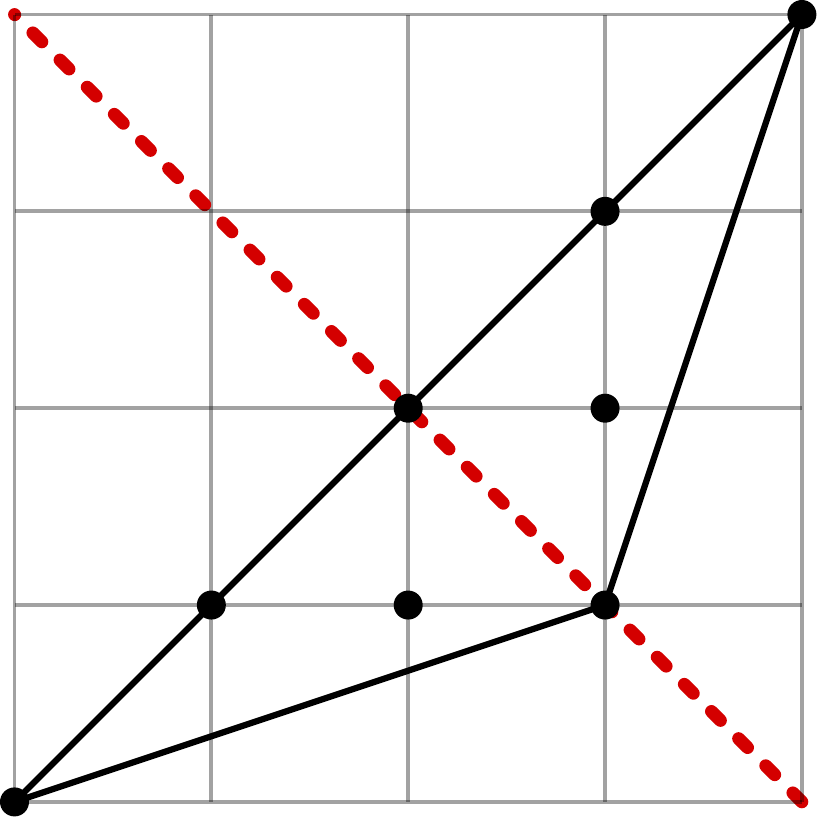}} \hskip 15pt 
	{\includegraphics[width = 1.6in]{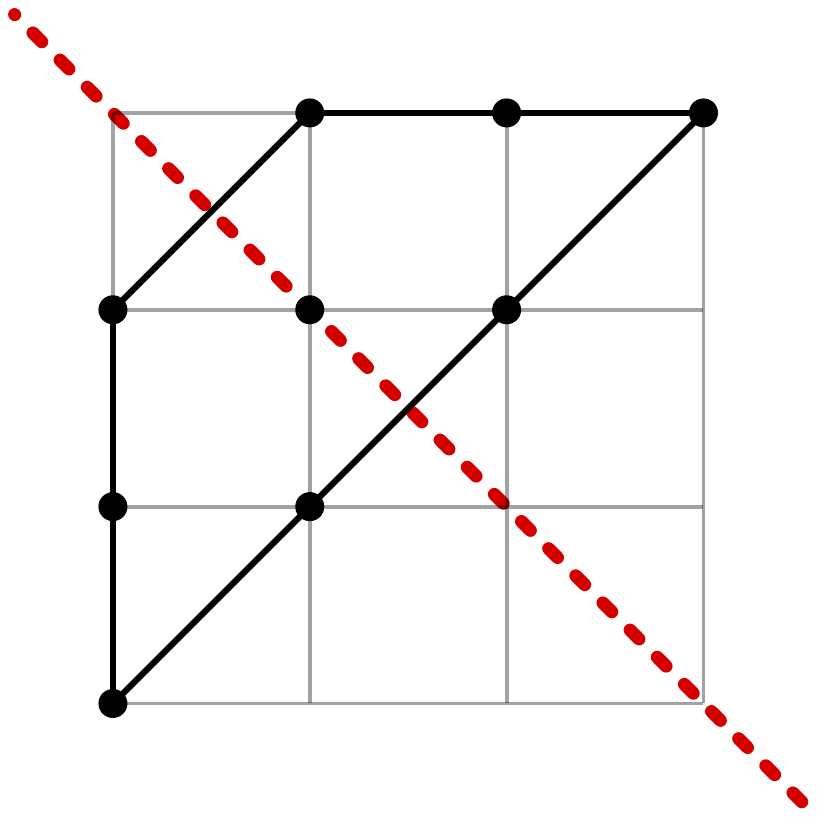}}
	\caption{Examples of trapezoids, which admit anomaly-free fixed line orientifolds. }
	\label{3trapezoids}
\end{figure}

\subsection{Horizontal/Vertical Line Orientifolds}\label{horizontalline}

In this section we consider horizontal fixed lines. The case of vertical lines is trivially related by rotation. The reasoning is essentially the same as the one described previously for the case of diagonal lines. This allows us to go fast to the main result for this class of orientifolds. In particular, we will not comment here about rectangles and faces with many tensors since the previous results are easily generalized.

Horizontal symmetry lines in the dimer correspond to a vertical symmetry in the toric diagram. The $\mathbb{Z}_2$ action maps a ZZP with winding $(p,q)$ to a ZZP with winding $(-p,q)$. Again, we distinguish two different types of ZZPs:
\begin{itemize}
	\item Pairs of ZZPs $\lbrace v_\alpha,v_{\overline \alpha}\rbrace$ for $\alpha=1,...,l$, where $v_\alpha$ and $v_{\overline \alpha}$ are exchanged under the symmetry, thus not parallel to the axis of symmetry.
	\item Self-identified ZZPs $\lbrace v_\gamma \rbrace$ for $\gamma=1,...,l_\parallel$, with winding numbers $(0,1)$ or $(0,-1)$.
\end{itemize}
A general illustration of this is depicted in \fref{horizontal}.

\begin{figure}[h!]
	\centering
	\includegraphics[scale=0.5]{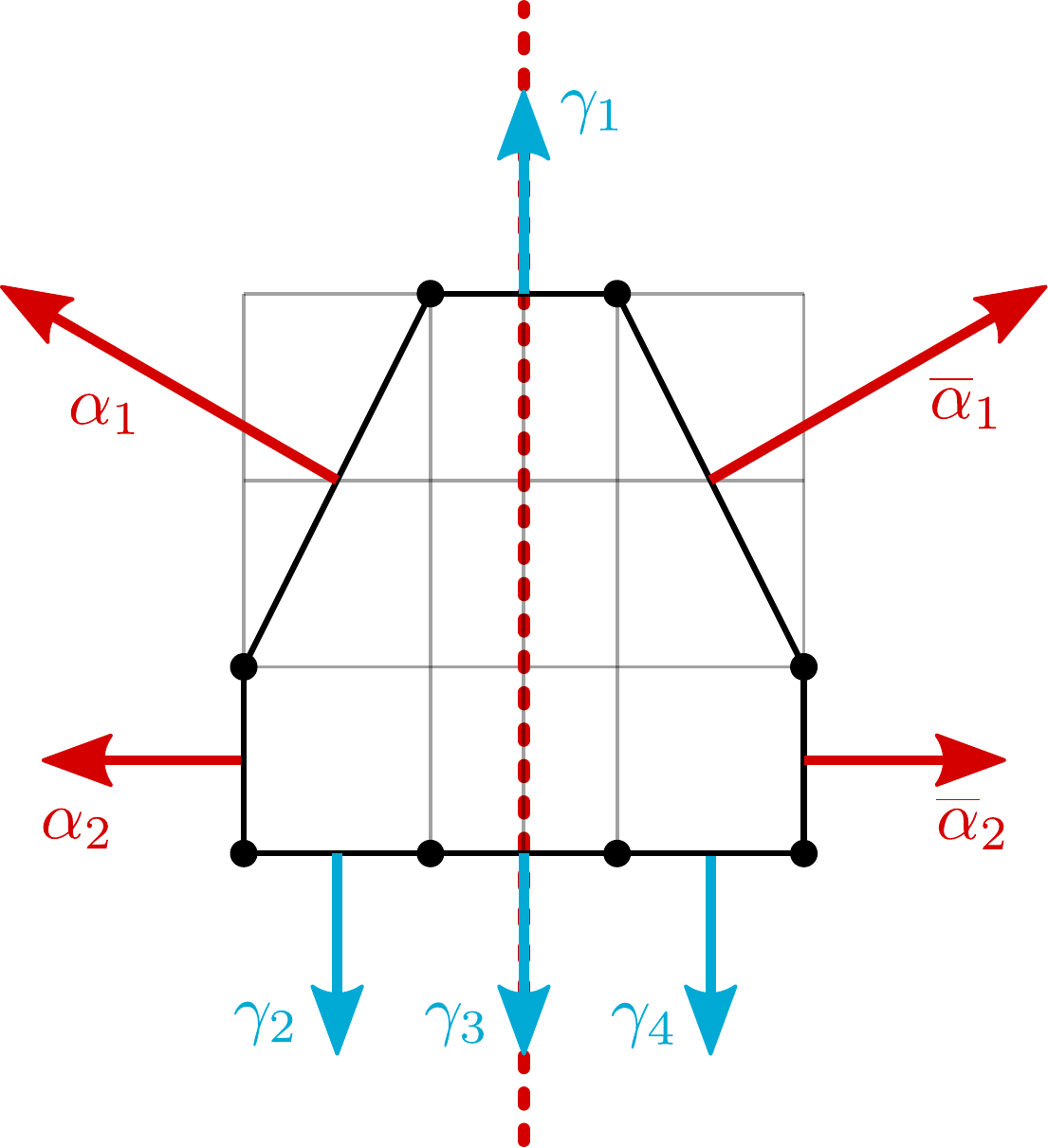}
	\caption{A generic singularity with a vertical axis symmetry.}\label{horizontal}
\end{figure}

In order to find the antisymmetric solutions to the ACC, we need to look at the antisymmetric value assignments of the ZZPs and impose the topological constraint
\begin{equation}\label{antiLambdahor}
\L=2\sum_{\alpha} v_{\alpha}p_{\alpha}=0 \ .
\end{equation}
Let us now consider the Rouch\'e-Capelli condition. A ZZP of type $\alpha$ with winding numbers $(p,q)$ crosses both fixed lines $-q$ times, counted with sign. The Rouch\'e-Capelli condition can be expressed as
\begin{equation}\label{RChor}
\sum_i N_i f_i= - \sum_{\alpha} v_{\alpha}q_{\alpha}(4\ \text{sign}(A)+4\ \text{sign}(B))=0 \,,
\end{equation}
where $\text{sign}(A)$ and $\text{sign}(B)$ indicate the signs of the two fixed lines. Unlike the case of diagonal lines, the Rouch\'e-Capelli condition in \eqref{RChor} becomes trivial as soon as $\text{sign}(A)$ and $\text{sign}(B)$ are different. In that case, the orientifold theory is always anomaly-free.

If the two fixed lines have the same sign, \eqref{antiLambdahor} allows us to express $v_{1}$ in terms of the remaining $v_{\alpha}$, as in the case of diagonal lines. Plugging this expression into \Cref{RChor} leads to
\begin{equation}
\sum_{\alpha \neq 1} v_{\alpha}\left(p_{\alpha} q_1-p_{1}q_{\alpha}\right)=0 \, . 
\end{equation}
With the same analysis of the previous section, we find that singularities with two horizontal lines of the same sign admit a solution to the ACC only if they are trapezoids, just as in the case of diagonal lines.  See \Cref{3trapezoidsv} for examples.
\begin{mdframed}[backgroundcolor=blue!20]
	\textbf{Horizontal/vertical line orientifold:} Toric diagrams symmetric with respect to a horizontal/vertical axis always lead to anomaly-free orientifold theories when the two O-lines have \textit{opposite signs}. When the signs are the same, instead, in order to yield a non-anomalous orientifold theory the toric diagram of the singularity must be a \textit{trapezoid}. 
\end{mdframed}
\begin{figure}[h!]
	\centering
	{\includegraphics[width = 1.6in]{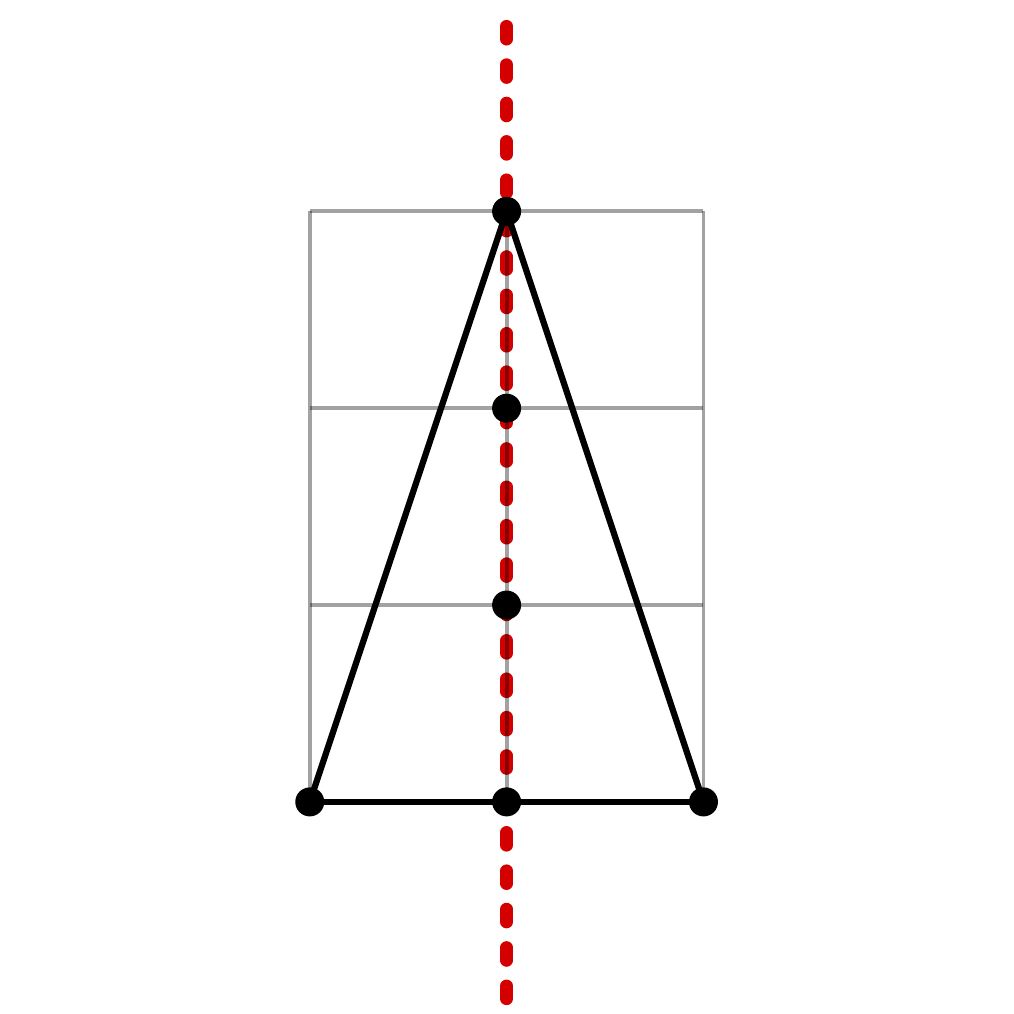}}\hskip 15pt 
	{\includegraphics[width = 1.6in]{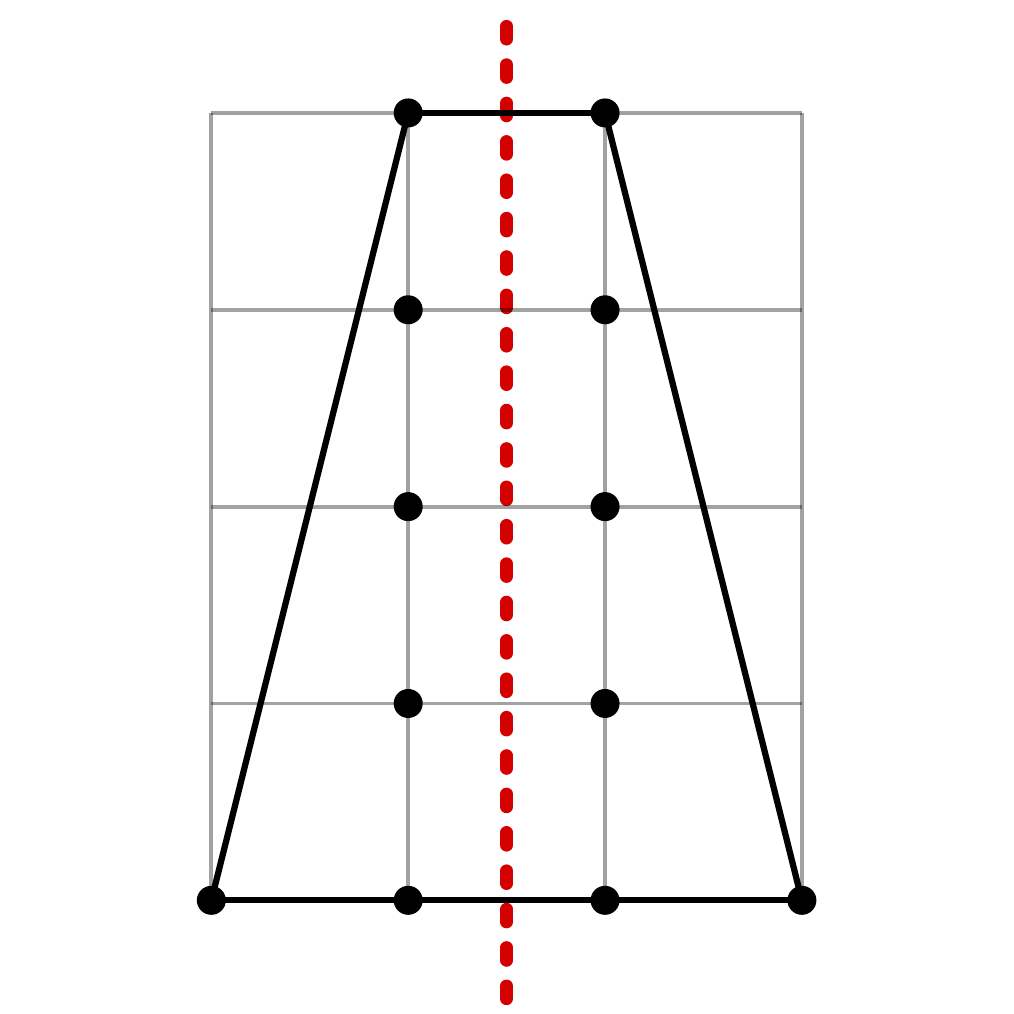}} \hskip 15pt 
	{\includegraphics[width = 1.6in]{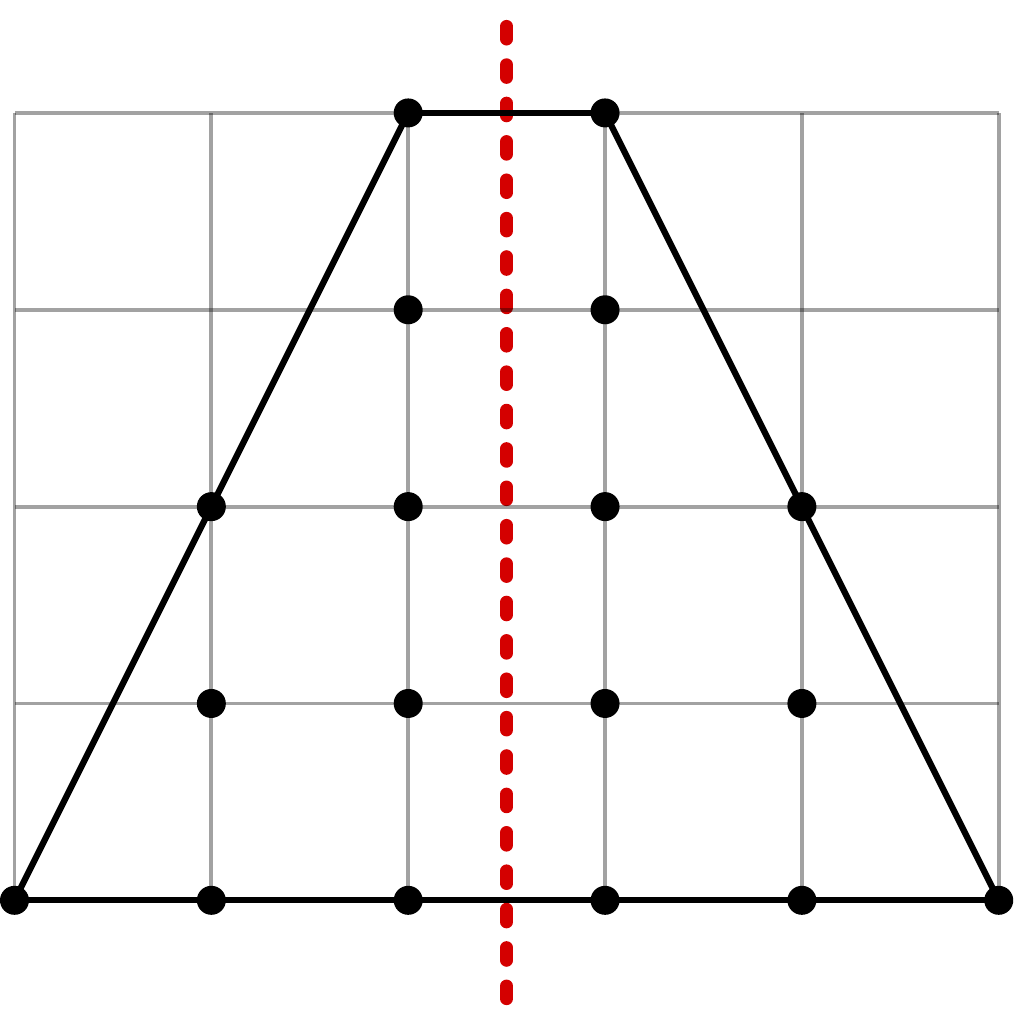}}
	\caption{Examples of trapezoids, which admit anomaly-free horizontal fixed line orientifolds. }
	\label{3trapezoidsv}
\end{figure}

\subsection{Fixed Point Orientifolds} \label{Subsec:FixedPoints}

Finally, we address the case of fixed point orientifolds. We 
should state right away that the results in this case are less conclusive than for fixed lines. Indeed, one can easily anticipate that having a larger number of signs to fix (at the four fixed points), it will be straightforward to satisfy the ACC just by a wise choice, similarly to the case with horizontal/vertical line. On the other hand, if one sticks with a `wrong' choice, the restriction on the allowed geometries is not as nicely characterizable as in the previous case.

As already explained in \sref{Sec:Algorithm}, the action of the orientifold on every ZZP is to map it either to itself or to another ZZP with the same winding numbers. We thus divide the ZZPs into two sets:
\begin{itemize}
	\item Pairs of distinct ZZPs $\lbrace v_\alpha, v_{\overline \alpha} \rbrace$ for $\alpha=1,\cdots, k $ that are exchanged. 	
	\item Self-identified ZZPs $\lbrace v_\gamma \rbrace$, for $\gamma=1,\cdots, l$. 
\end{itemize}
The total number of ZZPs is $n=2k + l$. 

In this kind of orientifolds, tensors arise whenever a pair of self-identified ZZPs intersect over a fixed point. Moreover, a ZZP going through a fixed point necessarily goes through a second fixed point \cite{Retolaza:2016alb}. As a result, it is easy to convince oneself that the number of tensors, if present at all, must be between 2 and 4, and it coincides with the total number of self-identified ZZPs that cross a fixed point.

In order to find the antisymmetric solutions to the ACC, we need to consider symmetric value assignments for the ZZPs, as explained in \sref{Sec:Algorithm}, subject to the topological constraints
\begin{equation}\label{toppoint}
\begin{array}{ccccl}
\L &= & 2 \sum_{\alpha}v_{\alpha}p_{\alpha}+\sum_{\gamma} v_{\gamma} p_{\gamma} &= &  0 \, , \\[.15 cm]
M & = & 2 \sum_{\alpha} v_{\alpha}q_{\alpha}+\sum_{\gamma} v_{\gamma} q_{\gamma} &= & 0 \,  .
\end{array}
\end{equation} 
The RC equation becomes
\begin{equation}\label{RCpoint}
\sum_i N_i f_i= \sum_{\gamma \neq \gamma'}(v_{\gamma}-v_{\gamma'})(\pm 4)=0 \, .
\end{equation}
where the sum in the middle runs over the tensors. The signs depend on the sign of the fixed points and on the orientations of the self-identified edges. Depending on which of the two faces adjacent to the edge we preserve in the projection, we get tensors or their conjugates, contributing with opposite signs to the ACC.

We recall that the signs of the fixed points, in contrast with fixed lines, are constrained by the \textit{sign rule} \cite{Franco:2007ii}. The rule prescribes that the product of the four signs is $(-1)^{n_W/2}$, with $n_W$ the number of superpotential terms.\footnote{Generically, it is not known whether the parity of $n_W/2$ can be deduced from the toric diagram.}

We now consider the different possibilities, i.e. $l=2$, 3 and 4 tensors. Our analysis is general and does not distinguish between faces with single or multiple tensors.

\begin{itemize}
	\item $l=2$: In this case we have two tensors, meaning that two ZZPs cross each other on two fixed points. \Cref{RCpoint} reads
	\begin{equation}\label{l=2}
	(v_{1}-v_{2})(\pm_1 4)\pm(v_1 - v_{2})(\pm_2 4) = 0\, ,
	\end{equation}
	where the $\pm_i$ indicate the signs of the fixed points, while the additional $\pm$ signs depends on whether the tensors are conjugated or not. 

Since only two fixed points are involved in this case, their signs can always be chosen such that this equation is trivially satisfied, while satisfying the sign rule. However, it is interesting to consider whether there are other ways to satisfy this constraint. We can impose $v_1=v_2$ using the two equations of the topological constraint. Expressing $v_1$ and $v_2$ as function of the other $v_\alpha$'s we get 
\begin{align}\label{toppoint}
v_{1} & =  \frac{2}{p_1 q_2 - p_2 q_1} (p_2  \sum_\alpha v_\alpha q_\alpha - q_2 \sum_\alpha v_\alpha p_\alpha)   \, , \nonumber \\ 
v_{2}  & =  \frac{2}{p_1 q_2 - p_2 q_1} (q_1  \sum_\alpha v_\alpha p_\alpha - p_1 \sum_\alpha v_\alpha q_\alpha)    \, ,
\end{align}
where we have assumed $p_1 q_2 - p_2 q_1\neq 0$. Equating $v_1$ and $v_2$, we obtain
\begin{align}
\sum_\alpha v_\alpha (p_\alpha(q_1+q_2)-q_\alpha (p_1+p_2))=0   \, .
\end{align}
Since this equation must hold for all $v_\alpha$, the only possibility is that all terms in the summation vanish, thus $p_\alpha(q_1+q_2)=q_\alpha (p_1+p_2)$ for all $\alpha$. Solutions are of the form $p_1=-p_2$ and $p_\alpha=0$, up to $SL(2, \mathbb{Z})$ transformations. Those correspond to trapezoids (not necessarily symmetric with respect to any axis) with an even number of ZZPs on each base and only one ZZP on each side.

If $p_1 q_2 - p_2 q_1= 0$, it means that $(p_1,q_1)=-(p_2,q_2)$, since the two ZZPs are parallel and, in order to intersect in a consistent way, they must have opposite winding numbers. In this case, the topological constraint imposes $v_1=v_2$ if $p_\alpha q_{\alpha^\prime} - q_\alpha p_{\alpha^\prime}=0$ where $\alpha \neq \alpha^\prime$. It means that all non self-identified ZZPs have to be either parallel or anti-parallel to each other. This condition is satisfied by all toric diagrams with the shape of a rectangle or a parallelogram where there is an even number of non self-identified ZZPs. Together with the solutions of the previous paragraph, they constitute a class of trapezoids for which any sign assignment for the fixed points leads to an anomaly free theory when two tensors are involved.

As an illustration, consider fixed point orientifolds of $\mathbb{C}^3/\mathbb{Z}_6$ with actions (1,1,4) and (1,2,3), whose toric diagrams are shown in \fref{Fig:Z6}. Both of them admit an orientifold with two tensors. 
Our analysis implies that only the first one admits tensors with any sign, as it can easily be checked by explicitly solving the ACC.
		\begin{figure}[h!]
			\centering
			\begin{subfigure}[t]{0.23\textwidth }
				\begin{center} 
					\includegraphics[width=\textwidth]{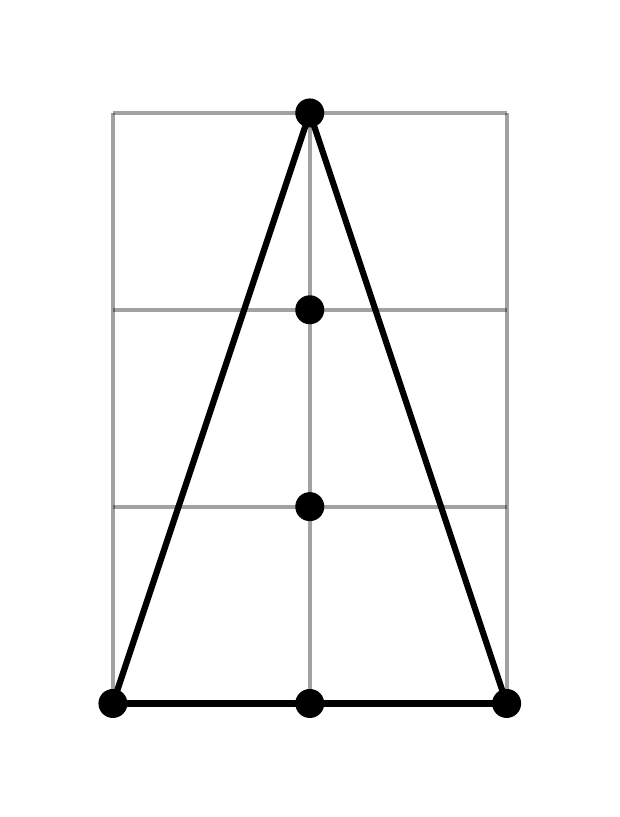}
					\caption{}
					\label{Fig:toric114}
				\end{center}
			\end{subfigure} \hspace{15mm}
			\begin{subfigure}[t]{0.23\textwidth } 
				\begin{center} 
					\includegraphics[width=\textwidth]{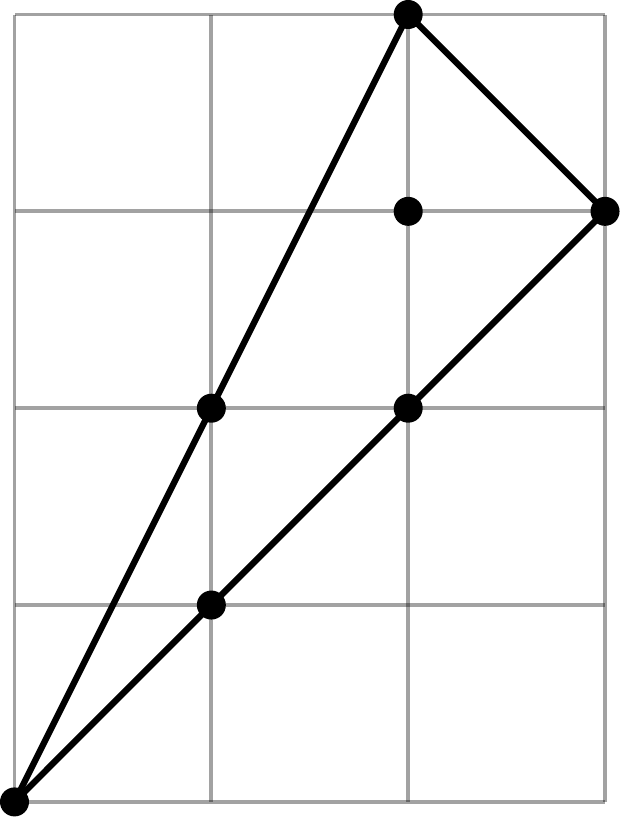}
					\caption{}
					\label{Fig:WebDiagramPdP4}
				\end{center}
			\end{subfigure}
			\caption{The toric diagrams for the $\mathbb{C}^3 / \mathbb{Z}_6$ orbifolds with actions: (a) (1,1,4) and (b) (1,2,3).}\label{Fig:Z6} 
		\end{figure}
		
An interesting scenario is when tensors arise from the orientifold projection of adjoints in the mother theory, namely from edges separating self-identified faces.  In this case, the ACC of the self-identified gauge group is trivially zero, since it is either $SO$ or $USp$. In this situation, the two self-identified ZZPs intersect all other ZZPs only once. This can be understood as follows. Let us consider a line passing through the fixed points under consideration.   
All the non self-identified ZZPs must be parallel to this line, since otherwise their intersections with the line would imply that they go through the self-identified face, which in turn would spoil the fact that it is self-identified. The $\mathbb{C}^2/\mathbb{Z}_{2m}$ orbifolds are examples in this class, see \fref{Fig:Z2m}.

		\begin{figure}[h!]
			\centering
			\includegraphics[scale=0.6]{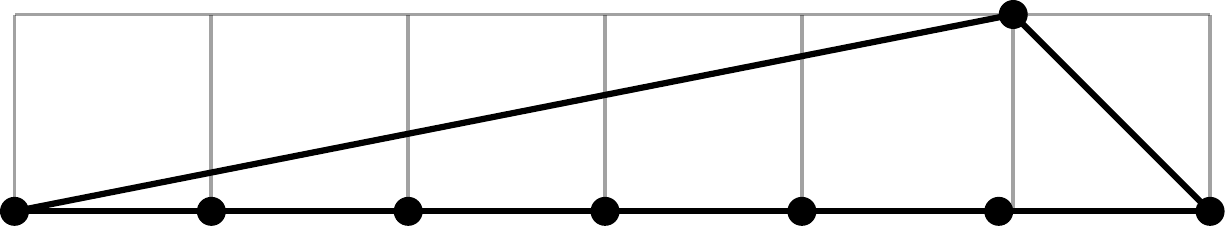}
			\caption{The toric diagram of $\mathbb{C}^2/ \mathbb{Z}_6$, as an example of the $\mathbb{C}^2/ \mathbb{Z}_{2m}$ family.}\label{Fig:Z2m}
		\end{figure}

	\item $l=3$: In this case we have three tensors, i.e. three ZZPs intersecting on three fixed points. \Cref{RCpoint} reads
	\begin{align}\label{l=3}
	(v_{1}-v_{2})(\pm_1 4)\pm(v_{2}-v_{3})(\pm_2 4) \pm (v_{3}-v_{1})(\pm_3 4)=0 \ .
	\end{align}
	Since only three of the fixed points are involved, it is possible to pick their signs such that this equation is trivially satisfied. These choices in turn determine the sign of the fourth fixed point due to the sign rule. 
	
	If instead we have a different combination of signs, we end up with an equation of the form
	\begin{equation}
	v_{\gamma} - v_{\gamma^\prime} = 0 \, ,
	\end{equation}
	with $\gamma$ and $\gamma'$ two of the three ZZPs above. The missing $v_{\gamma^{\prime\prime}}$ in the previous equation depends on the choice of fixed point signs in \eref{l=3}. Therefore, in order to have a solution for all possible fixed point sign assignments we need to impose $v_1=v_2=v_3$ with the topological constraint. This means that the ZZPs have winding numbers of the form $(p_1,0)$, $(-p_1,q_2)$ and $(0,-q_2)$, up to $SL(2,\mathbb{Z})$ transformations.  The only solution is $p_1=q_2=1$, corresponding to $\mathbb{C}^3$, i.e. flat space.

A face with multiple tensors imposes constraint(s) of the form $v_1-v_2=\pm (v_2-v_3)$, leading to an RC constraint of the form
	\begin{align}\label{l=3}
	(v_{1}-v_{2})(\pm_1 4)\pm(v_{1}-v_{2})(\pm_2 4) \pm (v_{3}-v_{1})(\pm_3 4)=0 \ .
	\end{align}
Again, the existence of solutions depends on the signs of the fixed points. Solutions for generic signs can be obtained only when $v_1=v_2=v_3$, i.e. for flat space.

	\item $l=4$: 
	
This case, in contrast with the previous ones, does not always admit a solution to the ACC. The reason for this is that the four fixed points are used, their signs are constrained by the sign rule and we no longer have the freedom of unused fixed points. 
	
	The RC equation can take two different forms, depending on the ZZP intersections:
	\begin{align}
	&(v_{1}-v_{2})(\pm_1 4)\pm (v_2 - v_{3})(\pm_2 4)\pm (v_3 - v_{4})(\pm_3 4)\pm (v_4 - v_{1})(\pm_4 4) = 0 \ , \nonumber \\
	&(v_{1}-v_{2})(\pm_1 4)\pm (v_1 - v_{2})(\pm_2 4)\pm (v_3 - v_{4})(\pm_3 4)\pm (v_3 - v_{4})(\pm_4 4) = 0 \ .
	\end{align}
Since the signs of the fixed points are constrained, it is not always possible to trivially solve the RC equation. 

Moreover, it is also impossible to find general non-trivial solutions by using the topological constraint to force some of the $v_i$ to be equal. For the first equation, we need all the $v_i$ to be equal. To do so, we need at least three equations, but the topological constraint provides only two. In the second case, we can impose $v_1=v_2$ and $v_3=v_4$ with the following ZZPs: $(1,0)$, $(-1,0)$, $(0,1)$ and $(0,-1)$, which define the conifold singularity. Unfortunately, the conifold gives rise to an RC of the first kind, not of the second one. 

\end{itemize}

To conclude, this partial analysis retained only one toric diagram that can accommodate any signs for its fixed points: flat space. We eventually found some particular trapezoids for which we can freely chose the signs of the tensors when only two are present, but those singularities also allow \textit{in principle} for fixed point orientifolds with four tensors, where our analysis showed its limits.  Thus, we cannot say in general that they provide every kind of anomaly-free orientifolds. As an illustrative example, one can check that the orientifold of \fref{Fig:toric114} with four tensors does not allow for every combination of signs satisfying the sign rule, although it does with only two tensors.

It would be interesting to investigate further whether it is possible to determine the solvability of the ACC from the toric diagram. We leave this question for future work. In the meantime, orientifolds with four self-identified ZZPs need to be studied in a case by case basis.

\section{Conclusions}
\label{Sec:Conc}

In this paper we studied anomalies in gauge theories living on D-branes probing orientifolds of toric singularities, focusing on pure D3-brane theories, namely without the addition of extra flavors. 

We introduced a new, geometric algorithm for finding anomaly-free solutions based on zig-zag paths. The main virtue of this procedure is not so much its practicality over the direct solution of the ACC in explicit examples, but the fact that it allows us to make general statements regarding anomalies directly from geometry. Indeed, we managed to derive stringent no-go theorems that establish the conditions for anomaly-free solutions in these orientifolds. Such results are extremely useful, since until now the cancellation of anomalies in this class of theories was analyzed on a case-by-case basis.

We can summarize our findings as follows, 
from the most stringent case to the less conclusive one:
\begin{itemize}
	\item 
	For orientifolds with a fixed diagonal line, for which one has to choose only one sign, we find that only singularities whose toric diagram is a trapezoid with respect to the diagonal axis of symmetry allow for a non-anomalous D-brane gauge theory. 
	\item
	For orientifolds with fixed horizontal lines, we have two signs to choose. All singularities lead to anomaly-free theories if the two signs are chosen to be opposite to each other. If the singularity has a toric diagram which is a trapezoid with respect to the vertical axis of symmetry, then the theory is non-anomalous also for equal signs. 
	\item
	For orientifolds with fixed points, there are four signs to choose, up to a constraint on their product. Moreover, the relation between the fixed points in the dimer and the toric diagram of the singularity is less direct. Because of these two facts, it is more difficult to summarize the few instances where a restriction is indeed obtained on the singularities that lead to non-anomalous theories. The particular cases have been detailed in \sref{Subsec:FixedPoints}.
\end{itemize}

As an illustration of the power of the ideas introduced in this work, they were exploited in \cite{Argurio:2020dkg,Argurio:2020npm} to guide the search of models of D-branes at singularities that display dynamical supersymmetry breaking. Such models necessarily involve orientifolds, but have a potential instability as soon as the singularity allows for a partial resolution which is non-isolated (in D-brane jargon, this translates to the presence of ${\cal N}=2$ fractional branes \cite{Buratti:2018onj}). In terms of the toric diagram, this property manifests itself through points within the external edges of diagram, or in other words, parallel ZZPs. It is straightforward to see that toric diagrams that fall in the class of trapezoids always include such points on the boundary, except for few very simple cases (namely $F_0$ and orbifolds with a toric diagram which is an isosceles triangle with a unit base). As a consequence, if one is to look for fixed line orientifolds which allow for anomaly free D-brane configuration, and with no non-isolated partial resolution, the only option one is left with is horizontal/vertical fixed lines with opposite signs. Indeed, the octagon singularity \cite{Argurio:2020dkg} is the simplest non-trivial singularity that satisfies these requirements.

\section*{Acknowledgements}

V.T. is very thankful to V. Fock for explaining him the methods of \cite{fock2015inverse}. These have been a great source of inspiration for the development of the techniques presented above. R.A., A.P. and E.G.-V. acknowledge support by IISN-Belgium (convention 4.4503.15) and by the F.R.S.-FNRS under the ``Excellence of Science" EOS be.h project n.~30820817, M.B. and S.M. by the MIUR PRIN Contract 2015 MP2CX4 ``Non-perturbative Aspects Of Gauge Theories And Strings" and by INFN Iniziativa Specifica ST\&FI. E.G.-V. was also partially supported by the ERC Advanced Grant ``High-Spin-Grav". The research of S.F. was supported by the U.S. National Science Foundation grants PHY-1820721 and DMS-1854179. R.A. is a Research Director and A.P. is a FRIA grantee of the F.R.S.-FNRS (Belgium).

\bibliographystyle{JHEP}
\bibliography{ref}


\end{document}